\documentclass[aps,prc,floatfix,showpacs,showkeys,nofootinbib]{revtex4}

\usepackage{amsmath,amssymb}
\usepackage{mathrsfs}
\usepackage[pdftitle={Algebraic Collective Model},
            pdfauthor={T.A. Welsh and D.J. Rowe}]{hyperref}

\DeclareRobustCommand{\Hb}{\mathbb{H}}    
\DeclareRobustCommand{\Hbt}{\mathbb{H}^{\text{trunc}}}    
\DeclareRobustCommand{\Rb}{\mathbb{R}}

\DeclareRobustCommand{\calQ}{\mathcal{Q}}
\DeclareRobustCommand{\hcalQ}{\hat{\mathcal{Q}}}

\DeclareRobustCommand{\scrD}{\mathscr{D}}
\newcommand{\sen}{v}           
\newcommand{\pot}{V}           
\newcommand{\rmu}{\mu}        
\newcommand{\rnu}{\nu}        
\newcommand{\sumer}{\xi}   
\newcommand{\summe}{\sigma}   

\def\afrac#1/#2{\leavevmode\kern.1em
\raise.5ex\hbox{\the\scriptfont0 #1}\kern-.1em
/\kern-.15em\lower.25ex\hbox{\the\scriptfont0 #2}}

\newcommand{\WaveFun}[7]{\Psi^{(#2,#1)}_{#3;#4 #5 #6 #7}}
\newcommand{\StateKet}[7]{{|(#2,#1)\,#3;#4 #5 #6 #7\rangle}}
\newcommand{\StateKetR}[6]{{|(#2,#1)\,#3;#4 #5 #6\rangle}}
\newcommand{\StateKetRR}[4]{{|(#2,#1)\,#3;#4\rangle}}
\newcommand{\StateBra}[7]{{\langle(#2,#1)\,#3;#4 #5 #6 #7|}}
\newcommand{\StateBrax}[7]{{\langle(#2,#1)\,#3;#4 #5 #6 #7}}
\newcommand{\StateBraR}[6]{{\langle(#2,#1)\,#3;#4 #5 #6|}}
\newcommand{\StateBraRR}[4]{{\langle(#2,#1)\,#3;#4|}}
\newcommand{\RadialKet}[3]{{|(#2,#1)\,#3\rangle}}
\newcommand{\RadialBra}[3]{{\langle(#2,#1)\,#3|}}
\newcommand{\RadialBrax}[3]{{\langle(#2,#1)\,#3}}

\newcommand{\SphericalKet}[4]{{|#1 #2 #3 #4\rangle}}
\newcommand{\SphericalKetR}[3]{{|#1 #2 #3\rangle}}
\newcommand{\SphericalKetRR}[1]{{|#1\rangle}}
\newcommand{\SphericalBra}[4]{{\langle #1 #2 #3 #4|}}
\newcommand{\SphericalBraR}[3]{{\langle #1 #2 #3|}}
\newcommand{\SphericalBraRR}[1]{{\langle #1|}}

\newcommand{\CG}[6]{{(#1 #2\,#3 #4 | #5 #6)}}

\newcommand{\Radial}[3]{{\mathcal R}^{(#2,#1)}_{#3}}  
\newcommand{\Wadial}[3]{{P^{(#2)}_{#3,#1}}}
\newcommand{\Sphericalu}[4]{{{\mathcal Y}^{#1}_{#2 #3 #4}}}
\newcommand{\hSphericalu}[4]{{\hat{\mathcal Y}^{#1}_{#2 #3 #4}}}

\newcommand{\hTphericalu}[4]{{\hat{T}^{#1}_{#2 #3 #4}}}

\newcommand{\SphericalHil}{\mathcal{L}^2(S_4,\sin3\gamma\,d\gamma\,d\Omega)}
\newcommand{\RadialHil}{\mathcal{L}^2(\Rb_+,\beta^4\,d\beta)}
\newcommand{\RadialHilc}{\mathcal{L}^2(\Rb_+,d\beta)}

\newcommand{\FOp}[3]{{F^{(#1)}_{#2;#3}}} 
\newcommand{\FOpb}[4]{{F^{(#1)}_{#2;#3}\left(#4\right)}} 
\newcommand{\SUOp}[3]{{\hat S^{(#2,#1)}_{#3}}}

\newcommand{\argsubsup}[3]{
        {\textit{#1}{\mspace{1mu}}^{#3}_{\mathrm{#2}}}}  
\newcommand{\argsub}[2]{
        {\textit{#1}{\mspace{1mu}}_{\mathrm{#2}}}}  
\newcommand{\argit}[1]{\textit{#1}}  

\newcommand{\Lvals}[1]{{\argit{Lvals}\mspace{0.5mu}[#1]}}
\newcommand{\Melements}[4]{{\argit{Melements}%
                                \mspace{0.5mu}[#1,#2]\mspace{0.5mu}[#3,#4]}}
\newcommand{\eigenvals}[2]{{\argit{eigenvals}%
                                \mspace{0.5mu}[#1]\mspace{0.5mu}[#2]}}
\newcommand{\countlab}{{n}} 

\newcommand{\Lini}{{\argsub{L}{i}}}
\newcommand{\sini}{{\argsub{$\sen$}{i}}}
\newcommand{\alphaini}{{\argsub{$\alpha$}{i}}}
\newcommand{\Mini}{{\argsub{M}{i}}}
\newcommand{\ctini}{{\argsub{\countlab}{i}}}
\newcommand{\kini}{{\argsub{k}{i}}}
\newcommand{\Lfin}{{\argsub{L}{f}}}
\newcommand{\sfin}{{\argsub{$\sen$}{f}}}
\newcommand{\alphafin}{{\argsub{$\alpha$}{f}}}
\newcommand{\Mfin}{{\argsub{M}{f}}}
\newcommand{\ctfin}{{\argsub{\countlab}{f}}}
\newcommand{\kfin}{{\argsub{k}{f}}}

\newcommand{\fivesupthree}{SO(5)$\,\supset\,$SO(3)}

\newcommand{\elrm}[1]{#1}      

\begin{document}

\title{A computer code for calculations in the\\
          algebraic collective model of the atomic nucleus}

\author{T.A.~Welsh}


\author{D.J.~Rowe}

\affiliation{%
  Department of Physics,
  University of Toronto, Toronto,
  Ontario M5S 1A7, Canada}

\begin{abstract}
A Maple code is presented for algebraic collective model (ACM) calculations.
The ACM is an algebraic version of the Bohr model of the atomic nucleus,
in which all required matrix elements are derived by exploiting the
model's SU(1,1)$\times$SO(5) dynamical group.
This paper reviews the mathematical formulation of the ACM,
and serves as a manual for the code.

The code enables a wide range of model Hamiltonians to be analysed.
This range includes essentially all Hamiltonians that are rational functions
of the model's quadrupole moments $\hat q_M$ and are at most quadratic in
the corresponding conjugate momenta $\hat\pi_N$ ($-2\le M,N\le 2$).
The code makes use of expressions for matrix elements derived elsewhere
and newly derived matrix elements of the operators
$[\hat\pi\otimes\hat q\otimes\hat\pi]_0$
and $[\hat\pi\otimes\hat\pi]_{LM}$.
The code is made efficient by use of an analytical expression for
the needed SO(5)-reduced matrix elements,
and use of \fivesupthree\ Clebsch-Gordan coefficients
obtained from precomputed data files provided with the code.
\end{abstract}





\maketitle


\section{Introduction}

The ACM (algebraic collective model) 
\cite{ACM1,ACM2,ACM3,RWC09}
is an algebraic version of the Bohr model \cite{Bohr52} 
based on a dynamical group $\mathrm{SU(1,1)}\times\mathrm{SO(5)}$
for which all the matrix elements needed in applications of the model
are calculated analytically.
It is a development of the \emph{computationally tractable version
of the collective model} \cite{ACM1} 
that enables collective model calculations to be
carried out efficiently by use of
wave functions that span modified oscillator series of SU(1,1) irreps
(irreducible representations)
\cite{CP77,CW02} and complementary SO(5) wave functions.
The availability of analytic SU(1,1) matrix elements
and SO(5) Clebsch-Gordan coefficients enable the calculations to
bypass expressions for the wave functions entirely,
avoiding, in particular, numerical integration.
A pedagogical treatment of the geometrical and algebraic foundations of
the ACM is given in the recent book by Rowe and Wood \cite{RowanWood}.

Earlier computer programs that implemented the Bohr
model utilised an U(5) $\supset$ O(5) $\supset$ SO(3) $\supset$ SO(2) basis
\cite{CMS76,CMS77,EG87}.
The basis was constructed by starting with a basis of SO(3) coupled
polynomials of a given degree in the quadrupole coordinates
and diagonalising the O(5) Casimir operator in this basis.
The computer program developed by Gneuss and Greiner \cite{GG71}
additionally employed a
U(5) $\supset$ Sp(4) $\supset$ SO(3)$\times$SO(3)
$\supset$ SO(2)$\times$SO(2) basis,
with transformations between the two bases 
carried out using the methods of \cite{Hecht65}.
On being further developed,
the calculations could be carried out fully in the 
U(5) $\supset$ O(5) $\supset$ SO(3) $\supset$ SO(2) basis,
culminating in the Frankfurt code \cite{HSMG80,HMG81,TMH91}.
This was able to analyse model 
SO(3)-invariant Hamiltonians
obtained from three kinetic energy terms and potential energy obtained
from various polynomials in $\beta$ and $\cos3\gamma$
(the collective model coordinates $\beta$ and $\gamma$
are described in Section \ref{Sec:Model} below).
In the Frankfurt code, the calculation of matrix elements was partially
carried out using numerical integration
(in the terminology of Section \ref{Sec:Model} below,
this numerical calculation was performed on the space $\SphericalHil$).

In the Maple \cite{Maple} code presented here,
the use of files of highly accurate precomputed
\fivesupthree\ Clebsch-Gordan coefficients together with an exact analytic
expression for the SO(5)-reduced matrix elements of SO(5) spherical harmonics
enable these computationally intensive methods to be avoided entirely.
These files of \fivesupthree\ Clebsch-Gordan coefficients were computed
\cite{CRW09,TAW08} using the algorithm developed in \cite{RTR04}.
This algorithm, which also calculates SO(5) spherical harmonics,
was based on the methods of \cite{ACM1}
for calculating model SO(5) wave functions.
A Mathematica code for the calculation of
\fivesupthree\ Clebsch-Gordan coefficients in exact arithmetic has been
published by Caprio \textit{et al.} \cite{CRW09}.

The code presented here also benefits enormously from the use of
modified oscillator SU(1,1) irreps,
as used previously by Davidson \cite{Davidson32} in molecular physics.
As a result, calculations for deformed nuclei converge much more
rapidly than in a conventional harmonic oscillator basis,
as illustrated in \elrm{Appendix }\ref{Sec:OpEx} and \cite{ACM2},
and more general Hamiltonians,
such as those involving negative powers of $\beta$, can be handled.
In addition, the pairing of each SO(5) irrep with a certain
modified SU(1,1) irrep enables
all matrix elements of the basic Bohr model observables,
and many others, to be computed algebraically.

The code is very versatile and can calculate the spectrum
and properties of virtually any Bohr model Hamiltonian
one might wish to consider, quickly and easily.
In addition to contributing to the stockpile of algebraic models
that can be used for exploratory studies of nuclear phenomena,
it is intended that this code will also serve as a resource for
extensions of the Bohr model.
For instance, it is of interest to develop models that include
extra degrees of freedom, such as vorticity degrees of freedom,
particle-core coupling models, the interacting boson model
as recently pursued in \cite{ThiamovaRowe12},
and other models in which SO(5) representations, their matrix elements,
and \fivesupthree\ Clebsch-Gordan coefficients,
can be used with advantage.

This article is organised such that Sections \ref{Sec:Model}--\ref{Sec:CMOMEs}
describe the theoretical framework of the ACM.
The code itself is described in Sections \ref{Sec:Code}--\ref{Sec:Components},
which serve as a manual for its use.
\elrm{Appendix }\ref{Sec:Rigid} describes how to use the code
to perform calculations in the rigid-$\beta$ limit of the ACM.
\elrm{Appendix }\ref{Sec:BasisStuff} discusses methods for obtaining
optimal values of the parameters that specify the bases in
which the calculations are performed (see Section \ref{Sec:Bases}).
\elrm{Appendices }\ref{Sec:Laguerre} and \ref{Sec:ExtraMEs} provide
derivations of certain matrix elements that are not available elsewhere.
\elrm{Appendix }\ref{Sec:Imp} gives a summary of the computer
implementation.
Some concluding remarks are given in Section \ref{Sec:Dis}.


\section{The model space}
\label{Sec:Model}

A number of ingredients contribute to the simplicity of the ACM relative
to the standard formulation of the Bohr model.
The first is the characterization of the nuclear shape by
quadrupole moments instead of surface deformation parameters.
This is important because quadrupole moments are well-defined
quantum mechanical observables; they also have well-defined
microscopic expressions in terms of nucleon coordinates.
Thus, the configuration space of the model is expressed as the real
five-dimensional space $\Rb^5$ of nuclear shapes, defined by complex
quadrupole moments $\{q_M,M=0,\pm1,\pm2\}$,
for which $q^*_M = (-1)^M q_{-M}$.
This space can be assigned a radial $\Rb_+$ coordinate $\beta\ge0$,
given by $\beta^2=\sum_{M=-2}^2 |q_M|^2$,
and $S_4$ spherical coordinates $(\gamma,\Omega)$,
where $S_4$ is the four-dimensional sphere of unit radius.
Here $0\le\gamma\le\pi/3$ is an angle coordinate,
and $\Omega$ labels an SO(3) element,
which may be expressed in terms of Euler angles in the standard way
(see \cite{Edmonds60}, for example).
The quadrupole moments are then expressed as products
$q_M=\beta \calQ_M$, with $\sum_{M=-2}^2 |\calQ_M|^2=1$
and
\begin{equation} \label{Eq:Qexp}
\calQ_M(\gamma,\Omega) =  \cos \gamma \, \scrD^2_{0M}(\Omega) 
+  \frac {1}{\sqrt{2}}\sin\gamma \left( \scrD^2_{2M}(\Omega) +
  \scrD^2_{-2M}(\Omega) \right),
\end{equation}
where $\scrD^L_{KM}(\Omega)$ is a Wigner $\scrD$-function
\cite{Edmonds60}.  

The volume element, $d^5x$, for $\Rb^5$ is a product of volume elements
for $\Rb_+$ and $S_4$ given by
\begin{equation}\label{Eq:d5x}
d^5x = (\beta^4\,d\beta) \times (\sin3\gamma \, d\gamma\, d\Omega),
\end{equation}
where the SO(3) volume element $d\Omega$ is normalised such that
$\int_{\mathrm{SO(3)}}d\Omega=8\pi^2$.
Thus, the Hilbert space $\Hb$ of the ACM is expressed as a tensor product
\begin{equation}\label{Eq:HilbertF}
\Hb = \RadialHil\otimes\SphericalHil ,
\end{equation}
where $\RadialHil$ is the Hilbert space of square-integrable
functions on $\Rb_+$ with {respect to the} volume element
$\beta^4\,d\beta$, and $\SphericalHil$ is the Hilbert space
of square-integrable functions on $S_4$ with {respect to the}
volume element $\sin3\gamma\,d\gamma\,d\Omega$.

The model becomes an algebraic model on introduction of orthonormal bases  
$\{\RadialKet{\lambda}{a}{\rnu}\}$ and $\{\SphericalKet{\sen}{\alpha}{L}{M}\}$
for the factors $\RadialHil$ and $\SphericalHil$ of \eqref{Eq:HilbertF},
whose elements are labelled by the quantum numbers
of the groups in the respective dynamical subgroup chains
\begin{equation}
\label{Eq:SU11chain}
\begin{array}{cccccccccc}
{\rm SU(1,1)}&\supset &{\rm U(1)} \\
\lambda&&\rnu
\end{array} ,
\end{equation}
and
\begin{equation}
\label{Eq:SO5chain}
\begin{array}{cccccccccc}
{\rm SO(5)}&\supset &{\rm SO(3)}&\supset& {\rm SO(2)}
 \\
\sen&\alpha&L&&M
\end{array} .
\end{equation}
The parameter $a$ in the basis $\{\RadialKet{\lambda}{a}{\rnu}\}$
for $\RadialHil$ is a useful scale parameter that implicitly defines the 
$\mathrm{U(1)}\subset\mathrm{SU(1,1)}$ subgroup
(see Section~\ref{Sec:RadMEs}).
The group SO(5) in the chain \eqref{Eq:SO5chain} is the group of
linear transformations of the five quadrupole moments $\{q_M\}$
that leave $\beta^2$ invariant, and SO(3) is the rotational subgroup
that transforms the quadrupole moments as a basis
for the 5-dimensional $L=2$ irrep.

An extra `missing label'  $\alpha$ in the range  $1\le\alpha\le d_{\sen L}$
is needed to distinguish the multiplicity, $d_{\sen L}$, of SO(3) irreps of the 
same angular momentum $L$ in an SO(5) irrep of seniority $\sen$
(seniority is the SO(5) analogue of angular momentum).
This multiplicity is given \cite{KishTamura71,CRW09} by
\begin{equation}\label{Eq:DimSO5>SO3}
d_{\sen L}
=(\lfloor\tfrac13(\sen-b)\rfloor +1)\theta_{\sen-b}
-\lfloor\tfrac13(\sen-L+2)\rfloor\theta_{\sen-L+2}\,,
\end{equation}
where $b=L/2$ for $L$ even and $b=(L+3)/2$ for L odd,
$\theta_k=1$ for $k\ge0$ and $\theta_k=0$ for $k<0$,
and $\lfloor x\rfloor$ is the largest integer not greater than $x$.

As we show in the following, the matrix elements of all operators
of interest are given, in the factored basis
$\{\StateKet{\lambda}{a}{\rnu}{\sen}{\alpha}{L}{M}
=\RadialKet{\lambda}{a}{\rnu}\otimes\SphericalKet{\sen}{\alpha}{L}{M}\}$
of $\Hb$,
in terms of easily calculated products of radial and SO(5) matrix elements.
However, it must be understood that whereas the set of states 
$\{\SphericalKet{\sen}{\alpha}{L}{M}\}$,
with seniority $\sen$ taking all integer values $\sen\ge0$,
is an orthonormal basis for $\SphericalHil$,
the set $\{\RadialKet{\lambda}{a}{\rnu}\}$
is an orthonormal basis for $\RadialHil$ for each pair $(a,\lambda)$.
Consequently, when $a'\ne a$ or $\lambda'\ne\lambda$, the overlaps
$\RadialBrax{\lambda'}{a'}{\rmu}\RadialKet{\lambda}{a}{\rnu}$
are not necessarily zero for $\rmu\ne\rnu$.
Thus, to obtain an orthonormal basis for the tensor product space $\Hb$,
and to facilitate exploitation of the SO(5) structure,
one must choose a set of states
$\{\StateKet{\lambda_\sen}{a_\sen}{\rnu}{\sen}{\alpha}{L}{M}\}$
such that $a$ and $\lambda$ take fixed values $a_\sen$ and $\lambda_\sen$
for all states of a given seniority $\sen$.
Such a basis then has the overlaps
\begin{equation}\label{Eq:onbasis}
\begin{split}
\StateBrax{\lambda_{\sen'}}{a_{\sen'}}{\rmu}{\sen'}{\alpha'}{L'}{M'}
\StateKet{\lambda_\sen}{a_\sen}{\rnu}{\sen}{\alpha}{L}{M}
&=
\RadialBrax{\lambda_\sen}{a_\sen}{\rmu}\RadialKet{\lambda_\sen}{a_\sen}{\rnu}\,
\delta_{\sen',\sen} \delta_{\alpha',\alpha} \delta_{L',L} \delta_{M',M}\\
&=
\delta_{\rmu,\rnu}\delta_{\sen',\sen} \delta_{\alpha',\alpha}
  \delta_{L',L} \delta_{M',M},
\end{split}
\end{equation}
and is therefore indeed an orthonormal basis.
In fact, it will be convenient to fix $a_\sen=a$ to have a
single $\sen$-independent value in any given calculation.
The ACM then uses bases
\begin{equation}\label{Eq:BasisVar}
\{\,\StateKet{\lambda_\sen}{a}{\rnu}{\sen}{\alpha}{L}{M}
=\RadialKet{\lambda_\sen}{a}{\rnu}\otimes\SphericalKet{\sen}{\alpha}{L}{M},\,
\rnu\ge0,\sen\ge0,L\ge0,\,
1\le\alpha\le d_{\sen L},\,
-L\le M\le L\,
\}
\end{equation}
of orthonormal states for $\Hb$.


\section{SU(1,1) matrix elements}
\label{Sec:RadMEs}

The basis states
$\{\RadialKet{\lambda}{a}{\rnu}, \rnu = 0,1,2, \dots\}$
for the Hilbert space $\RadialHil$ correspond to wave functions given by
\begin{equation}\label{Eq:DefBeta}
\Big\{
\Wadial{\lambda}{a}{\rnu}(\beta)=
\frac1{\beta^2}\Radial{\lambda}{a}{\rnu}(\beta),
      \rnu=0,1,2,3,\ldots \Big\},
\end{equation}
where
\begin{equation}\label{Eq:DefRadial}
\Radial{\lambda}{a}{\rnu}(\beta)
 = (-1)^\rnu\sqrt{\frac{2\rnu!\, a}{\Gamma(\lambda+\rnu)}} \:
 (a\beta)^{\lambda-1/2} \: e^{-a^2\beta^2 /2}
   \: {\text L}_\rnu^{(\lambda-1)}(a^2\beta^2),
\end{equation}
and 
${\text L}_\rnu^{(\lambda-1)}$ is a generalised
Laguerre polynomial \cite{AbramowitzStegun68}.
When $\lambda-5/2$ is a  non-negative integer
and $a$ is the inverse of a harmonic oscillator unit of length,
the wave functions
$\Wadial{\lambda}{a}{\rnu}(\beta)$
are standard radial wave functions for an isotropic
five-dimensional harmonic oscillator.%
\footnote{More generally, for $n\ge1$ and $\lambda-n/2$ a
non-negative integer, the functions
$\beta^{(1-n)/2}\Radial{\lambda}{a}{\rnu}(\beta)$
are radial wave functions for an isotropic $n$-dimensional
harmonic oscillator.}
However, for  arbitrary $\lambda>0$, they define ordered bases of
modified oscillator radial wave functions in terms of which,
for optimal choices of $a$ and $\lambda$, the expansion of
collective model wave functions for deformed nuclei converge
more rapidly, often considerably (see \elrm{Appendix }\ref{Sec:OpEx}).

The normalization of the functions $\Radial{\lambda}{a}{\rnu}(\beta)$
in \eqref{Eq:DefRadial} is such that
\begin{equation}\label{Eq:RadOrtho}
\RadialBrax{\lambda}{a}{\rmu}
\RadialKet{\lambda}{a}{\rnu}
=
\int_0^\infty
\Wadial{\lambda}{a}{\rmu}(\beta)\, \Wadial{\lambda}{a}{\rnu}(\beta)\,
 \beta^4\, d\beta
=
\int_0^\infty
\Radial{\lambda}{a}{\rmu}(\beta)\, \Radial{\lambda}{a}{\rnu}(\beta)\,
 d\beta
= \delta_{\rmu,\rnu}\,.
\end{equation}
For each $a,\lambda>0$,
the wave functions (\ref{Eq:DefBeta}) then form an orthonormal basis
with respect to the volume element $\beta^4\,d\beta$ for $\RadialHil$,
and $\{\RadialKet{\lambda}{a}{\rnu}, \rnu = 0,1,2,3,\ldots\}$
are orthonormal basis states.
Correspondingly, for each $a,\lambda>0$,
the functions (\ref{Eq:DefRadial}) form an orthonormal basis with
respect to the volume element $d\beta$ for the Hilbert space $\RadialHilc$.

Matrix elements of an operator $\hat X$ acting on $\RadialHil$ are given by
\begin{equation}\label{Eq:FOpXX}
\RadialBra{\lambda'}{a}{\rmu}\, \hat X\, \RadialKet{\lambda}{a}{\rnu} 
= \int_0^\infty
\Wadial{\lambda'}{a}{\rmu}(\beta)\,
[\hat X\, \Wadial{\lambda}{a}{\rnu}(\beta)]\, \beta^4\, d\beta
= \FOpb{a}{\lambda'\rmu}{\lambda\rnu}{\beta^2 \hat X\frac1{\beta^2}},
\end{equation}
where 
\begin{equation}\label{Eq:DefFOp}
\FOp{a}{\lambda'\rmu}{\lambda\rnu} (\hat Z)
= \int_0^\infty \Radial{\lambda'}{a}{\rmu}(\beta)\,
[\hat Z\,\Radial{\lambda}{a}{\rnu}(\beta)]\, d\beta .
\end{equation}
Note that the map
$\Wadial{\lambda}{a}{\rnu}(\beta)
\mapsto \Radial{\lambda}{a}{\rnu}(\beta)
= \beta^2 \Wadial{\lambda}{a}{\rnu}(\beta)$
between basis wave functions generates an isomorphic mapping
between the Hilbert spaces $\RadialHil \to \RadialHilc$. 
This mapping induces the mapping
$\hat X\mapsto\hat Z=\hat\beta^2\hat X\hat\beta^{-2}$
between operators on these spaces.

The space $\RadialHilc$ carries representations of the
Lie algebra of SU(1,1), whose complexification has basis elements
$S_{0}$, $S_{+}$ and $S_{-}$ that satisfy the commutation relations
\begin{equation}
[S_{+},S_{-}]=-2S_{0},\qquad
[S_{0},S_{\pm}]=\pm S_{\pm},
\end{equation}
and which, in a unitary representation, are realised as
operators $\hat S_{0}$, $\hat S_{+}$ and $\hat S_{-}$ respectively,
that satisfy the Hermiticity relations
\begin{equation}
\hat S_0^\dag=\hat S_0,\qquad
\hat S_{\pm}^\dag=\hat S_{\mp}.
\end{equation}
In fact, for each fixed pair $a,\lambda>0$,
there is a realisation of this Lie algebra in which $S_{0}$ and $S_{\pm}$
map respectively to operators
$\SUOp{\lambda}{a}{0}$ and $\SUOp{\lambda}{a}{\pm}$ on $\RadialHilc$,
whose matrix elements are
\begin{align}
\label{Eq:S0_ME}
\FOp{a}{\lambda\rmu}{\lambda\rnu}(\SUOp{\lambda}{a}{0})
&=\frac12(\lambda+2\rnu)\,\delta_{\rmu,\rnu},\\
\label{Eq:Sp_ME}
\FOp{a}{\lambda\rmu}{\lambda\rnu}(\SUOp{\lambda}{a}{+})
&=\sqrt{(\lambda+\rnu)(\rnu+1)}\;\delta_{\rmu,\rnu+1},\\
\label{Eq:Sm_ME}
\FOp{a}{\lambda\rmu}{\lambda\rnu}(\SUOp{\lambda}{a}{-})
&=\sqrt{(\lambda+\rnu-1)\rnu}\;\delta_{\rmu,\rnu-1},
\end{align}
relative to the basis $\{ \Radial{\lambda}{a}{\rmu}\}$.
Explicitly, these realisations are given by \cite{ACM2}%
\footnote{For operators that are clearly multiplicative,
we omit their `hat's when it is typographically convenient.}
\begin{align}
\label{Eq:S0_realise}
\SUOp{\lambda}{a}{0}
&= \frac14 \left[-
  \frac{1}{a^2}\frac{d^2}{d\beta^2} +\frac{(\lambda
  -\afrac3/2)(\lambda -\afrac1/2)}{(a\beta)^2} + (a\beta)^2  \right],\\
\label{Eq:Sp_realise}
\SUOp{\lambda}{a}{\pm}
&= \frac14 \left[
  \frac{1}{a^2}\frac{d^2}{d\beta^2} -\frac{(\lambda
  -\afrac3/2)(\lambda -\afrac1/2)}{(a\beta)^2} +
  (a\beta)^2 \mp\left(2\beta \frac{d}{d\beta} +1\right)\right].
\end{align}
Using these, the following matrix elements can be derived
(see \cite[Section 4.2.2]{RowanWood}):
{\allowdisplaybreaks
\begin{align}
a^2\FOpb{a}{\lambda\rmu}{\lambda\rnu}{\beta^2}
&= \delta_{\rmu,\rnu+1} \sqrt{(\lambda + \rnu)(\rnu+1)}
    + \delta_{\rmu,\rnu-1} \sqrt{(\lambda +\rnu-1)\rnu}
    + \delta_{\rmu,\rnu} (\lambda+2\rnu),
\label{Eq:beta1}\\*
\frac1{a^2}\FOpb{a}{\lambda\rmu}{\lambda\rnu}{\frac{1}{\beta^2}}
&=
\frac1{a^2}\FOpb{a}{\lambda\rnu}{\lambda\rmu}{\frac{1}{\beta^2}}
=  \frac{(-1)^{\rmu-\rnu}}{\lambda-1}
   \sqrt{\frac{\rmu!\, \Gamma(\lambda+\rnu)}{\rnu!\, \Gamma(\lambda+\rmu)}}
   \qquad {\rm for\;} \rmu \geq\rnu ,\; \lambda > 1,
\label{Eq:beta2}\\
\frac{1}{a^2}\FOpb{a}{\lambda\rmu}{\lambda\rnu}{\frac{d^2}{d\beta^2}}
&= \delta_{\rmu,\rnu+1}\sqrt{(\lambda +\rnu)(\rnu+1)} +
   \delta_{\rmu,\rnu-1}  \sqrt{(\lambda +\rnu-1)\rnu}
   \nonumber\\*
&\qquad\qquad\qquad
  - \delta_{\rmu,\rnu} (\lambda+2\rnu)
  + \frac1{a^2}
    \Big(\lambda-\frac32\Big)\Big(\lambda-\frac12\Big)
        \FOpb{a}{\lambda\rmu}{\lambda\rnu}{\frac{1}{\beta^2}} ,  
\label{Eq:beta3}\\
\FOpb{a}{\lambda\rmu}{\lambda\rnu}{\beta\frac{d}{d\beta}}
&= -\delta_{\rmu,\rnu+1} \sqrt{(\lambda + \rnu)(\rnu+1)} +
    \delta_{\rmu,\rnu-1} \sqrt{(\lambda +\rnu-1)\rnu} -\frac12 \delta_{\rmu,\rnu}.
\label{Eq:beta4}  
\end{align}
}%
The dependence on $a$, given by the factor premultiplying
each $\FOp{a}{\lambda\rmu}{\lambda\rnu}$, exhibits the fact
that $a$ is a scale factor in $\beta$.

The following matrix elements were derived \cite{ACM3} by expressing 
$\SUOp{\lambda}{a}{0}$ and $\SUOp{\lambda}{a}{\pm}$
in terms of products of the operators
\begin{equation}\label{Eq:Aops}
A(\kappa)
:=\frac{1}{a}\frac{d}{d\beta}+\frac{\kappa}{a\beta} + a\beta, \quad 
 A^{\dagger}(\kappa) :=
-\frac{1}{a}\frac{d}{d\beta}+\frac{\kappa}{a\beta} + a\beta,
\end{equation}
with $\kappa= \pm (\lambda - \tfrac12)$ and
$\kappa = \pm (\lambda - \tfrac32)$:
{\allowdisplaybreaks
\begin{align}
%
a\FOp{a}{\lambda+1,\rmu}{\lambda\rnu}(\beta)
&= \delta_{\rmu,\rnu} \sqrt{\lambda+\rnu}
    + \delta_{\rmu,\rnu-1} \sqrt{\rnu},
\label{Eq:beta5}\\*
a\FOp{a}{\lambda-1,\rmu}{\lambda\rnu}(\beta)
&= \delta_{\rmu,\rnu} \sqrt{\lambda+\rnu-1}
    + \delta_{\rmu,\rnu+1} \sqrt{\rnu+1},
\label{Eq:beta6}\\
%
\frac1a \FOpb{a}{\lambda+1,\rmu}{\lambda\rnu}{\frac{1}{\beta}}
&= \begin{cases}
      0 & \text{if $\rmu <\rnu$} ,\cr
      \displaystyle (-1)^{\rmu-\rnu} \sqrt{\frac{\rmu!\, \Gamma(\lambda+\rnu )}
      {\rnu!\, \Gamma(\lambda+\rmu +1)}} & \text{if $\rmu\geq\rnu$},
   \end{cases} 
\label{Eq:beta7}\\*
\frac1a \FOpb{a}{\lambda-1,\rmu}{\lambda\rnu}{\frac{1}{\beta}}
&= \begin{cases}
      0 & \text{if $\rmu >\rnu$}, \cr
      \displaystyle (-1)^{\rmu -\rnu} \sqrt{\frac{\rnu!\, \Gamma(\lambda+\rmu -1)}
      {\rmu!\, \Gamma(\lambda+\rnu)}} & \text{if  $\rmu \leq\rnu$},
   \end{cases}
\label{Eq:beta8}\\
\frac1a \FOpb{a}{\lambda+1,\rmu}{\lambda\rnu}{\frac{d}{d\beta}}\;
&= - \delta_{\rmu,\rnu}\sqrt{\lambda +\rnu}
   + \delta_{\rmu,\rnu-1} \sqrt{\rnu }
   + \frac1a \Big(\lambda-\frac12\Big)
         \FOpb{a}{\lambda+1,\rmu}{\lambda\rnu}{\frac{1}{\beta}},
\label{Eq:beta9}\\*
\frac1a \FOpb{a}{\lambda-1,\rmu}{\lambda\rnu}{\frac{d}{d\beta}}\;
&= - \delta_{\rmu,\rnu+1} \sqrt{\rnu +1}
    + \delta_{\rmu,\rnu}\sqrt{\lambda +\rnu-1}
    - \frac1a \Big(\lambda-\frac32\Big)
         \FOpb{a}{\lambda-1,\rmu}{\lambda\rnu}{\frac{1}{\beta}}.
\label{Eq:beta10}
\end{align}
}

Note that, with respect to the inner product \eqref{Eq:DefFOp}
for the functions $\Radial{\lambda}{a}{\rnu}(\beta)$, the
operators $\hat\beta^{\pm1}$, $\hat\beta^{\pm2}$
and $d^2/d\beta^2$
are Hermitian, whereas $d/d\beta$ is skew Hermitian.
Thus,  we have the identities
\begin{subequations}\label{Eq:Hermitian}
\begin{align}
\label{Eq:Hermitian_beta}
\FOp{a}{\lambda'\rmu}{\lambda\rnu}\big(\beta^{\pm1}\big)
&= \FOp{a}{\lambda\rnu}{\lambda'\rmu }\big(\beta^{\pm1}\big),
&\FOp{a}{\lambda'\rmu}{\lambda\rnu}\big(\beta^{{\pm}2}\big)
&= \FOp{a}{\lambda\rnu}{\lambda'\rmu}\big(\beta^{{\pm}2} \big),\\
\label{Eq:Hermitian_Dbeta}
\FOpb{a}{\lambda'\rmu}{\lambda\rnu}{\frac{d}{d\beta}}
&= -\FOpb{a}{\lambda\rnu}{\lambda'\rmu }{\frac{d}{d\beta}},
&\FOpb{a}{\lambda'\rmu}{\lambda\rnu}{\frac{d^2}{d\beta^2}}
&= \FOpb{a}{\lambda\rnu}{\lambda'\rmu}{\frac{d^2}{d\beta^2}}.
\end{align}
\end{subequations}

For maximum flexibility in the choice of $\lambda$ and $\lambda'$,
the program also makes use of the matrix elements of the
identity operator $\hat1$ between states
$\RadialKet{\lambda}{a}{\rnu}$ and $\RadialKet{\lambda'}{a}{\rmu}$
for which $\lambda'-\lambda$ is an even integer.
For $r>0$, the calculation given in \elrm{Appendix }\ref{Sec:Laguerre} yields
\begin{equation}\label{Eq:Radial_id}
\begin{split}
\FOp{a}{\lambda+2r,\rmu}{\lambda,\rnu}\big(\hat1\big)
 = (-1)^{\rnu-\rmu}
  \sqrt{\frac{\rmu!\,\Gamma(\lambda+\rnu)}{\rnu!\,\Gamma(\lambda+\rmu+2r)}}\:
   c^{(2r)}_{\rmu,\rnu},
\end{split}
\end{equation}
where
\begin{equation}\label{Eq:Radial_idC}
c^{{(2r)}}_{\rmu,\rnu}
=
\sum_{j=\max\{0,\rnu-\rmu\}}^r
(-1)^j \binom{r}{j}
\frac
{\Gamma(\lambda+\rmu+2r)\,\Gamma(\rmu+j+1)}
{\Gamma(\lambda+\rmu+r+j)\,\Gamma(\rmu+1)}
\binom{\rmu-\rnu+j+r-1}{\rmu-\rnu+j}.
\end{equation}
Note that $c^{{(2r)}}_{\rmu,\rnu}$ is a polynomial in $\lambda$
with integer coefficients and degree at most $\min\{r,r-\rnu+\rmu\}$.
In particular, $c^{(2r)}_{\rmu,\rnu}=0$ if $\rnu>\rmu+r$.

For an arbitrary operator $\hat Z$ 
acting on the Hilbert space $\RadialHil$
and a positive integer $r$,
the matrix elements \eqref{Eq:Radial_id} enable matrix elements
$\FOp{a}{\lambda'\pm2r,\rmu}{\lambda,\rnu}\big(\hat Z\big)$
to be obtained from those of
$\FOp{a}{\lambda',\rmu}{\lambda,\rnu}\big(\hat Z\big)$
using
\begin{subequations}\label{Eq:Radial_idsft}
\begin{align}
\label{Eq:Radial_idsft1}
\FOp{a}{\lambda'+2r,\rmu}{\lambda,\rnu}\big(\hat Z\big)
&=
\sum_{\sumer\ge0}
\FOp{a}{\lambda'+2r,\rmu}{\lambda',\sumer}\big(\hat1\big)\,
\FOp{a}{\lambda',\sumer}{\lambda,\rnu}\big(\hat Z\big);\\
\label{Eq:Radial_idsft2}
\FOp{a}{\lambda'-2r,\rmu}{\lambda,\rnu}\big(\hat Z\big)
&=
\sum_{\sumer\ge0}
\FOp{a}{\lambda'-2r,\rmu}{\lambda-2r,\sumer}\big(\hat Z\big)\,
\FOp{a}{\lambda-2r,\sumer}{\lambda,\rnu}\big(\hat1\big)
=
\sum_{\sumer\ge0}
\FOp{a}{\lambda'-2r,\rmu}{\lambda-2r,\sumer}\big(\hat Z\big)\,
\FOp{a}{\lambda,\rnu}{\lambda-2r,\sumer}\big(\hat1\big),
\end{align}
\end{subequations}
where the final identity follows because $\hat1$ is an Hermitian operator.
Then, because $c^{(2r)}_{\rmu,\sumer}=0$ for $\sumer>\rmu+r$, it follows
that a given matrix element
$\FOp{a}{\lambda'\pm2r,\rmu}{\lambda,\rnu}\big(\hat Z\big)$
is obtained precisely by restricting the sum in \eqref{Eq:Radial_idsft}
to $0\le\sumer\le\rmu+r$ or $0\le\sumer\le\rnu+r$ as appropriate.
Therefore, when the sum is carried out through the multiplication
of matrices, which are necessarily finite-dimensional in a
computer implementation, the matrix element
$\FOp{a}{\lambda'\pm2r,\rmu}{\lambda,\rnu}\big(\hat Z\big)$
is computed precisely for matrix dimensions exceeding
$\max\{\rmu,\rnu\}+r$.


\section{\texorpdfstring
          {\fivesupthree\ matrix elements and Clebsch-Gordan coefficients}
          {SO(5)>SO(3) matrix elements and Clebsch-Gordan coefficients}}
\label{Sec:SO5MEs}

Consider an SO(5) tensor operator $\hat T^\sen$.
This operator has components $\hTphericalu{\sen}{\alpha}{L}{M}$ 
that transform under SO(5) in the same way as the basis states 
$\SphericalKet{\sen}{\alpha}LM$.
The Wigner-Eckart theorem for SO(3)
\cite{Edmonds60}
implies that the matrix elements of these operators can be expressed
\begin{equation}\label{Eq:WE_SO3}
\SphericalBra{\sfin}{\alphafin}{\Lfin}{\Mfin}
 \hTphericalu{\sen}{\alpha}{L}{M}
\SphericalKet{\sini}{\alphaini}{\Lini}{\Mini}
 = \CG{\Lini}{\Mini}{L}{M}{\Lfin}{\Mfin}\,
\frac{
\SphericalBraR{\sfin}{\alphafin}{\Lfin}
 | \hTphericalu{\sen}{\alpha}{L}{} |
\SphericalKetR{\sini}{\alphaini}{\Lini}
}{
\sqrt{2\Lfin+1}},
\end{equation}
where $\CG{\Lini}{\Mini}{L}{M}{\Lfin}{\Mfin}$
is an SO(3) Clebsch-Gordan coefficient, and 
$\SphericalBraR{\sfin}{\alphafin}{\Lfin}
 | \hTphericalu{\sen}{\alpha}{L}{} |
\SphericalKetR{\sini}{\alphaini}{\Lini}$
is an SO(3)-reduced matrix element. 
Note that the factor $\sqrt{2L_f+1}$ in (\ref{Eq:WE_SO3}) is included
for symmetry reasons, 
and some authors use different phase conventions, neither of which are
essential to the Wigner-Eckart theorem.
In fact, in what follows, we often find it convenient to use
the following alternative definition of SO(3)-reduced matrix elements:
\begin{equation}\label{Eq:WE_SO3alt}
\SphericalBraR{\sfin}{\alphafin}{\Lfin}
 | \hTphericalu{\sen}{\alpha}{L}{} |
\SphericalKetR{\sini}{\alphaini}{\Lini}^\natural =
 \frac{
\SphericalBraR{\sfin}{\alphafin}{\Lfin}
 | \hTphericalu{\sen}{\alpha}{L}{} |
\SphericalKetR{\sini}{\alphaini}{\Lini}
}{\sqrt{2\Lfin+1}}.
\end{equation}
A similar use of the Wigner-Eckart theorem for SO(5) implies that
\begin{equation}\label{Eq:WE_SO5}
\SphericalBra{\sfin}{\alphafin}{\Lfin}{\Mfin}
 \hTphericalu{\sen}{\alpha}{L}{M}
\SphericalKet{\sini}{\alphaini}{\Lini}{\Mini}
=
(\sini\alphaini\Lini\Mini\,\sen\alpha LM| \sfin\alphafin\Lfin\Mfin)\,
\SphericalBraRR{\sfin} || \hTphericalu{\sen}{}{}{} || \SphericalKetRR{\sini},
\end{equation}
where $(\sini\alphaini\Lini\Mini\,\sen\alpha LM| \sfin\alphafin\Lfin\Mfin)$
is an SO(5) Clebsch-Gordan coefficient,
and 
$\SphericalBraRR{\sfin} || \hTphericalu{\sen}{}{}{} || \SphericalKetRR{\sini}$
is an SO(5)-reduced matrix element. 

Racah's factorisation lemma \cite{Racah49} combines
\eqref{Eq:WE_SO3} and \eqref{Eq:WE_SO5} to give the identity
\begin{equation}\label{Eq:WE_SO5>3a}
\SphericalBraR{\sfin}{\alphafin}{\Lfin}
 | \hTphericalu{\sen}{\alpha}{L}{} |
\SphericalKetR{\sini}{\alphaini}{\Lini}
= \sqrt{2\Lfin+1}\,
(\sini\alphaini\Lini\,\sen\alpha L || \sfin\alphafin\Lfin)\,
\SphericalBraRR{\sfin}
|| \hTphericalu{\sen}{}{}{} ||
\SphericalKetRR{\sini},
\end{equation}
 where 
\begin{equation}\label{Eq:WE_SO5>3b}
(\sini\alphaini\Lini\,\sen\alpha L || \sfin\alphafin\Lfin)
=
\frac{(\sini\alphaini\Lini\Mini\,\sen\alpha LM |\sfin\alphafin\Lfin\Mfin)}
{\CG{\Lini}{\Mini}{L}{M}{\Lfin}{\Mfin}},
\end{equation}
which is independent of $\Mini$, $M$ and $\Mfin$.
The quantity $(\sini\alphaini\Lini\,\sen\alpha L || \sfin\alphafin\Lfin)$
has been called an \fivesupthree\ Clebsch-Gordan coefficient \cite{CRW09}
(it is also known as an SO(3)-reduced SO(5) Clebsch-Gordan coefficient
and as an \emph{isoscalar factor}).

The \fivesupthree\ Clebsch-Gordan coefficients have been calculated
\cite{TAW08,CRW09} using the algorithm of \cite{RTR04}.
In this algorithm, SO(5) spherical harmonics
$\{\Sphericalu{\sen}{\alpha}{L}{M}(\gamma,\Omega)\}$
are calculated first.
These form an orthonormal basis of $\SphericalHil$,
and are wave functions for the basis states
$\{\SphericalKet{\sen}{\alpha}{L}{M}\}$.
SO(5) spherical harmonics of particular importance are 
\begin{equation}\label{Eq:Y112XX}
 \Sphericalu112M = \frac{\sqrt{15}}{4\pi} \, \calQ_M,
\end{equation}
and
\begin{equation}\label{Eq:SphHL=0}
\Sphericalu{3n}{1}{0}{0}(\gamma,\Omega)
=\frac{1}{4\pi }\sqrt{3(2n+1)} \, P_n(\cos3\gamma),
\end{equation}
for $n\ge0$,
where $P_n$ is a Legendre polynomial \cite{AbramowitzStegun68}.
For example,
\begin{equation}\label{Eq:Y310Y620}
\Sphericalu{0}{1}{0}{0}(\gamma,\Omega) = \frac{\sqrt3}{4\pi }\,,
\qquad
\Sphericalu{3}{1}{0}{0}(\gamma,\Omega) = \frac{3}{4\pi }\, \cos3\gamma\,,
\qquad
\Sphericalu{6}{1}{0}{0}(\gamma,\Omega) = \frac{\sqrt{15}}{8\pi }\, 
(3\cos^2 3\gamma -1)\,.
\end{equation}

To each spherical harmonic there is then a component of a tensor operator 
$\hSphericalu{\sen}{\alpha}{L}{M}$  such that the state
$\hSphericalu{\sen}{\alpha}{L}{M}\SphericalKet{\sini}{\alphaini}{\Lini}{\Mini}$
has wave function
$\Sphericalu{\sen}{\alpha}{L}{M}(\gamma,\Omega)
 \Sphericalu{\sini}{\alphaini}{\Lini}{\Mini}(\gamma,\Omega)$.
It follows from \eqref{Eq:WE_SO3} and \eqref{Eq:WE_SO5>3a} that evaluation of
integrals of the form
\begin{equation}\label{eq:SO5ME}
\SphericalBra{\sfin}{\alphafin}{\Lfin}{\Mfin}
\hSphericalu{\sen}{\alpha}{L}{M}
\SphericalKet{\sini}{\alphaini}{\Lini}{\Mini}
=\int_{S_4}
\Sphericalu{\sfin}{\alphafin}{\Lfin}{\Mfin}(\gamma,\Omega)^{*}\,
\Sphericalu{\sen}{\alpha}{L}{M}(\gamma,\Omega)\,
\Sphericalu{\sini}{\alphaini}{\Lini}{\Mini}(\gamma,\Omega)\,
\sin3\gamma \, d\gamma\, d\Omega
\end{equation}
enables both the \fivesupthree\ Clebsch-Gordan
coefficients and the SO(5)-reduced matrix elements
$\SphericalBraRR{\sfin} || \hSphericalu{\sen}{}{}{} || \SphericalKetRR{\sini}$
to be determined.
In the process of such calculations, the empirical expression
\begin{equation}\label{Eq:SO5red_ME}
\begin{split}
\SphericalBraRR{\sfin} || \hSphericalu{\sen}{}{}{} || \SphericalKetRR{\sini}
&= \frac{1}{4\pi}\,
\frac{(\frac{\summe}2+1)!}
{(\frac{\summe}2-\sini)!(\frac{\summe}2-\sen)!(\frac{\summe}2-\sfin)!}\,
 \sqrt{\frac{(2\sini+3)(2\sen+3)}{(\sfin+2)(\sfin+1)}}\\
&\qquad\times \sqrt{
\frac{({\summe}+4)({\summe}-2\sini+1)!({\summe}-2\sen+1)!({\summe}-2\sfin+1)!}
     {({\summe}+3)!}},
\end{split}
\end{equation}
with ${\summe}=\sini+\sen+\sfin$, was obtained
by recognising the pattern of results obtained numerically.
This expression is consistent with such reduced matrix elements as
have been obtained by algebraic methods \cite{ACM1,ACM2,ACM3}%
; for example,
\begin{equation}\label{Eq:QME}
\SphericalBraRR{\sfin} || \hcalQ || \SphericalKetRR{\sini}
= \delta_{\sfin,\sini+1}\,\sqrt{\frac{\sini+1}{2\sini+5}}\\
  +\delta_{\sfin,\sini-1}\,\sqrt{\frac{\sini+2}{2\sini+1}}.
\end{equation}

It will be shown in the following that the SO(5) tensors  
that arise in the ACM have matrix elements related
to those of corresponding spherical harmonics.
Thus, these results prove to be sufficient for present needs.
Consequently, the use of \eqref{Eq:SO5red_ME} and the available
tabulations of \fivesupthree\ Clebsch-Gordan coefficients obviate
the need to evaluate integrals such as \eqref{eq:SO5ME}.


\section{Collective model observables and their matrix elements}
\label{Sec:CMOMEs}

\subsection{%
  \texorpdfstring
    {Operators acting on the tensor product space
                         $\Hb = \RadialHil\otimes\SphericalHil$}%
    {Operators acting on the tensor product space H=RxS4}%
}
\label{Sec:Xspace_Ops}

In the bases defined by \eqref{Eq:BasisVar}, the matrix elements
of any operator $\hat W = \hat X \hat T$
that is a product of operators, 
$\hat X$ and $ \hat T$, that act independently
on the component spaces $\RadialHil$ and $\SphericalHil$ of $\Hb$,
respectively, are given by the products
\begin{equation}\label{Eq:WME}
\begin{split}
\StateBra{\lambda_{\sfin}}{a}{\rmu}{\sfin}{\alphafin}{\Lfin}{\Mfin}
\hat W
\StateKet{\lambda_{\sini}}{a}{\rnu}{\sini}{\alphaini}{\Lini}{\Mini}
&=
\RadialBra{\lambda_{\sfin}}{a}{\rmu}
\hat X
\RadialKet{\lambda_{\sini}}{a}{\rnu}
\;
\SphericalBra{\sfin}{\alphafin}{\Lfin}{\Mfin}
\hat T
\SphericalKet{\sini}{\alphaini}{\Lini}{\Mini}\\
&=
\FOpb{a}{\lambda_{\sfin}\rmu}{\lambda_{\sini}\rnu}
        {\beta^2 \hat X \frac1{\beta^2}}
\,
\SphericalBra{\sfin}{\alphafin}{\Lfin}{\Mfin}
\hat T
\SphericalKet{\sini}{\alphaini}{\Lini}{\Mini},
\end{split}
\end{equation}
after using \eqref{Eq:FOpXX}.
For observables acting on $\Hb$ that are products of
SU(1,1)$\,\times\,$SO(5) tensors of the form
\begin{equation}\label{Eq:Wform}
\hat W^{\sen}_{\alpha LM}
=\hat X
\hSphericalu{\sen}{\alpha}{L}{M},
\end{equation}
the Wigner-Eckart theorem for each of SO(5) and SO(3) implies that
\begin{subequations}\label{Eq:WigEkX}
\begin{align}
&\StateBra{\lambda_{\sfin}}{a}{\rmu}{\sfin}{\alphafin}{\Lfin}{\Mfin}
\hat W
\StateKet{\lambda_{\sini}}{a}{\rnu}{\sini}{\alphaini}{\Lini}{\Mini}
\nonumber
\\
\label{Eq:WigEkX1}
&\hspace{35mm}=
(\sini\alphaini\Lini\Mini\,\sen\alpha LM| \sfin\alphafin\Lfin\Mfin)\;
\StateBraRR{\lambda_{\sfin}}{a}{\rmu}{\sfin}
|| \hat W^\sen ||
\StateKetRR{\lambda_{\sini}}{a}{\rnu}{\sini}\\
&\hspace{35mm}=
\label{Eq:WigEkX2}
\CG{\Lini}{\Mini}{L}{M}{\Lfin}{\Mfin}\;
\StateBraR{\lambda_{\sfin}}{a}{\rmu}{\sfin}{\alphafin}{\Lfin}
| \hat W^{\sen}_{\alpha L} |
\StateKetR{\lambda_{\sini}}{a}{\rnu}{\sini}{\alphaini}{\Lini}^\natural,
\end{align}
\end{subequations}
where, using \eqref{Eq:WME} and \eqref{Eq:WE_SO5},
the SO(5)-reduced matrix elements of $\hat W^{\sen}_{\alpha LM}$ are given by
\begin{equation}\label{Eq:XWE_SO5}
\StateBraRR{\lambda_{\sfin}}{a}{\rmu}{\sfin}
|| \hat W^\sen ||
\StateKetRR{\lambda_{\sini}}{a}{\rnu}{\sini}
  = \FOpb{a}{{\lambda_{\sfin}}\rmu}{\lambda_{\sini}\rnu} 
  {\beta^2\hat X\frac1{\beta^2}}\;
\SphericalBraRR{\sfin} || \hSphericalu{\sen}{}{}{} || \SphericalKetRR{\sini},
\end{equation}
and, using first \eqref{Eq:WME}, \eqref{Eq:WE_SO3} and \eqref{Eq:WE_SO3alt},
and then \eqref{Eq:WE_SO3alt}, \eqref{Eq:WE_SO5>3a} and \eqref{Eq:XWE_SO5},
the adjusted SO(3)-reduced matrix elements of $\hat W^{\sen}_{\alpha LM}$
are given by
\begin{equation}\label{Eq:XWE_SO3}
\begin{split}
&\StateBraR{\lambda_{\sfin}}{a}{\rmu}{\sfin}{\alphafin}{\Lfin}
| \hat W^{\sen}_{\alpha L} |
\StateKetR{\lambda_{\sini}}{a}{\rnu}{\sini}{\alphaini}{\Lini}^\natural
=
\FOpb{a}{{\lambda_{\sfin}}\rmu}{\lambda_{\sini}\rnu} 
  {\beta^2\hat X\frac1{\beta^2}}\;
\SphericalBraR{\sfin}{\alphafin}{\Lfin}
| \hSphericalu{\sen}{\alpha}{L}{} |
\SphericalKetR{\sini}{\alphaini}{\Lini}^\natural\\
&\hskip 40mm
=
(\sini\alphaini\Lini\,\sen\alpha L || \sfin\alphafin\Lfin)\;
\StateBraRR{\lambda_{\sfin}}{a}{\rmu}{\sfin}
|| \hat W^\sen ||
\StateKetRR{\lambda_{\sini}}{a}{\rnu}{\sini}.
\end{split}
\end{equation}

Because $q=\beta\calQ$, with each $\calQ_M$ proportional to the
SO(5) spherical harmonic $\Sphericalu112M$, as in \eqref{Eq:Y112XX},
it follows from \eqref{Eq:QME} and \eqref{Eq:XWE_SO5}
that the non-zero SO(5)-reduced matrix elements of the basic collective
model observables $\hat q_M$ are given by
\begin{subequations}\label{Eq:XQME}
\begin{align}
\StateBraRR{\lambda_{\sen+1}}{a}{\rmu}{\sen+1}
  || \hat q ||
  \StateKetRR{\lambda_\sen}{a}{\rnu}{\sen}
&=\FOp{a}{\lambda_{\sen+1}\rmu}{\lambda_\sen\rnu}(\beta)\sqrt{\frac{\sen+1}{2\sen+5}},
\label{Eq:XQME1}\\ 
\StateBraRR{\lambda_{\sen-1}}{a}{\rmu}{\sen-1}
  || \hat q ||
  \StateKetRR{\lambda_\sen}{a}{\rnu}{\sen}
&=\FOp{a}{\lambda_{\sen-1}\rmu}{\lambda_\sen\rnu}(\beta)\sqrt{\frac{\sen+2}{2\sen+1}}.
\label{Eq:XQME2}
\end{align}
\end{subequations}
It has also been shown \cite{RowanWood} that the non-zero
SO(5)-reduced matrix elements of the observables $\hat\pi_N$,
the momenta conjugate to $\hat q_M$, are given by
\begin{subequations}\label{Eq:XPiME}
\begin{align}
\StateBraRR{\lambda_{\sen+1}}{a}{\rmu}{\sen+1}
  || \hat\pi ||
  \StateKetRR{\lambda_{\sen}}{a}{\rnu}{\sen}
&= -{\rm i}\hbar \FOpb{a}{\lambda_{\sen+1}\rmu}{\lambda_\sen\rnu}
                        {\frac{d}{d\beta} - \frac{\sen+2}{\beta}} \,
\sqrt{\frac{\sen+1}{2\sen+5}},
\label{Eq:XPiME1}\\ 
\StateBraRR{\lambda_{\sen-1}}{a}{\rmu}{\sen-1}
  || \hat\pi ||
  \StateKetRR{\lambda_\sen}{a}{\rnu}{\sen}
&= -{\rm i}\hbar \FOpb{a}{\lambda_{\sen-1}\rmu}{\lambda_\sen\rnu}
                         {\frac{d}{d\beta} + \frac{\sen+1}{\beta}} \,
\sqrt{\frac{\sen+2}{2\sen+1}}.
\label{Eq:XPiME2}
\end{align}
\end{subequations}
Thus, if we choose the $\sen$-dependence of $\lambda_\sen$ such that
\begin{equation}\label{Eq:lambdapm1}
\lambda_{\sen+1} = \lambda_\sen \pm 1
\end{equation}
for all $\sen$,
explicit algebraic expressions for these collective model matrix elements
\eqref{Eq:XQME} and \eqref{Eq:XPiME}
are immediately obtained using the expressions for
$\FOp{a}{\lambda\pm 1,\rmu}{\lambda\rnu}(\beta)$,
$\FOp{a}{\lambda\pm1,\rmu}{\lambda\rnu}(1/\beta)$
and
$\FOp{a}{\lambda\pm1,\rmu}{\lambda\rnu}(d/d\beta)$
given by \eqref{Eq:beta5}--\eqref{Eq:beta10}.
More generally, if the $\sen$-dependence of $\lambda_\sen$ is such that
\begin{equation}\label{Eq:lambdapm2}
\lambda_{\sen+1} - \lambda_\sen \text{ is odd}
\end{equation}
for all $\sen$, then algebraic expressions for \eqref{Eq:XQME}
and \eqref{Eq:XPiME} are obtained by using \eqref{Eq:Radial_id}
in addition to \eqref{Eq:beta5}--\eqref{Eq:beta10}.
One immediate possibility for
\eqref{Eq:lambdapm2} (and \eqref{Eq:lambdapm1})
is to choose
\begin{equation}\label{Eq:lambda5sh}
\lambda_{\sen} = \sen+\frac52,
\end{equation}
which corresponds to the five-dimensional harmonic oscillator states.
Another convenient choice is discussed in Section \ref{Sec:Bases} below.

In $\Rb^5$, the Laplacian $\nabla^2$ can be expressed
(see \cite[Section 2.2]{RowanWood})
\begin{equation}\label{Eq:LaplacianDef}
\nabla^2=
\frac1{\beta^4}\frac{\partial}{\partial\beta}
\beta^4\frac{\partial}{\partial\beta}
-
\frac1{\beta^2}\hat\Lambda^2,
\end{equation}
where $\hat\Lambda^2$ is the SO(5) Casimir operator
whose eigenfunctions are the SO(5) spherical harmonics
$\Sphericalu{\sen}{\alpha}{L}{M}$, with
\begin{equation}\label{Eq:CasimirDef}
\hat\Lambda^2
\Sphericalu{\sen}{\alpha}{L}{M}
=\sen(\sen+3)
\Sphericalu{\sen}{\alpha}{L}{M}.
\end{equation}
By applying \eqref{Eq:XWE_SO5} to the two terms
of \eqref{Eq:LaplacianDef},
we obtain the SO(5)-reduced matrix elements
\begin{equation}\label{Eq:Laplacian_ME}
\StateBraRR{\lambda_{\sfin}}{a}{\rmu}{\sfin}
  || \nabla^2 ||
  \StateKetRR{\lambda_{\sini}}{a}{\rnu}{\sini}
=\FOpb{a}{\lambda_{\sini}\rmu}{\lambda_{\sini}\rnu}
         {\frac{d^2}{d\beta^2}  -
          \frac{\sini(\sini+3)+2}{\beta^2}}\,
\delta_{\sfin,\sini}\,.
\end{equation}
Explicit analytic expressions for these matrix elements are 
 immediately obtained from those of
$\FOp{a}{\lambda\rmu}{\lambda\rnu}(d^2/d\beta^2)$ and
$\FOp{a}{\lambda\rmu}{\lambda\rnu}(1/\beta^2)$
given in Section \ref{Sec:RadMEs}.

\subsection{Collective model observables in the ACM}
\label{Sec:CMOS}

In our implementation of the ACM, observables of interest are formed
by taking sums of products of the following generating operators:
\begin{subequations}\label{Eq:generators}
\begin{gather}
\label{Eq:generators1}
\hat\beta^2; \quad
\hat \beta^{-2}; \quad
\frac{d^2}{d\beta^2}; \quad
\beta\frac{d}{d\beta}; \quad
\nabla^2; \quad
[\hat\pi\otimes\hat q\otimes\hat\pi]_0; \quad
\hSphericalu6100; \quad
\hat 1;
\\
\label{Eq:generators2}
\hat\beta; \quad
\hat\beta^{-1}; \quad
\frac{d}{d\beta}; \quad
\cos 3\gamma; \quad
\\
\label{Eq:generators3}
\hat\pi_M; \quad
[\hat\pi\otimes\hat\pi]_{2M}; \quad
[\hat\pi\otimes\hat\pi]_{4M}; \quad
\hSphericalu{\sen}{\alpha}LM \text{ for } \sen\in\{2,4,6\} \text{ and } L>0;
\\[4pt]
\label{Eq:generators4}
\hSphericalu{\sen}{\alpha}LM \text{ for } \sen\in\{1,3,5\} \text{ and } L>0.
\end{gather}
\end{subequations}
Using \eqref{Eq:XWE_SO3},
SO(3)-reduced matrix elements of $\hat\pi_M$ and $\nabla^2$
are obtained from the SO(5)-reduced matrix elements given
in \eqref{Eq:XPiME} and \eqref{Eq:Laplacian_ME} respectively.
Those of the spherical harmonics
$\hSphericalu \sen\alpha LM$ are
obtained from \eqref{Eq:WE_SO5>3a} and \eqref{Eq:SO5red_ME},
with those of $\cos3\gamma$ similarly obtained,
by virtue of \eqref{Eq:Y310Y620},
from those of the spherical harmonic $\hSphericalu3100$.
The reduced matrix elements of the remaining operators are 
obtained from the algebraic expressions of Section \ref{Sec:RadMEs}.
In the case of the operators $[\hat\pi\otimes\hat q\otimes\hat\pi]_0$
and $[\hat\pi\otimes\hat\pi]_{LM}$,
previously unpublished expressions for their SO(3)-reduced matrix elements
are derived in \elrm{Appendix }\ref{Sec:ExtraMEs}.
For the operators involving $\hat\pi$, the ACM code calculates,
for convenience, the renormalised operators
$\hbar^{-2}[\hat\pi\otimes\hat q\otimes\hat\pi]_0$,
${\rm i}\hbar^{-1}\hat\pi_M$, and
$\hbar^{-2}[\hat\pi\otimes\hat\pi]_{LM}$ for $L\in\{2,4\}$.

By design, the ACM code analyses Hamiltonians that are SO(3) invariant
(i.e.~scalar).
Thus, for convenience, the list \eqref{Eq:generators} of operators
is partitioned such that only those in \eqref{Eq:generators1} and
\eqref{Eq:generators2} are SO(3) invariant.
The partitioning is also such that there is some redundancy in that
some operators in \eqref{Eq:generators1}
can be obtained by multiplying two of those from \eqref{Eq:generators2}.
However, the matrix elements of the former are obtained directly
from a single SU(1,1) representation, thereby avoiding the inefficiency
of a sum over intermediate states.

Further SO(3) invariant operators may be obtained by extracting the
scalar-coupled product of two operators of equal (non-zero) angular momentum.
This enables, in particular, reduced matrix elements of the
scalar components of operators
$\hat\pi\otimes\hat q\otimes\hat q\otimes\cdots\otimes\hat q\otimes \hat\pi$
to be obtained
(details are given in Section \ref{Sec:Op_Racah}).

The list \eqref{Eq:generators} of operators is also partitioned
such that only those in \eqref{Eq:generators1} and \eqref{Eq:generators3}
are rational in the quadrupole moments
$\{\hat q_M\}$ and their conjugate momenta $\{\hat\pi_N\}$.
In addition, products of operators \eqref{Eq:generators} which
contain an even number of factors from
\eqref{Eq:generators2} and \eqref{Eq:generators4} are also rational.
That this is so follows because
(i) $\hat\beta^2 = \hat q \cdot\hat q \equiv \sum_M(-1)^M \hat q_M\hat q_{-M}$, 
$-{\rm i}\hbar\hat\beta \partial/\partial\beta
= \hat q \cdot\hat\pi \equiv \sum_M (-1)^M \hat q_M \hat\pi_{-M}$,
and $-\hbar^2\nabla^2 =
         \hat\pi \cdot\hat\pi\equiv\sum_M (-1)^M \hat\pi_M \hat\pi_{-M}$
are each quadratic in $\{\hat q_M,\hat\pi_N\}$; and
(ii) any $\hat\beta^\sen \hSphericalu{\sen}{\alpha}{L}{M}$ is a polynomial
in the quadrupole moments and hence 
$\hat\beta^{\sen-2n} \hSphericalu{\sen}{\alpha}{L}{M}$ is a rational function
of the quadrupole moments for any integer $n$.

The significance of this is that any observable of the ACM that is rational 
in $\{\hat q_M\}$ and $\{\hat\pi_N\}$ 
is expressed efficiently in the code,
being readily obtained from the analytical expressions
in Sections~\ref{Sec:RadMEs} and \ref{Sec:SO5MEs}.
Noting that the matrix elements 
$F^{(a)}_{\lambda \pm 1,\rmu; \lambda \rnu}(\hat\beta^{\pm1})$ and
$F^{(a)}_{\lambda \pm 1,\rmu; \lambda \rnu}(d/d \beta)$
have analytic expressions, we obtain analytic matrix elements 
for all rational observables in each basis \eqref{Eq:BasisVar} for which
\eqref{Eq:lambdapm2} holds.
The specific choices for \eqref{Eq:lambdapm2}
that are used in the ACM are discussed in Section \ref{Sec:Bases} below.

Observables that are not rational in $\hat q_M$ and $\hat\pi_N$ can
also be generated.
For example, one can include potential energy terms in a Hamiltonian
with odd powers of $\beta$.
However, the matrix elements of such observables are not obtained
analytically, and require a larger truncated Hilbert space to
achieve accurate results.

The matrix elements of observables that are polynomial in the above
generating operators \eqref{Eq:generators},
as well as the scalar-coupled products,
are now obtained by combining those for the generating operators
calculated in a finite-dimensional subspace $\Hbt$ of $\Hb$ spanned
by a subset of the SU(1,1)$\times$SO(5) coupled states
\eqref{Eq:BasisVar}, specified by the user.
In particular, we are able to generate algebraic expressions for
any ACM Hamiltonian that is SO(3) and time-reversal invariant,
is a polynomial function of the basic $\hat q_M$, $\hat\pi_M$ observables
and $\beta^{-2}$,
and at most quadratic in the momentum operators.

With respect to a truncated basis $\Hbt$, the SO(3)-reduced matrix elements
of all the operators in \eqref{Eq:generators}, apart from
$[\hat\pi\otimes\hat q\otimes\hat\pi]_0$ and
$[\hat\pi\otimes\hat\pi]_{LM}$,
are determined precisely.
However, the SO(3)-reduced matrix elements of operators obtained by
combining these operators as described above, as well as the operators
$[\hat\pi\otimes\hat q\otimes\hat\pi]_0$ and
$[\hat\pi\otimes\hat\pi]_{LM}$,
are subject to truncation errors.
One consequence is that these matrix elements
do not precisely satisfy the required Hermiticity relationships.
However, the discrepancy only occurs close to the truncation boundary.
Thus, to ensure that diagonalisation of a model Hamiltonian
leads to real eigenvalues and eigenvectors in the ACM code,
we retain only the Hermitian component of the truncated matrix
by taking the average of the matrix and its Hermitian conjugate.
This leaves most of the matrix elements unchanged, and doesn't
affect the low-lying eigenvalues or any of their calculated properties.
This is ensured by working with a space sufficiently large
such that a further increase in size produces no significant
change in the calculated quantities.

\subsection{Bases in the ACM}
\label{Sec:Bases}

The specification of a particular basis \eqref{Eq:BasisVar}
for the Hilbert space $\Hb$
requires the values of $a$ and $\lambda_\sen$ to be set.
As discussed in the previous section, analytic expressions for
the matrix elements of rational observables are efficiently obtained
when $\lambda_\sen$ is such that \eqref{Eq:lambdapm2} holds.

In our main implementation of the ACM, two solutions to
\eqref{Eq:lambdapm2}
are immediately available.
In the first, we set
\begin{equation}\label{Eq:lambdaDefSHO}
\lambda_\sen = \lambda_0 + \sen,
\end{equation}
for a specified value of $\lambda_0$.
Such a basis generalises \eqref{Eq:lambda5sh} which pertains to
the five-dimensional harmonic oscillator basis.
In the second, we set
\begin{equation}\label{Eq:lambdaDef}
\lambda_\sen = 
  \begin{cases} \lambda_0 & \text{if $\sen$ is even,}\\
                \lambda_0 + 1 & \text{if $\sen$ is odd.}
  \end{cases} 
\end{equation}
For these bases, the matrix elements of rational observables are obtained
especially efficiently because there is no need 
for the code to employ \eqref{Eq:Radial_id}.

The ACM code also allows the user to set a constant
\begin{equation}\label{Eq:lambdaDefConst}
\lambda_\sen=\lambda_0.
\end{equation}
In this case, analytic expressions are still obtained for the
matrix elements of parity-preserving rational observables,
where an observable $\hat W$ is said to be parity-preserving if
\begin{equation}\label{Eq:ParityDef}
\StateBraR{\lambda_{\sfin}}{a}{\rmu}{\sfin}{\alphafin}{\Lfin}
|\hat W|
\StateKetR{\lambda_{\sini}}{a}{\rnu}{\sini}{\alphaini}{\Lini}
=0
\end{equation}
in all instances where $\sfin-\sini$ is odd.
However, if $\hat W$ is not parity-preserving then its matrix
elements are obtained by the less precise non-analytic means
described in Section \ref{Sec:NonRat} below.

Having set $\lambda_\sen$ to be a function of $\lambda_0$ and $\sen$,
it remains to set the two values $a$ and $\lambda_0$ to determine
a suitable basis \eqref{Eq:BasisVar} for the Hilbert space $\Hb$
on which a particular model Hamiltonian acts.
Any positive values for these parameters define a basis.
However, given that in a computer implementation,
it is necessary to use a truncation $\Hbt$ of $\Hb$,
and that the calculation time increases dramatically with
the dimension of $\Hbt$, it is important to choose values
of $a$ and $\lambda_0$ that give convergent results for
as few basis states as possible.
Methods for optimising the parameters $a$ and $\lambda_0$
are discussed in \elrm{Appendix }\ref{Sec:BasisStuff}.
Note that the values of these two parameters have a much more
significant effect on the efficiency of the calculation than the
choice of basis type \eqref{Eq:lambdaDefSHO},
\eqref{Eq:lambdaDef} or \eqref{Eq:lambdaDefConst}.

\subsection{Non-rational observables}
\label{Sec:NonRat}

As discussed above, the use of bases \eqref{Eq:BasisVar} for which
\eqref{Eq:lambdapm2} holds
enables the matrix elements of all
observables that are rational functions of the quadrupole and
conjugate momentum operators to be obtained precisely.
Non-rational observables such as $\hat\beta^{\pm1}$ and $d/d\beta$
may also be used, with the
$\lambda$-conserving matrix elements
$F^{(a)}_{\lambda\rmu;\lambda\rnu}(\hat\beta)$
obtained from the equation
\begin{equation}\label{Eq:ME_betalambda}
F^{(a)}_{\lambda\rmu;\lambda\rnu}(\hat\beta^2)
= \sum_\sumer F^{(a)}_{\lambda\rmu;\lambda\sumer}(\hat \beta)
F^{(a)}_{\lambda\sumer;\lambda\rnu}(\hat\beta),
\end{equation}
by taking the positive square root of the matrix
$F^{(a)}_{\lambda\rmu;\lambda\rnu}(\hat\beta^2)$
with the knowledge that the operator $\hat\beta$ is positive definite.

The matrix elements
$F^{(a)}_{\lambda\rmu;\lambda\rnu}(\hat\beta^{-1})$
can then be obtained by taking the inverse of the matrix
$F^{(a)}_{\lambda\rmu;\lambda\rnu}(\hat\beta)$
obtained above.
%
The matrix elements of $d/d\beta$ cannot be obtained by taking
the square root of the matrix representing
$d^2/d\beta^2$ because the latter is not positive definite.
Instead, they can be determined by use of the identity
\begin{equation}\label{Eq:ME_Dbetalambda}
F^{(a)}_{\lambda\rmu;\lambda\rnu}(d/d\beta) =
\sum_\sumer  F^{(a)}_{\lambda\rmu;\lambda\sumer}(\hat\beta^{-1}) 
F^{(a)}_{\lambda\sumer;\lambda\rnu}(\hat\beta\,d/d\beta).
\end{equation}

Matrix elements of other non-rational observables may be obtained
by combining those for $\hat\beta^{\pm1}$ and $d/d\beta$,
obtained as described above, with those for the rational observables,
including those for the identity operator $\hat1$
given by \eqref{Eq:Radial_id}.
The matrix elements of non-parity-preserving rational operators
in the basis \eqref{Eq:lambdaDefConst}
may also be obtained from such combinations.

It should be noted that matrix elements obtained non-analytically,
as described in this section, suffer somewhat from truncation effects.
However, when an operator that yields such matrix elements
is used as a term in a Hamiltonian,
then in general the truncation doesn't significantly affect the
low-lying eigenstates,
and thus is not detrimental to obtaining converged results.

\subsection{Reduced E2 transition rates}

The standard definition (see \cite[Section 2.3.1]{RowanWood})
of reduced E2 transition rates between sets of levels in nuclear physics
is to take the sum of the squared matrix elements of the E2 transition
operator $\hat Q^{(E)}_m = (Ze/ A)\hat q_m$
over the final set of states and then average over the initial states.
Thus, for a transition between sets of states of angular momenta
$\Lini$ and $\Lfin$, the SO(3)-reduced E2 transition rate is given by
\begin{equation}\label{Eq:BE2_SO3}
B({\rm E2}; \alphaini\Lini \to \alphafin\Lfin)
 = \left(\frac{Ze}{A}\right)^2
 \frac{ \big|\SphericalBraR{}{\alphafin}{\Lfin}
        | \hat q |
        \SphericalKetR{}{\alphaini}{\Lini} \big|^2
      }{2\Lini+1}.
\end{equation}
Here, the multiplicity labels $\alphaini$ and $\alphafin$ serve to
distinguish between different sets of states with identical angular
momenta.
The ACM code allows the user to multiply the expression \eqref{Eq:BE2_SO3}  
by any convenient factor so that the $B({\rm E}2)$ values can be obtained
in any convenient units.

Applying the above definition
to initial and final sets of states that carry SO(5) irreps gives
SO(5)-reduced E2 transition rates $\bar B({\rm E2}; \sini \to \sfin)$.  
Using (\ref{Eq:WE_SO5>3a}) then results in the following relationship
between these transition rates and the above standard SO(3)-reduced
transition rates \cite[Section VIIC]{RWC09}:
\begin{equation}\label{Eq:BE2_SO5}
B({\rm E2}; \sini\alphaini\Lini \to \sfin\alphafin\Lfin)
=(\sfin\alphafin \Lfin\, 112 || \sini\alphaini \Lini)^2\,
\bar B({\rm E2}; \sini\to\sfin).
\end{equation}


\section{Using the code}
\label{Sec:Code}

In this section, we describe the basic usage of the ACM code.
Although basic, the procedures described here enable a vast
range of Hamiltonians to be analysed,
with their eigenenergies, transition rates and amplitudes
calculated and displayed.
With a few minor adjustments to the default settings,
the procedures of this section may also be used to carry out
calculations in the rigid-$\beta$ limit of the ACM.
How this is done is described in \elrm{Appendix }\ref{Sec:Rigid}.

It is anticipated that many users will find the flexibility
offered by the procedures of this section to be
more than adequate for their purposes.
For those who wish to further process the data calculated,
use other Hamiltonians, or examine the transition rates
of other operators,
various ways in which the functionality of the procedures
described in this section may be extended are described in
Section \ref{Sec:Extend}.

More fundamental procedures of the ACM code are described in
Sections \ref{Sec:BasicACM} and \ref{Sec:Components}:
these might be of use in models beyond
the ACM that make use of the SO(5) and/or SU(1,1) matrix elements
or the \fivesupthree\ Clebsch-Gordan coefficients.
However, it is suggested that the reader skip
Sections \ref{Sec:Extend}, \ref{Sec:BasicACM} and \ref{Sec:Components}
on a first reading, and return to tackle these sections after
gaining some experience in using the procedures of the current section.

A Maple worksheet \texttt{acm-examples.mw}, supplied with the code,
illustrates use of most of the procedures
that are described in this and the following sections.

\subsection{Preliminaries}
\label{Sec:Code_Prelim}

\subsubsection{Installation}

Before the code can be used, it is necessary to install
the files that contain the \fivesupthree\
Clebsch-Gordan coefficients onto the host computer.
These files are supplied in three zipped files {named}
\texttt{so5cg-data13.zip},
\texttt{so5cg-data24.zip} and
\texttt{so5cg-data56.zip}.
The first of these zipped
files contains \fivesupthree\ Clebsch-Gordan coefficients
$(\sen_1\alpha_1 L_1,\,\sen_2\alpha_2 L_2 || \sen_3\alpha_3 L_3)$
for $\sen_2=1$ and $\sen_2=3$;
the second contains those for $\sen_2=2$ and $\sen_2=4$;
and the third contains those for $\sen_2=5$ and $\sen_2=6$.
(For the basic usage described in this section,
only the data from the first two files is required,
with that from the second then only required if the operator
$[\hat\pi\otimes\hat q\otimes\hat\pi]_0$ is present in the Hamiltonian.)

Unzipping the file \texttt{so5cg-data13.zip} will create the
directory \texttt{so5cg-data/} (if it doesn't already exist)
and in that directory create subdirectories \texttt{v2=1/} and \texttt{v2=3/}
containing further subdirectories and data files.
Likewise, unzipping the file \texttt{so5cg-data24.zip}
will create subdirectories \texttt{v2=2/} and \texttt{v2=4/}
of \texttt{so5cg-data/}, and unzipping \texttt{so5cg-data56.zip}
will create subdirectories \texttt{v2=5/} and \texttt{v2=6/}
of \texttt{so5cg-data/}.%
\footnote{In some instances, such as when the files are automatically
unzipped on downloading, or when a file is unzipped by clicking on it,
the subdirectories \texttt{v2=1/}, \texttt{v2=2/},
\texttt{v2=3/}, \texttt{v2=4/}, \texttt{v2=5/} and \texttt{v2=6/}
might not appear in the desired common \texttt{so5cg-data/} directory.
It will then be necessary to move those subdirectories.}

Once unzipped, and the location of the initial directory \texttt{so5cg-data/}
is specified to the program (see below), this subdirectory structure is
invisible to the user, and all the required \fivesupthree\ Clebsch-Gordan
coefficients will be automatically available to the program.
However, it may be useful to know that each of the subdirectories
\texttt{v2=$\sen_2$/} contains subdirectories \texttt{SO5CG\_$\sen_1$\_$\sen_2$\_$\sen_3$/}
for various values of $\sen_1$ and $\sen_3$, and
each subdirectory \texttt{SO5CG\_$\sen_1$\_$\sen_2$\_$\sen_3$/}
contains files named \texttt{SO5CG\_$\sen_1$\_$\sen_2$-$\alpha_2$-$L_2$\_$\sen_3$},
for various values of $\alpha_2$ and $L_2$, that contain the
\fivesupthree\ Clebsch-Gordan coefficients.

\subsubsection{Initialisation}

Whenever the code is used, it is necessary to specify the
location of the files that contain the \fivesupthree\
Clebsch-Gordan coefficients.
This is achieved by specifying the location of the directory
\texttt{so5cg-data/} whose subdirectories, as described above,
contain the data.
This location is specified in the global variable \texttt{SO5CG\_directory}.
A typical declaration for a Unix, Linux or Macintosh system
would be%
\footnote{Note that the final ``/'' is necessary.}
\begin{equation}\label{Eq:SO5cgsDIR}
\texttt{
SO5CG\_directory:="/home/username/maple/acm/so5cg-data/":
}
\end{equation}
For a Windows system, a typical declaration would be%
\footnote{The slash \lq\lq/\rq\rq may be used to separate the
subdirectory names here, even though, on Windows systems,
this is usually done with the backslash \lq\lq$\backslash$\rq\rq.
However, in Maple strings, the backslash is an \emph{escape sequence},
and therefore in order to use the usual Windows form,
it is necessary to write
\texttt{SO5CG\_directory:=
"C:$\backslash\backslash$Users$\backslash\backslash$%
Username$\backslash\backslash$%
maple$\backslash\backslash$%
acm$\backslash\backslash$so5cg-data$\backslash\backslash$":}
instead of \eqref{Eq:SO5cgsDIRwindows}.}
\begin{equation}\label{Eq:SO5cgsDIRwindows}
\texttt{
SO5CG\_directory:="C:/Users/Username/maple/acm/so5cg-data/":}
\end{equation}

The specification of \texttt{SO5CG\_directory}, using a command
of the form \eqref{Eq:SO5cgsDIR} or \eqref{Eq:SO5cgsDIRwindows},
may be done in the Maple session, immediately after reading
the main ACM code from \texttt{acm.mpl}.
Alternatively, it may be more convenient to make this specification
in a settings file.
An example of such a file, \texttt{acm-user.mpl}, is supplied
with the ACM code.
The supplied file \texttt{acm-user.mpl} also specifies various
default values that affect the output of eigenvalues and transition rates,
as explained in Sections \ref{Sec:Code_Rates} and
\ref{Sec:Code_Control} below.
These can be altered to suit the user's requirements.

If used, the settings file should be called by using
Maple's \texttt{read} command immediately after the main code
in \texttt{acm.mpl} is called:
\begin{equation}
\begin{split}
&\texttt{read "acm.mpl":}\\
&\texttt{read "acm-user.mpl":}
\end{split}
\end{equation}
These two files \emph{should not} be opened
as Maple worksheets. Doing so is likely to corrupt them.
They should only be used by calling them from a
Maple session or Maple worksheet using the \texttt{read} command
as described above.
In addition, the file \texttt{"acm-user.mpl"} should only be edited
using a text editor.
The file \texttt{"acm.mpl"} should not be edited at all.

\subsection{The main functionality}
\label{Sec:Code_Main}

This subsection describes the framework in which calculations are
made with the ACM code.
Firstly, it explains a simple way of specifying Hamiltonians.
Secondly, it explains how to specify the truncated Hilbert spaces
$\Hbt$ in which the calculations are carried out.
It then describes two procedures,
\texttt{ACM\_Scale} and \texttt{ACM\_Adapt},
which carry out the diagonalisation of a Hamiltonian on a particular $\Hbt$,
and displays the resulting eigenvalues.
These two procedures differ only in the way that the results
are scaled before being displayed.

The procedures \texttt{ACM\_Scale} and \texttt{ACM\_Adapt}
are also able to calculate and display E2 transition rates and
amplitudes of the quadrupole operator $\hat q$.
This is described in Section \ref{Sec:Code_Rates}.

In Section \ref{Sec:Code_Control}, various ways to configure the output
of \texttt{ACM\_Scale} and \texttt{ACM\_Adapt} are described.
Section \ref{Sec:Code_Basis} shows how to specify the dependence
of $\lambda_\sen$ on $\sen$  for $\Hbt$,
while Section \ref{Sec:Code_Params} describes a procedure
which enables optimal values of the adjustable parameters $a$ and
$\lambda_0$ to be obtained for certain types of Hamiltonians.

\subsubsection{Specifying Hamiltonians}
\label{Sec:Ham_Encode}

The Hamiltonians that can be analysed with this code are polynomials
in the operators \eqref{Eq:generators1} and \eqref{Eq:generators2}.
The coefficients in these polynomials may be arbitrary real numbers,
or they may be functions of the quantum numbers $\rnu$, $\sen$ and $L$
of the states $\StateKet{\lambda_\sen}{a}{\rnu}{\sen}{\alpha}{L}{M}$ on
which they operate.

In the ACM code, Hamiltonians and other operators are encoded using
a particular list structure, which is described in detail in
Section \ref{Sec:Op_Encode} below.
However, the user does not need to know this encoding
when the Hamiltonian is of the form%
\footnote{These Hamiltonians are rational in the
basic observables $\hat q_M$ and $\hat\pi_M$, and therefore,
as explained in Section \ref{Sec:CMOS}, their matrix elements are
obtained efficiently and accurately in the code via analytic expressions.}
\begin{equation}\label{Eq:HamACM}
\begin{split}
&
x_1\nabla^2 +x_2 +x_3\beta^2 +x_4\beta^4 + \frac{x_5}{\beta^2}
+x_6\beta\cos 3\gamma +x_7\beta^3\cos 3\gamma +x_8\beta^5\cos 3\gamma
+\frac{x_9}{\beta}\cos 3\gamma
+x_{10}\cos^2 3\gamma\\
&\hskip 22mm
+x_{11}\beta^2\cos^2 3\gamma +x_{12}\beta^4\cos^2 3\gamma
+\frac{x_{13}}{\beta^2}\cos^2 3\gamma
+\frac{x_{14}}{\hbar^2}\,
[\hat\pi\otimes\hat q\otimes\hat\pi]_0.
\end{split}
\end{equation}
For such a Hamiltonian, the encoding is generated by the procedure call
\begin{equation}\label{Eq:Ham_OpLC}
\texttt{ACM\_Hamiltonian(}
x_1,x_2,x_3,x_4,x_5,x_6,x_7,x_8,x_9,x_{10},x_{11},x_{12},x_{13},x_{14}
\texttt{):}
\end{equation}
In typical usage, the value returned by this procedure
will be assigned to a Maple variable, such as in
\begin{equation}  \label{eq:HOp}
\texttt{HOp := ACM\_Hamiltonian(0.5,0,3.5)}:
\end{equation}
This variable may then be used as the first argument to the
procedures described in Sections \ref{Sec:ACM_Scale} and
\ref{Sec:ACM_Adapt} below.
Note that, in using the procedure \texttt{ACM\_Hamiltonian},
arguments that are not specified are taken to be 0.
Thus, in the example (\ref{eq:HOp}), all $x_i$ for
$4\le i\le 14$ are set to zero.

As mentioned above, each of the arguments $x_i$
in \eqref{Eq:Ham_OpLC}
is either a real number or a real function of $\rnu$, $\sen$ or $L$.
This latter case is accommodated by using the symbolic names
\texttt{NUMBER}, \texttt{SENIORITY} and \texttt{ANGMOM},
which are then assigned the values $\rnu$, $\sen$ and $L$,
respectively, of the states acted upon by the Hamiltonian.

This is useful, for example, in examining the rigid-$\beta$
Wilets-Jean model \cite{WJ56}, which has Hamiltonian
proportional to $\hat\Lambda^2$. 
In view of \eqref{Eq:CasimirDef}, the ACM code's encoding of this
Hamiltonian is the value returned by
\begin{equation}
\texttt{ACM\_Hamiltonian(0,SENIORITY*(SENIORITY+3))}:
\end{equation}
How to use the ACM code to perform rigid-$\beta$
calculations is discussed in \elrm{Appendix }\ref{Sec:Rigid}.

\subsubsection{Specifying the truncated Hilbert space}
\label{Sec:Xspace_Params}

In using the ACM code, it is necessary to truncate the Hilbert
space $\Hb$ on which the Hamiltonian and other operators act
to a finite-dimensional subspace $\Hbt$.
To do this, we specify non-negative integers
$\argsub{$\rnu$}{min}$, $\argsub{$\rnu$}{max}$,
$\argsub{$\sen$}{min}$, $\argsub{$\sen$}{max}$,
$\argsub{L}{min}$, $\argsub{L}{max}$,
in addition to the parameters $\lambda_0$ and $a$.
The subspace $\Hbt$ is then that spanned by
\begin{equation}\label{Eq:Hilbert_Trunc}
\{
\StateKet{\lambda_\sen}{a}{\rnu}{\sen}{\alpha}{L}{M}\,,
\argsub{$\rnu$}{min} \le\rnu\le \argsub{$\rnu$}{max},
1\le\alpha\le d_{\sen L},
\argsub{$\sen$}{min} \le \sen\le \argsub{$\sen$}{max},
\argsub{L}{min} \le L\le \argsub{L}{max},
-L\le M\le L
\},
\end{equation}
where, in the default case,
$\lambda_\sen$ is determined by $\lambda_0$ using \eqref{Eq:lambdaDef}.
However, by using reduced matrix elements, the
code avoids reference to the quantum number $M$ entirely.

The range of states \eqref{Eq:Hilbert_Trunc} available to the code is
restricted only by the availability of
\fivesupthree\ Clebsch-Gordan coefficients.
There is thus an upper limit on the maximum seniority
$\argsub{$\sen$}{max}$ of $\Hbt$,
with this limit dependent on the Hamiltonian being analysed.
Specific values are given in the \emph{Program summary} section.

Techniques to determine good values of $a$ and $\lambda_0$
are discussed in \elrm{Appendix }\ref{Sec:BasisStuff}.
Procedures that implement those techniques for certain types of
Hamiltonians are described in Section \ref{Sec:Code_Params} below.

\subsubsection{Diagonalisation using \texttt{ACM\_Scale}}
\label{Sec:ACM_Scale}

For a Hamiltonian $\hat H$ encoded in the Maple variable \texttt{HOp},
its action on the truncated Hilbert space $\Hbt$,
spanned by \eqref{Eq:Hilbert_Trunc}, is diagonalised by the procedure call
\begin{equation}\label{Eq:ACM_Scale}
\texttt{ACM\_Scale(HOp},
a, \lambda_0,
\argsub{$\rnu$}{min}, \argsub{$\rnu$}{max},
\argsub{$\sen$}{min}, \argsub{$\sen$}{max},
\argsub{L}{min}, \argsub{L}{max}
\texttt{):}
\end{equation}
The final argument $\argsub{L}{max}$ of \texttt{ACM\_Scale} may be omitted,
in which case its value is taken to be $\argsub{L}{max}=\argsub{L}{min}$.%
\footnote{This will be the case for all procedures in this package for
which the final two arguments are $\argsub{L}{max}$ and $\argsub{L}{min}$.
This will not be mentioned henceforth.}
The call \eqref{Eq:ACM_Scale} diagonalises the action of $\hat H$
separately on each subspace of $\Hbt$ of angular momentum $L$,
for $\argsub{L}{min} \le L\le \argsub{L}{max}$.
For each such $L$-space, the procedure prints a list of
eigenvalues in increasing order.
These are given relative to their minimum value
across all the $L$-spaces.
This minimal value is displayed.
The number of eigenvalues output for each $L$, as well as
their displayed precision may be altered, as described in
Section \ref{Sec:Code_Output} below.

The procedure \texttt{ACM\_Scale} is also able to calculate
and display transition rates and amplitudes
of the quadrupole operator $\hat q$.
As described in Section \ref{Sec:Code_Rates} below,
these are produced by specifying, beforehand, two lists of pairs
of states between which transition rates and amplitudes are required.

The eigenvalues, transition rates and amplitudes are automatically
scaled by certain factors stored by the code before being displayed.
These scaling factors may be set explicitly using the procedure
\texttt{ACM\_set\_scales} described in Section \ref{Sec:Code_Scale} below.
Thus, by setting each of these values to 1
(the default),
the raw eigenvalues, transition rates and amplitudes are
displayed by \texttt{ACM\_Scale}.
Note that these stored scaling factors change whenever the procedure
\texttt{ACM\_Adapt}, described next, is used.

As is usually the case with Maple procedures, the procedure
\texttt{ACM\_Scale} has a return value.
This return value contains the information from which the displayed data 
is obtained, as well as much more.
In basic usage, where it is adequate for the user simply to view
the energy eigenvalues, transition rates and amplitudes
displayed by \texttt{ACM\_Scale}, the return value may be ignored.
However, if it is required to further process the displayed information
or to obtain other results from the calculation,
the value returned by \eqref{Eq:ACM_Scale} should be assigned to
a Maple variable.
The format of this return value is described in Section \ref{Sec:Returns}
below, but because of its voluminous and somewhat cryptic nature,
it is probably not usefully displayed directly.
Instead, the data it contains may be displayed by using procedures
described in Section \ref{Sec:Display} below.

\subsubsection{Diagonalisation using \texttt{ACM\_Adapt}}
\label{Sec:ACM_Adapt}

The procedure call
\begin{equation}\label{Eq:ACM_Adapt}
\texttt{ACM\_Adapt(HOp},
a, \lambda_0,
\argsub{$\rnu$}{min}, \argsub{$\rnu$}{max},
\argsub{$\sen$}{min}, \argsub{$\sen$}{max},
\argsub{L}{min}, \argsub{L}{max}
\texttt{):}
\end{equation}
performs a similar function to that of \texttt{ACM\_Scale} above.
The difference here is that a scaling factor is applied to the
relative eigenvalues so that 
a specified eigenvalue takes a prescribed value.
All the other relative eigenvalues are scaled accordingly.
The quadrupole transition rates and amplitudes
are treated similarly, in that a certain transition rate is
set to a specified value, and the required scaling is then
used for all other transition rates.
A corresponding scaling is applied to the amplitudes.

It is explained in Section \ref{Sec:Code_Adapt} below how the
specific values used here are set.
With the default settings, the scaling factor for the relative eigenvalues
obtained from \eqref{Eq:ACM_Adapt} is set so that the
lowest for angular momentum $L=2$ takes the value $6.0$.
This enables the results to be quickly compared with those
of a rigid rotor which has rotational energies proportional to $L(L+1)$.
In the case of the quadrupole transition rates, the scaling factor
is set so that the transition rate $B({\rm E2}; 2(1)\to 0(1))$
takes the value $100.0$, where $2(1)$ and $0(1)$ label the
lowest eigenvalue states of angular momenta 2 and 0 respectively
(cf.~\eqref{Eq:RateEspace_q} below).

The scaling factors obtained by \texttt{ACM\_Adapt} for the eigenvalues,
transition rates and amplitudes are retained so that they 
are used in subsequent calls to the procedure \texttt{ACM\_Scale}.

The value returned by the procedure \texttt{ACM\_Adapt} has
the same format as that returned by \texttt{ACM\_Scale} and,
as in that case, may be disregarded in basic usage.

\subsection{Calculation and display of transition rates and amplitudes}
\label{Sec:Code_Rates}

The procedure calls
\begin{equation}\label{Eq:ACM_SetRats}
\begin{split}
&\texttt{ACM\_set\_rat\_lst(}
\argit{ratelist}
\texttt{):}\\
&\texttt{ACM\_add\_rat\_lst(}
\argit{ratelist}
\texttt{):}
\end{split}
\end{equation}
determine which quadrupole transition rates are displayed by subsequent
calls to the procedures \texttt{ACM\_Scale} and \texttt{ACM\_Adapt}.
The first of these specifies a list of transition rate designators,
and the second augments it.
The argument \argit{ratelist} is a list in which each element
itself is a list of up to five integers.
A four element designator
$[\Lini,\Lfin,\ctini,\ctfin]$
indicates that the SO(3)-reduced transition rate defined by
\begin{equation}\label{Eq:RateEspace_q}
B({\rm E2}; \Lini(\ctini)\to\Lfin(\ctfin))
=
\frac{\langle{\ctfin\Lfin}|
|\hat q|
|{\ctini}{\Lini}\rangle^2}
{{2\Lini+1}}
\end{equation}
(omitting the overall factor $(Ze/A)^2$ from \eqref{Eq:BE2_SO3}),
is to be displayed,
where $|{\ctini}{\Lini}\rangle$ (resp.~$|{\ctfin}{\Lfin}\rangle$)
denotes the $\ctini$th (resp.~$\ctfin$th) lowest energy
eigenstate of angular momentum $\Lini$ (resp.~$\Lfin$).
The zero, one, two and three element designators
$[\,]$,
$[\Lfin]$,
$[\Lini,\Lfin]$,
$[\Lini,\Lfin,\ctfin]$
indicate that particular lists of values \eqref{Eq:RateEspace_q}
are to be displayed.
For these, the stipulated indices are constant while the other
indices from $\{\Lini,\Lfin,\ctini,\ctfin\}$ take a range.
For $\ctini$ and $\ctfin$, this range
is as described in Section \ref{Sec:Code_Output} below.
The values of $\Lini$ and $\Lfin$ range over all values
between $\argsub{L}{min}$ and $\argsub{L}{max}$
(specified in \eqref{Eq:ACM_Scale} and \eqref{Eq:ACM_Adapt})
with $|\Lini-\Lfin|\le 2$.
For the five element designator
$[\argsubsup{L}{i}{0},\argsubsup{L}{f}{0},\ctini,\ctfin,\argsub{L}{mod}]$,
a sequence of the transition rates
\eqref{Eq:RateEspace_q} is displayed for fixed values of $\ctini$ and
$\ctfin$, and all available pairs $\Lini$ and $\Lfin$ given by
$\Lini =\argsubsup{L}{i}{0}+k \argsub{L}{mod}$ and
$\Lfin =\argsubsup{L}{f}{0}+k \argsub{L}{mod}$,
for $k=0,1,2,\ldots\,$.
Note that $\argsub{L}{mod}<0$ is permitted.

The transition amplitudes to be displayed are determined in an analogous way.
Thus, the procedure calls
\begin{equation}\label{Eq:ACM_SetAmps}
\begin{split}
&\texttt{ACM\_set\_amp\_lst(}
\argit{amplist}
\texttt{):}\\
&\texttt{ACM\_add\_amp\_lst(}
\argit{amplist}
\texttt{):}
\end{split}
\end{equation}
specify a list of designators, which is
independent of that specified by \eqref{Eq:ACM_SetRats}.
Each element of \argit{amplist} is a list of up to five integers.
A four element designator $[\Lini,\Lfin,\ctini,\ctfin]$
indicates that the transition amplitude given by
\begin{equation}\label{Eq:AmpsXspace_q}
\CG{\Lini,}{\Lini,}{2,}%
{\Lfin-\Lini}{\Lfin,}{\Lfin}\,
\frac{\langle{\ctfin\Lfin}|
|\hat q|
|{\ctini}{\Lini}\rangle}
{\sqrt{2\Lfin+1}},
\end{equation}
where the first factor is a standard SO(3) Clebsch-Gordan coefficient,
is to be displayed 
(in the most important case where 
$|{\ctfin}{\Lfin}\rangle=|{\ctini}{\Lini}\rangle$,
this reduces to the usual expression for the static quadrupole moment
$\langle\ctini\Lini\Lini
|\hat q|
\ctini\Lini\Lini\rangle$).
The zero, one, two, three and five element designators in \argit{amplist}
indicate that lists of transition amplitudes \eqref{Eq:AmpsXspace_q}
are to be displayed,
with these lists determined as for \argit{ratelist} above.

The procedure calls
\begin{equation}\label{Eq:ACM_ShowRats}
\begin{split}
&\texttt{ACM\_show\_rat\_lst(%
\argit{show}%
);}\\
&\texttt{ACM\_show\_amp\_lst(%
\argit{show}%
);}
\end{split}
\end{equation}
return, respectively, the currently stored lists of
transition rate designators and transition amplitude designators.
If $\argit{show}>0$ or the argument is omitted then a formatted
printout of the designators is output.
For $\argit{show}\le 0$, the procedures acts silently.

Note that the values \eqref{Eq:RateEspace_q} and \eqref{Eq:AmpsXspace_q}
are displayed after being scaled.
These scaling factors are set whenever \texttt{ACM\_Adapt} is used.
Alternatively, they may be set explicitly using the procedure
\texttt{ACM\_set\_scales}
described in Section \ref{Sec:Code_Scale} below.

Also note that the formulae by which the displayed values
\eqref{Eq:RateEspace_q} and \eqref{Eq:AmpsXspace_q}
are determined from the SO(3)-reduced matrix elements
of $\hat q$ may be changed.
How to do this is described later in Section \ref{Sec:ACM_alter}.

\subsection{Controlling the output}
\label{Sec:Code_Control}

\subsubsection{Setting display precision and datum}
\label{Sec:Code_Output}

The format of the values displayed by the procedures
\texttt{ACM\_Scale} and \texttt{ACM\_Adapt} can be altered
by using the procedure
\begin{equation}
\texttt{ACM\_set\_output(}
\argit{precision},\argit{width},\argit{precision0},\argit{show}
\texttt{):}
\end{equation}
Here, \argit{precision} is the maximal number of digits to be used
beyond the decimal point to display floating point values;
\argit{width} is the maximal total number of digits to be used
to display floating point values;
\argit{precision0} is the maximal number of digits to be used
beyond the decimal point to display the minimal eigenvalue.
Here, and for all the other procedures in the ACM code whose names begin
\texttt{ACM\_set},
if the final parameter \argit{show} is positive then the
procedure prints a brief summary of the result of its invocation.
This is also the case if this parameter is omitted.
If $\argit{show}\le0$ then the procedure acts silently.

By default, eigenvalues are displayed relative to their
overall minimum value. This behaviour can be changed using
the procedure
\begin{equation}
\texttt{ACM\_set\_datum(}
\argit{datum},\argit{show}
\texttt{):}
\end{equation}
For $\argit{datum}=0$, this specifies that absolute eigenvalues
are displayed instead.
The default behaviour is restored by using $\argit{datum}\ge1$.

The number of values that appear in the horizontal lists of
eigenvalues and quadrupole transition rates displayed by
the procedures \texttt{ACM\_Scale} and \texttt{ACM\_Adapt}
can be constrained by issuing the procedure call
\begin{equation}\label{Eq:ACM_SetListln}
\texttt{
ACM\_set\_listln(}
\argsub{count}{eigs},\argsub{count}{rats},\argit{show}
\texttt{):}
\end{equation}
Thereafter, the list of eigenvalues that appears for each value of the
angular momentum will contain at most the lowest $\argsub{count}{eigs}$
eigenvalues.
As explained in Section \ref{Sec:Code_Rates},
a list of transition rates \eqref{Eq:RateEspace_q} is produced whenever
the \argit{ratelist} argument of \eqref{Eq:ACM_SetRats} contains
a designator that has three or fewer indices.
If $\ctini$ or $\ctfin$ is not specified in the designator, then,
following the call \eqref{Eq:ACM_SetListln},
a list of transition rates is displayed for these indices
in the ranges
$1\le\ctini\le\argsub{count}{rats}$
or
$1\le\ctfin\le\argsub{count}{rats}$
respectively.
The output of transition amplitudes \eqref{Eq:AmpsXspace_q} is
determined from the list \argit{amplist} of transition amplitude
designators in exactly the same way.

\subsubsection{Explicit setting of display scaling factors}
\label{Sec:Code_Scale}

It is often useful to scale the eigenvalues, transition rates
and transition amplitudes calculated by the ACM code.
For the procedure \texttt{ACM\_Scale},
such scaling factors can be specified by using the procedure call
\begin{equation}\label{Eq:ACM_set_scales}
\texttt{ACM\_set\_scales(}
\argsub{scale}{eigs}, \argsub{scale}{rats}, \argit{show}
\texttt{):}
\end{equation}
Then, after the eigenvalues are calculated across all the
$L$-spaces, and their minimal value determined,
the relative eigenvalues with respect to this minimal value
are each divided by $\argsub{scale}{eigs}$ before being displayed
(the minimal value is displayed unscaled).
Similarly, each transition rate \eqref{Eq:RateEspace_q} is divided
by the scaling factor $\argsub{scale}{rats}$ before being displayed.

This call \eqref{Eq:ACM_set_scales} also sets a scaling factor
$\argsub{scale}{amps}$ which applies similarly to the
transition amplitudes \eqref{Eq:AmpsXspace_q}.
This is determined by $\argsub{scale}{amps}=\sqrt{\argsub{scale}{rats}}$.

The scaling factors set by \texttt{ACM\_set\_scales}
apply to all subsequent invocations of \texttt{ACM\_Scale},
until reset by \texttt{ACM\_set\_scales},
or the procedure \texttt{ACM\_Adapt} is used.
In this latter case, \texttt{ACM\_Adapt} resets the
values of $\argsub{scale}{eigs}$ and $\argsub{scale}{rats}$,
and sets $\argsub{scale}{amps}=\sqrt{\argsub{scale}{rats}}$,
and applies these to the eigenvalues, transition rates and
transition amplitudes that it calculates.
How these scaling factors are determined in this case is explained in
Section \ref{Sec:Code_Adapt} below.

The scaling factors that are in force at a given time may be obtained
using the procedure call
\begin{equation}\label{Eq:ACM_show_scales}
\texttt{ACM\_show\_scales(}
\argit{show}
\texttt{):}
\end{equation}
This returns a list
$[\argsub{scale}{eigs},\argsub{scale}{rats},\argsub{scale}{amps}]$
of the three current factors by which eigenvalues, transition rates
and transition amplitudes are divided.
If $\argit{show}>0$ or the argument is omitted then a brief description
of the values and their purpose is output.
For $\argit{show}\le 0$, the procedures acts silently.

\subsubsection{Settings for implicit determination of display scaling factors}
\label{Sec:Code_Adapt}

Here we describe two procedures that specify values that are used
by the procedure \texttt{ACM\_Adapt} to adjust the scaling factors
$\argsub{scale}{eigs}$ and $\argsub{scale}{rats}$ so that a certain
eigenvalue and a certain transition rate then take specific values.

The procedure call
\begin{equation}\label{Eq:ACM_set_eig_fit}
\texttt{ACM\_set\_eig\_fit(}
\argsub{val}{eig}, L, \countlab, \argit{show}
\texttt{):}
\end{equation}
specifies that if the procedure \texttt{ACM\_Adapt} is invoked
subsequently, then the scaling factor $\argsub{scale}{eigs}$
applied to the relative eigenvalues would be chosen such that
the $\countlab$th lowest eigenvalue in the space of
angular momentum $L$ takes the value $\argsub{val}{eig}$.
All the other relative eigenvalues would be scaled accordingly,
with the minimal value unscaled.

If either of the two final arguments in \eqref{Eq:ACM_set_eig_fit}
is omitted, then the value 1 is used.

The procedure call
\begin{equation}\label{Eq:ACM_set_rat_fit}
\texttt{ACM\_set\_rat\_fit(}
\argsub{val}{rat}, \Lini, \Lfin, \ctini, \ctfin, \argit{show}
\texttt{):}
\end{equation}
specifies that if the procedure \texttt{ACM\_Adapt} is invoked
subsequently, then the scaling factor $\argsub{scale}{rats}$
applied to the transition rates would be chosen
such that the specific transition rate
$B({\rm E2}; \Lini(\ctini)\to\Lfin(\ctfin))$
scales to the value $\argsub{val}{rat}$.
All the other transition rates would be scaled accordingly,
as would the transition amplitudes.

If either of the three final arguments in \eqref{Eq:ACM_set_rat_fit}
is omitted, then the value 1 is used.

As described above, each invocation of the procedure \texttt{ACM\_Adapt}
sets three scaling factors: one for the eigenvalues, one for
the transition rates, and one for the transition amplitudes.
These scaling factors are retained for use by
any subsequent calls to the procedure \texttt{ACM\_Scale}.
Note that these scaling factors may be reset also by calling
\texttt{ACM\_set\_scales}, as described in
Section \ref{Sec:Code_Scale} above.

\subsection{Setting the basis type}
\label{Sec:Code_Basis}

As discussed in Section \ref{Sec:Bases}, specifying the basis
\eqref{Eq:BasisVar} requires the dependence of $\lambda_\sen$
on $\sen$ to be set.
This is done by using the procedure call
\begin{equation}\label{Eq:ACM_set_basis_type}
\texttt{ACM\_set\_basis\_type(}
\argit{basistype},
\argsub{pot}{min},
\argit{show}
\texttt{):}
\end{equation}
Here the argument \argit{basistype} takes either of the four values 0, 1, 2 or 3.
For all but the last of these, the argument $\argsub{pot}{min}$ has no effect.
For \argit{basistype} set to 1, this call specifies that
the harmonic oscillator basis type \eqref{Eq:lambdaDefSHO} is to be used.
For \argit{basistype} set to 2, this call specifies that
the parity basis type \eqref{Eq:lambdaDef} is to be used.
This latter is the default.
The constant-$\lambda$ basis type \eqref{Eq:lambdaDefConst}
is selected using this procedure when \argit{basistype} is set to 0.

The use of \eqref{Eq:ACM_set_basis_type} with \argit{basistype} set to 3,
specifies the dependence of $\lambda_\sen$ on $\sen$
given by \eqref{Eq:lambdaDefDavi} with $\beta_*=\argsub{pot}{min}$.
As with \eqref{Eq:lambdaDefSHO} and \eqref{Eq:lambdaDef},
this dependence respects the condition
\eqref{Eq:lambdapm1}
and thus enables analytic matrix elements to be used for rational operators.
The utility of the basis type given by \eqref{Eq:lambdaDefDavi}
is explained in \elrm{Appendix }\ref{Sec:BasisStuff}.

Note that to completely determine a basis \eqref{Eq:BasisVar},
it is necessary, in addition to specifying a basis type,
also to specify values of $a$ and $\lambda_0$.
These two values are passed directly as arguments to the procedures
\texttt{ACM\_Scale} and \texttt{ACM\_Adapt}, and also to the more
basic procedures described later in Section \ref{Sec:ACM_Diag}.
A procedure for estimating optimal values of these parameters is
given in the following section.

\subsection{Optimizing the parameters}
\label{Sec:Code_Params}

The accuracy of the results obtained using the ACM code is dependent
not only on the size of the truncated Hilbert space $\Hbt$ on which
the Hamiltonians operate, but also on the values of the two
parameters $a$ and $\lambda_0$.
In fact, for optimal or near-optimal values of these
two parameters, a much smaller truncated Hilbert space
is required to achieve accurate results, and the
calculation can be completed more quickly and reliably.

In \elrm{Appendix }\ref{Sec:BasisStuff},
a method is described for estimating optimal values
of the parameters $a$ and $\lambda_0$ for Hamiltonians
of the general form \eqref{Eq:HamGen}.
The ACM code implements this method for the five-parameter Hamiltonians
$\hat H_{\text{RWC}} (B,c_1,c_2,\chi,\kappa)$ defined by \eqref{Eq:HamRWC},
and which were considered in \cite{RWC09}.
Such Hamiltonians provide a simple alternative to the more general
Hamiltonian \eqref{Eq:HamACM}, and their encoding is obtained
using the procedure call
\begin{equation}
\texttt{RWC\_Ham(}
B,c_1,c_2,\chi,\kappa
\texttt{):}
\end{equation}
For these Hamiltonians,
the method for estimating the optimal values of $a$ and $\lambda_0$
is detailed in \eqref{Eq:HamRWC}--\eqref{Eq:ExptRWC},
and carried out using the procedure call
\begin{equation}\label{Eq:RWC_alam}
\texttt{RWC\_alam(}
B,c_1,c_2
\texttt{):}
\end{equation}
This returns a pair $[a_{\text opt},\lambda_{\text opt}]$,
whose components are the estimated optimal values of $a$
and $\lambda_0$, respectively
(they are independent of $\chi$ and $\kappa$).

Refinements of these estimates, if desired, may be obtained by
inspection of the expectation values \eqref{Eq:ExptRWC} of the
Hamiltonian $\hat H_{\text{RWC}} (B,c_1,c_2,\chi,\kappa)$ 
on the $\rnu=\sen=0$ basis state.
These expectation values are obtained,
for given values of $a$ and $\lambda_0$, using the procedure call
\begin{equation}
\texttt{RWC\_exp(}
B,c_1,c_2,\kappa,a,\lambda_0
\texttt{):}
\end{equation}
(The expectation values are independent of $\chi$.)

In fact, as explained in \elrm{Appendix }\ref{Sec:BasisStuff},
the method used by the procedure \texttt{RWC\_alam}
to obtain the optimal values
$a_{\text opt}$ and $\lambda_{\text opt}$
treats $\lambda_0$ as the function of $a$ given in \eqref{Eq:EigLambda0},
when $\beta_0$ is obtained from $c_1$ and $c_2$ using \eqref{Eq:MinRWC}.
With $\lambda_0$ depending on $a$, $c_1$ and $c_2$ in this way,
the expectation value of
$\hat H_{\text{RWC}} (B,c_1,c_2,\chi,\kappa)$
on the $\rnu=\sen=0$ basis state is returned by the procedure call
\begin{equation}
\texttt{RWC\_exp\_link(}
B,c_1,c_2,\kappa,a
\texttt{):}
\end{equation}


\section{Extending the basic functionality}
\label{Sec:Extend}

The procedures \texttt{ACM\_Scale} and \texttt{ACM\_Adapt},
described in the previous section,
provide a simple means to analyse Hamiltonians in the ACM,
calculating and displaying their eigenvalues, transition rates
and amplitudes.
In this section, we describe means to extend the functionality
of these two procedures,
and describe the format of the data they return.
Procedures are described which can use the return value to readily
display eigenvalues, transition rates or amplitudes from the
calculation, further to those automatically displayed
by \texttt{ACM\_Scale} and \texttt{ACM\_Adapt}.
Moreover, the data in the return value can be readily accessed
and further processed in any way the user sees fit.

\subsection{Return values}
\label{Sec:Returns}

In this section, we describe the return values of the
procedures \texttt{ACM\_Scale} and \texttt{ACM\_Adapt}.
These contain the calculated eigenvalues on the full truncated
Hilbert space $\Hbt$ and all quadrupole amplitudes between the
corresponding eigenvectors.
These values are the raw unscaled values.

The procedures \texttt{ACM\_Scale} and \texttt{ACM\_Adapt}, invoked
as in \eqref{Eq:ACM_Scale} and \eqref{Eq:ACM_Adapt} respectively,
for the Hamiltonian $\hat H$ encoded in $\texttt{HOp}$,
both return a list of three values:
\begin{equation}\label{Eq:ACM_Scale_RET}
[\argit{eigenvals}, \argit{Melements}, \argit{Lvals}].
\end{equation}
The first of these, \argit{eigenvals}, contains the
energy eigenvalues for $\hat H$ acting on
the Hilbert space $\Hbt$ spanned by \eqref{Eq:Hilbert_Trunc}.
Specifically, \argit{eigenvals} is a list of lists, with
$\eigenvals{k}{n}$ the eigenvalue of $|n\,L\rangle$,
the $n$th lowest energy eigenstate of angular momentum $L$, where
$L$ is obtained from the list \argit{Lvals} through $L=\Lvals{k}$.
The component \argit{Lvals} of \eqref{Eq:ACM_Scale_RET} contains,
in ascending order, all values of $L$
in the range $\argsub{L}{min}\le L\le\argsub{L}{max}$,
specified by \eqref{Eq:ACM_Scale} or \eqref{Eq:ACM_Adapt},
that have non-zero dimension in $\Hbt$.
Note that these values are not necessarily consecutive.

The second component \argit{Melements} of \eqref{Eq:ACM_Scale_RET}
is a Matrix each of whose elements is itself a Matrix.%
\footnote{In our descriptions of ACM procedures, we use a capitalised
``Matrix'' to refer to the type that Maple uses for matrices
processed by its ``LinearAlgebra'' package.}
Each internal Matrix corresponds to a pair
$(\Lfin,\Lini)$ of angular momenta from the list \argit{Lvals}.
If $\kini$ and $\kfin$ are such that
$\Lini=\Lvals{\kini}$ and $\Lfin=\Lvals{\kfin}$,
then
$\Melements{\kfin}{\kini}{\ctfin}{\ctini}$
is the alternative SO(3)-reduced matrix element
\begin{equation}\label{Eq:RepEspace_TOp}
\langle{\ctfin}{\Lfin}\|\hat W\|{\ctini}{\Lini}\rangle^\natural
\equiv
\frac{\langle{\ctfin}{\Lfin}|
|\hat W|
|{\ctini}{\Lini}\rangle}
{\sqrt{2\Lfin+1}}.
\end{equation}
In default usage,
the procedures \texttt{ACM\_Scale} and \texttt{ACM\_Adapt}
calculate the matrix elements \eqref{Eq:RepEspace_TOp} for
$\hat W=\hat q$, the quadrupole operator.
However, this operator can be exchanged for another as explained
in Section \ref{Sec:ACM_alterOps} below.

The data contained in the returned values \eqref{Eq:ACM_Scale_RET}
is most conveniently
displayed using the procedures described in the next section.

Note that if \texttt{ACM\_Scale} or \texttt{ACM\_Adapt} is invoked with
the lists \argit{ratelist} and \argit{amplist} of designators both empty,
then the component \argit{Melements} of \eqref{Eq:ACM_Scale_RET}
contains no data
(because those procedures don't then need to calculate it for their
displayed output).

\subsection{Display procedures}
\label{Sec:Display}

In this section, we describe three procedures
which conveniently display eigenvalues, transition rates and amplitudes.
These display procedures are designed to make direct use of
the elements \argit{eigenvals}, \argit{Melements} and \argit{Lvals}
of the value \eqref{Eq:ACM_Scale_RET} returned by
\texttt{ACM\_Scale} and \texttt{ACM\_Adapt}.
This is most conveniently done by setting a Maple variable,
say \texttt{acmvals}, to the value \eqref{Eq:ACM_Scale_RET},
and using \texttt{acmvals[1]}, \texttt{acmvals[2]} and
\texttt{acmvals[3]} respectively for the arguments \argit{eigenvals},
\argit{Melements} and \argit{Lvals} required in the display procedures below.
In the descriptions of these procedures,
the arguments \argit{eigenvals}, \argit{Melements} and \argit{Lvals}
have, therefore, the same format as that described in
Section \ref{Sec:Returns} for the elements having the same names.
However, the display procedures here are ignorant of the origin of the
data passed to them, being concerned, only, that it has the correct format.
Thus these procedures may be used to display data produced by other means,
or perhaps by further processing the data \eqref{Eq:ACM_Scale_RET}
returned by \texttt{ACM\_Scale} and \texttt{ACM\_Adapt}.
They may also be used to display the data calculated by
procedures described later in Section \ref{Sec:ACM_Diag}.

\subsubsection{Displaying of eigenvalues}
\label{Sec:Show_Eigs}

The procedure call
\begin{equation}\label{Eq:Show_Eigs}
\texttt{Show\_Eigs(}
\argit{eigenvals}, \argit{Lvals}, \argsub{count}{eigs},
\argsub{L}{min}, \argsub{L}{max}
\texttt{):}
\end{equation}
displays, in a convenient format, the eigenvalues stored in the
argument \argit{eigenvals}.
As in Section \ref{Sec:Returns}, \argit{eigenvals} is a list,
each element of which is a list of eigenvalues that pertains to
the corresponding angular momentum value in the list \argit{Lvals}.
For each value of the angular momentum $L$ in the list \argit{Lvals}
for which $\argsub{L}{min}\le L\le\argsub{L}{max}$,
the eigenvalues are displayed on a horizontal line,
with, at most, the lowest $\argsub{count}{eigs}$ eigenvalues shown.
In the default usage,
the eigenvalues are displayed relative to the lowest value
across all the angular momenta in \argit{Lvals}.
By using \texttt{ACM\_set\_datum}, as described in
Section \ref{Sec:Code_Output}, absolute values can be displayed instead.
These eigenvalues, whether relative or absolute,
are also divided by the value $\argsub{scale}{eigs}$
specified in the most recent call to \texttt{ACM\_set\_scales}
as in \eqref{Eq:ACM_set_scales},
or as set by a more recent call to \texttt{ACM\_Adapt}.
The format of the values output is affected by the most recent call to
\texttt{ACM\_set\_output} as described in Section \ref{Sec:Code_Output}.

If the final three arguments are omitted, then $\argsub{count}{eigs}$
is taken to be the value set by the most recent call to the procedure
\texttt{ACM\_set\_listln} described in Section \ref{Sec:Code_Output}.
If the final two arguments are omitted, then eigenvalues for
all values of $L$ in \argit{Lvals} are displayed.
If only the final argument is omitted then only the eigenvalues for
angular momentum $\argsub{L}{min}$ are displayed, and then only if
this value is present in \argit{Lvals}.
The value returned by the procedure is the lowest (unscaled)
eigenvalue across all the angular momenta in \argit{Lvals}.

\subsubsection{Displaying of transition rates}
\label{Sec:Show_Rats}

The procedure call
\begin{equation}\label{Eq:Show_Rats}
\texttt{Show\_Rats(}
\argit{Melements}, \argit{Lvals},
\argit{ratelist},
\argsub{count}{rats}
\texttt{):}
\end{equation}
displays, in a convenient format, selected transition rates
obtained from the values stored in \argit{Melements},
with these values taken to be alternative
SO(3)-reduced matrix elements
$\langle{\ctfin}{\Lfin}\|\hat W\|{\ctini}{\Lini}\rangle^\natural$.
In the default usage, the transition rates that are displayed
are the values
\begin{equation}\label{Eq:RateEspace_TOp}
\frac{\langle{\ctfin\Lfin}|
|\hat W|
|{\ctini}{\Lini}\rangle^2}
{{2\Lini+1}}.
\end{equation}
These values \eqref{Eq:RateEspace_TOp}
are displayed after being divided by the current value of
$\argsub{scale}{rats}$.

In the procedure call \eqref{Eq:Show_Rats},
the first argument \argit{Melements} is precisely of the
form specified in Section \ref{Sec:Returns}, in that it
is a Matrix, each element of which is itself a Matrix.
Its elements
$\Melements{\kfin}{\kini}{\ctfin}{\ctini}$
are taken to be alternative SO(3)-reduced matrix elements
$\langle{\ctfin}{\Lfin}\|\hat W\|{\ctini}{\Lini}\rangle^\natural$,
where
$\Lini=\Lvals{\kini}$ and $\Lfin=\Lvals{\kfin}$.

The transition rates obtained from these matrix elements are displayed
only for those pairs of states specified in the argument \argit{ratelist}.
This argument has precisely the format described
in Section \ref{Sec:Code_Rates}, in that it is a list of designators,
each of which is a list of up to five integers.
A four element designator
$[\Lini,\Lfin,\ctini,\ctfin]$
indicates that the SO(3)-reduced transition rate
\eqref{Eq:RateEspace_TOp}
is to be displayed.
The zero, one, two and three element designators
$[\,]$,
$[\Lfin]$,
$[\Lini,\Lfin]$,
$[\Lini,\Lfin,\ctfin]$
indicate that particular lists of values \eqref{Eq:RateEspace_TOp}
are to be displayed.
For these, the stipulated indices are constant while the other
indices from $\{\Lini,\Lfin,\ctini,\ctfin\}$ take a range.
For $\ctini$ and $\ctfin$, transition rates
(if available in \argit{Melements}) are displayed for indices in the range
$1\le\ctini,\ctfin\le\argsub{count}{rats}$.
The indices $\Lini$ and $\Lfin$ range over all values in \argit{Lvals},
but subject to $|\Lini-\Lfin|\le L_{\hat W}$,
where $L_{\hat W}=2$ in the default usage.
For the five element designator
$[\argsubsup{L}{i}{0},\argsubsup{L}{f}{0},\ctini,\ctfin,\argsub{L}{mod}]$,
a sequence of transition rates
\eqref{Eq:RateEspace_TOp} is displayed for fixed values of
$\ctini$ and $\ctfin$, and all available pairs $\Lini$ and $\Lfin$ given by
$\Lini =\argsubsup{L}{i}{0}+k \argsub{L}{mod}$ and
$\Lfin =\argsubsup{L}{f}{0}+k \argsub{L}{mod}$,
for $k\ge0$.

When invoking the procedure \texttt{Show\_Rats},
either of the final two arguments may be omitted.
If the third argument \argit{ratelist} is omitted,
the current list of transition rate designators,
described in Section \ref{Sec:Code_Rates}, is used in its place.
If the fourth argument $\argsub{count}{rats}$ is omitted,
the value specified in the most recent call to the procedure
\texttt{ACM\_set\_listln} is used in its place.


It is possible to change the formula that is used to calculate
the displayed values from the Matrix elements
$\Melements{\kfin}{\kini}{\ctfin}{\ctini}$.
How this is done is explained in Section \ref{Sec:ACM_alterRats} below.

\subsubsection{Displaying of transition amplitudes}
\label{Sec:Show_Amps}

The procedure call
\begin{equation}\label{Eq:Show_Amps}
\texttt{Show\_Amps(}
\argit{Melements}, \argit{Lvals},
\argit{amplist},
\argsub{count}{amps}
\texttt{):}
\end{equation}
displays, in a convenient format, selected transition amplitudes
obtained from the values stored in
\argit{Melements}, with these values assumed to be
alternative SO(3)-reduced matrix elements
$\langle{\ctfin\Lfin}\|\hat W\|{\ctini}{\Lini}\rangle^\natural$.
In the default usage, the transition amplitudes that are displayed
are the values
\begin{equation}\label{Eq:AmpsXspace}
\CG{\Lini,}{\Lini,}{L_{\hat W},}
{\Lfin-\Lini}{\Lfin,}{\Lfin}\,
\langle{\ctfin\Lfin}|
|\hat W|
|{\ctini}{\Lini}\rangle^\natural ,
\end{equation}
where $L_{\hat W}=2$.
These values \eqref{Eq:AmpsXspace}
are displayed after being divided by the current value of
$\argsub{scale}{amps}$ which, as described in Section \ref{Sec:Code_Scale},
is given by $\argsub{scale}{amps}=\sqrt{\argsub{scale}{rats}}$.

The arguments \argit{Melements} and \argit{Lvals} are exactly as
for \texttt{Show\_Rats} in the previous section,
while the arguments \argit{amplist} and $\argsub{count}{amps}$
determine the transition amplitudes that are displayed in the
same way as the corresponding arguments of \eqref{Eq:Show_Rats}
determine the values displayed there.

As for \texttt{Show\_Rats},
either of the final two arguments of \texttt{Show\_Amps} may be omitted.
If the third argument \argit{amplist} is omitted,
the current list of transition amplitude designators,
described in Section \ref{Sec:Code_Rates}, is used in its place.
If the fourth argument $\argsub{count}{amps}$ is omitted,
the value $\argsub{count}{rats}$ specified in the most recent call
to the procedure \texttt{ACM\_set\_listln} is used in its place.


It is possible to change the formula that is used to calculate
the displayed values from the Matrix elements
$\Melements{\kfin}{\kini}{\ctfin}{\ctini}$.
How this is done is explained in Section \ref{Sec:ACM_alterAmps} below.

\subsection{Internal representation of Hamiltonians and other operators}
\label{Sec:Op_Encode}

Here we describe the  format used to encode all
operators on the Hilbert space $\Hb$ that are available in the ACM.
In particular, this enables \texttt{ACM\_Scale} and
\texttt{ACM\_Adapt} to analyse the full range of Hamiltonians
available in the ACM, beyond those supplied by the procedure
\texttt{ACM\_Hamiltonian} of Section \ref{Sec:Ham_Encode}.
It also enables the specification of other operators for which
the transition rates and amplitudes are to be calculated.

The Hamiltonian and other operators that are used by the
ACM code are encoded using nested Maple lists.
Each such operator is formed from the operators
given in Tables~\ref{Tab:Op_list_rad}, \ref{Tab:Op_list_sph},
and \ref{Tab:Op_list_xsp}.
\begin{table}[ht]
\begin{center}
\begin{tabular}{|c|l|c|c|}\hline
\rule{0pt}{13pt}%
\:Operator\: & \:Symbolic Name\: &\:Parity\: &\:Comment\\[3pt]
\hline
\rule{0pt}{13pt}%
$\hat\beta^2$ & \texttt{ Radial\_b2}&$+$&\\
$\hat\beta^{-2}$ & \texttt{ Radial\_bm2}&$+$&\\
$\displaystyle\frac{d^2}{d\beta^2}$ & \texttt{ Radial\_D2b}&$+$&\\
$\displaystyle{\hat\beta}\frac{d}{d\beta}$ & \texttt{ Radial\_bDb}&$+$&\\
$\hat\beta$ & \texttt{ Radial\_b} &$-$& \:Non-rational\:\\
$\hat\beta^{-1}$ & \texttt{ Radial\_bm} &$-$& \:Non-rational\:\\
$\displaystyle\frac{d}{d\beta}$& \texttt{ Radial\_Db} &$-$&
       \:Non-rational\:\\
$\hat S_{0}$& \texttt{ Radial\_S0}&$+$& \:SU(1,1) operator\:\\
$\hat S_{+}$& \texttt{ Radial\_Sp}&$+$& \:SU(1,1) operator\:\\
$\hat S_{-}$& \texttt{ Radial\_Sm}&$+$& \:SU(1,1) operator\:\\
[3pt]
\hline
\end{tabular}
\caption{Radial generating operators}
\label{Tab:Op_list_rad}
\end{center}
\end{table}
\begin{table}[htp]
\begin{center}
\begin{tabular}{|c|l|c|c|l|}\hline
\rule{0pt}{13pt}%
\:Tensor operator\:&\:Symbolic Name\:&\:A.M.\:&\:Parity\:&\:Comment\\[3pt]
\hline
\rule{0pt}{13pt}%
$\hSphericalu010{}$& \texttt{ SpHarm\_010} &0&$+$&
 \:$\hSphericalu0100=\frac{\sqrt{3}}{4\pi}$\\
$\hSphericalu112{}$& \texttt{ SpHarm\_112} &2&$-$&
 \:$\hSphericalu112M=\frac{\sqrt{15}}{4\pi}\hcalQ_M$\\
$\hSphericalu212{}$& \texttt{ SpHarm\_212} &2&$+$&
 \:$\hSphericalu212M=
         -\frac1{4\pi}\sqrt{\frac{105}{2}}[\hcalQ\otimes\hcalQ]_{2M}$\\
$\hSphericalu214{}$& \texttt{ SpHarm\_214} &4&$+$&
 \:$\hSphericalu214M=
         \frac1{4\pi}\sqrt{\frac{105}{2}}[\hcalQ\otimes\hcalQ]_{4M}$\\
$\hSphericalu310{}$& \texttt{ SpHarm\_310} &0&$-$&
 \:$\hSphericalu3100=\frac{3}{4\pi}\cos 3\gamma$\\
$\hSphericalu313{}$& \texttt{ SpHarm\_313} &3&$-$&\\
$\hSphericalu314{}$& \texttt{ SpHarm\_314} &4&$-$&\\
$\hSphericalu316{}$& \texttt{ SpHarm\_316} &6&$-$&
 \:$\hSphericalu316M=
    \frac3{4\pi}\sqrt{\frac{35}{2}}[\hcalQ\otimes \hcalQ\otimes\hcalQ]_{6M}$\\
$\hSphericalu412{}$& \texttt{ SpHarm\_412} &2&$+$&\\
$\hSphericalu414{}$& \texttt{ SpHarm\_414} &4&$+$&\\
$\hSphericalu415{}$& \texttt{ SpHarm\_415} &5&$+$&\\
$\hSphericalu416{}$& \texttt{ SpHarm\_416} &6&$+$&\\
$\hSphericalu418{}$& \texttt{ SpHarm\_418} &8&$+$&\\
$\hSphericalu512{}$& \texttt{ SpHarm\_512} &2&$-$&\\
$\hSphericalu514{}$& \texttt{ SpHarm\_514} &4&$-$&\\
$\hSphericalu515{}$& \texttt{ SpHarm\_515} &5&$-$&\\
$\hSphericalu516{}$& \texttt{ SpHarm\_516} &6&$-$&\\
$\hSphericalu517{}$& \texttt{ SpHarm\_517} &7&$-$&\\
$\hSphericalu518{}$& \texttt{ SpHarm\_518} &8&$-$&\\
$\hSphericalu51{,10}{}$& \texttt{ SpHarm\_51A} &10&$-$&\\
$\hSphericalu610{}$& \texttt{ SpHarm\_610} &0&$+$&
 \:$\hSphericalu6100=\frac{\sqrt{15}}{8\pi}(3\cos^2 3\gamma-1)$\\
$\hSphericalu613{}$& \texttt{ SpHarm\_613} &3&$+$&\\
$\hSphericalu614{}$& \texttt{ SpHarm\_614} &4&$+$&\\
$\hSphericalu616{}$& \texttt{ SpHarm\_616} &6&$+$&\\
$\hSphericalu626{}$& \texttt{ SpHarm\_626} &6&$+$&\\
$\hSphericalu617{}$& \texttt{ SpHarm\_617} &7&$+$&\\
$\hSphericalu618{}$& \texttt{ SpHarm\_618} &8&$+$&\\
$\hSphericalu619{}$& \texttt{ SpHarm\_619} &9&$+$&\\
$\hSphericalu61{,10}{}$& \texttt{ SpHarm\_61A} &10&$+$&\\
$\hSphericalu61{,12}{}$& \texttt{ SpHarm\_61C} &12&$+$&\\
[3pt]
\hline
\end{tabular}
\caption{Spherical tensor operators}
\label{Tab:Op_list_sph}
\end{center}
\end{table}
\begin{table}[ht]
\begin{center}
\begin{tabular}{|c|l|c|c|l|}\hline
\rule{0pt}{13pt}%
\:Tensor operator\:&\:Symbolic Name\:&\:A.M.\:&\:Parity\:&\:Comment\\[3pt]
\hline
\rule{0pt}{13pt}%
${\rm i}\hbar^{-1}\, \hat\pi$& \texttt{ Xspace\_Pi} &2&$-$&
\:{see \eqref{Eq:XPiME}}\\
$\hbar^{-2}\,
 [\hat\pi\otimes\hat\pi]_{2}$& \texttt{ Xspace\_PiPi2} &2&$+$&
\:{see \eqref{Eq:PixPip}}\\
$\hbar^{-2}\,
 [\hat\pi\otimes\hat\pi]_{4}$& \texttt{ Xspace\_PiPi4} &4&$+$&
\:{see \eqref{Eq:PixPip}}\\
$\hbar^{-2}\,
 [\hat\pi\otimes\hat q\otimes\hat\pi]_0$& \texttt{ Xspace\_PiqPi} &0&$+$&
\:{see \eqref{Eq:qpipi3}, \eqref{Eq:qpipi1}}\\[3pt]
\hline
\end{tabular}
\caption{Tensor operators on product space $\Hb$}
\label{Tab:Op_list_xsp}
\end{center}
\end{table}

\subsubsection{Products of scalar operators}
\label{Sec:Op_Ham}

Consider a tensor operator of the form
\begin{equation}\label{Eq:Op_list}
\hat W=
\sum_{k=1}^{N}
c_k \hat W_{k1}\otimes\hat W_{k2}\otimes\hat W_{k3}\otimes\cdots
    \otimes\hat W_{kM_k},
\end{equation}
where $N\ge0$, each $M_k\ge0$, each $c_k$ is a constant,
and each $\hat W_{ki}$ is
a zero angular momentum operator from Table~\ref{Tab:Op_list_rad},
\ref{Tab:Op_list_sph} or \ref{Tab:Op_list_xsp}
(each operator in Table~\ref{Tab:Op_list_rad} has zero angular momentum,
while for operators in Tables~\ref{Tab:Op_list_sph} and \ref{Tab:Op_list_xsp},
the angular momenta are given in the column labelled by A.M.).
The restriction to operators of zero angular momentum
ensures the operator $\hat W$ is itself of zero angular momentum,
and thus SO(3)-invariant,
permitting its use as a Hamiltonian in the ACM.
In the ACM code, this tensor operator $\hat W$ is encoded by
the nested list
\begin{equation}\label{Eq:Op_encode}
[\: [co_1, [\argsub{op}{11},op_{12},\ldots,op_{1M_1}]],
 \quad [co_2, [op_{21},op_{22},\ldots,op_{2M_2}]],
 \quad\ldots,
 \quad [co_N, [op_{N1},op_{N2},\ldots,op_{NM_N}]]
 \:],
\end{equation}
where $op_{ki}$ is the symbolic name in the tables
that corresponds to the operator $\hat W_{ki}$,
and $co_{k}$ corresponds to the constant $c_k$ in \eqref{Eq:Op_list}.
These constants can be simple numerical values or they can be functions
of the radial quantum number $\rnu$, the seniority $\sen$ and the angular
momentum $L$ of the state on which the operator (eventually) acts.
The constant $c_k$ is encoded in $co_k$ by using the
symbolic names \texttt{NUMBER}, \texttt{SENIORITY} or \texttt{ANGMOM}
to represent these three values respectively.
For example, the Hamiltonian
\begin{equation}
-\frac1{80}\nabla^2+20(-\beta^{2}+\beta^{5})
+\frac32\beta^2\cos^3 3\gamma
\end{equation}
is encoded
\begin{equation}
\begin{split}
&[\:\: [-1/80, [\texttt{Radial\_D2b}]\,],
 \: [(\texttt{SENIORITY}*(\texttt{SENIORITY}+3)+2)/80,
                   [\texttt{Radial\_bm2}]\,],\\
&
 \qquad [-20, [\texttt{Radial\_b2}]\,],
 \: [20, [\texttt{Radial\_b2,Radial\_b2,Radial\_b}]\,],\\
&
 \qquad [3/2*\texttt{Convert\_310}\symbol{"5E}3,
   [\texttt{Radial\_b2,SpHarm\_310,SpHarm\_310,SpHarm\_310}]\,]\:\:]\,,
\end{split}
\end{equation}
having used the expression \eqref{Eq:Laplacian_ME} for the
matrix elements of the Laplacian $\nabla^2$.
Here, \texttt{Convert\_310} is a symbolic name which,
as indicated in Table \ref{Tab:Con_list},
evaluates to $4\pi/3$,
this being the factor that, according to \eqref{Eq:Y310Y620},
converts from $\Sphericalu3100$ to $\cos 3\gamma$.
The particular constants listed in Table \ref{Tab:Con_list} are
the inverses of those that appear in the last column of
Table \ref{Tab:Op_list_sph}.

Note that the procedure \texttt{ACM\_Hamiltonian}, described in
Section \ref{Sec:Ham_Encode} above, produces the encoding
\eqref{Eq:Op_encode} of rational Hamiltonians in many cases of interest.

\begin{table}[ht]
\begin{center}
$
\begin{array}{|l|c|}
\hline
\rule{0pt}{13pt}%
\text{Symbolic Name}& \text{Constant}\\[3pt]
\hline
\rule{0pt}{13pt}%
\texttt{ Convert\_112}
 &\frac{4}{\sqrt{15}}\pi\\
\texttt{ Convert\_212}
 &-4\pi\sqrt{\frac{2}{105}}\\
\texttt{ Convert\_310}
 &\frac43\pi\\
\texttt{ Convert\_316}
 &\frac43\sqrt{\frac{2}{35}}\,\pi\\
\texttt{ Convert\_610}
 &\frac{8}{\sqrt{15}}\pi\\
\texttt{ Convert\_red}
 &\frac1{4\pi}\\[3pt]
\hline
\end{array}
$
\caption{Predefined constants}
\label{Tab:Con_list}
\end{center}
\end{table}

\subsubsection{Products involving one non-scalar operator}
\label{Sec:Op_Other}

Spherical tensor operators $\hat W$ of non-zero angular momentum $L_1$
are encoded in the ACM in a similar way.
Here, $\hat W$ is of the form \eqref{Eq:Op_list} as above,
with each $\hat W_{ki}$ an operator from Table~\ref{Tab:Op_list_rad},
\ref{Tab:Op_list_sph} or \ref{Tab:Op_list_xsp},
restricted such that for each $k$, there is exactly
one operator $\hat W_{ki}$ having angular momentum $L_1$, 
with all the remaining $\hat W_{ki}$ having zero angular momentum.
This operator $\hat W$ is then also encoded using the nested
list \eqref{Eq:Op_encode}.

A simple example is provided by the quadrupole operator $\hat q$,
for which the ACM code contains the assignment
\begin{equation}\label{Eq:q_Encode1}
\texttt{quad\_op:=}
[\, [\texttt{Convert\_112},\,
[\,\texttt{Radial\_b,SpHarm\_112}\,]\,]\,]:
\end{equation}
As indicated in Table \ref{Tab:Con_list},
the symbolic name \texttt{Convert\_112} evaluates to $4\pi/\sqrt{15}$,
this being the factor appearing in the expression 
$q_M = (4\pi/\sqrt{15}) \beta \Sphericalu112M$
 (see \eqref{Eq:Y112XX}).
Internally, the ACM code makes use of this encoding
\eqref{Eq:q_Encode1} of the quadrupole operator $\hat q$ to
obtain values of its transition rates.

The user may define other operators in this way, and then
determine their transition rates and amplitudes
as described in Section \ref{Sec:ACM_alter} below.

\subsubsection{Scalar-coupled products}
\label{Sec:Op_Racah}

The specification of operators described above permits at most one
operator of non-zero angular momentum in each summand of \eqref{Eq:Op_list}.
This is because, on the one hand,
a spherical tensor operator $\hat W_{ki}$ is represented in the
computer as a matrix of alternative reduced elements
$\langle{\ctfin\Lfin}\|\hat W_{ki}\|{\ctini}{\Lini}\rangle^\natural$
and, on the other, the matrix representation of the tensor product of two
non-scalar tensors is not generally a product of their reduced matrices,
whether  alternative or not.
However, with a small adjunct, it is possible to extend the
construction to include scalar-coupled products
of pairs of tensors of the same angular momentum.
The extension makes use of the expression for the matrix elements of a
scalar-coupled product of operators $[\hat A_{L_1}\otimes \hat B_{L_1}]_0$
in an orthonormal basis $\{\,|\eta LM\rangle\,\}$:%
\footnote{This is a re-expression of eqn.~(A.48) of \cite{RowanWood}.
          It also follows from eqn.~(7.1.1) of \cite{Edmonds60}.}%
%
\begin{equation}\label{Eq:NewRacah}
{\langle \xi\Lfin || \, [\hat A_{L_1}\otimes \hat B_{L_1}]_0 \, ||
    \eta\Lini\rangle}^\natural
=
 \;\delta_{\Lfin,\Lini} \frac{(-1)^{L_1}}{\sqrt{2{L_1}+1}} 
 \sum_{\zeta L'} \frac{(-1)^{\Lfin}}{\sqrt{2\Lfin+1}} 
{\langle \xi\Lfin || \hat A_{L_1} || \zeta L'\rangle}^\natural
(-1)^{L'}\sqrt{2L'+1}\, 
{\langle\zeta L'|| \hat B_{L_1} || \eta\Lini\rangle}^\natural,
\end{equation}
which is seen to be a matrix element in the product of a constant
and four matrices, two of which are diagonal.
For example, the operator
$\big[\,[\hat\pi\otimes\hat\pi]_2 \otimes \hSphericalu412{}\,\big]_0$,
which is represented as a matrix with elements
$ \StateBra{\lambda_{\sfin}}{a}{\rmu}{\sfin}{\alphafin}{L}{}
|\big[\,[\hat\pi\otimes\hat\pi]_2 \otimes \hSphericalu412{}\,\big]_0|
\StateKet{\lambda_{\sini}}{a}{\rnu}{\sini}{\alphaini}{L}{}^{\natural}$
on each $L$-space,
is encoded by the Maple list
\begin{equation}\label{Eq:RacahList}
[\, [\, 1/\texttt{sqrt}(5),\, [\,
\texttt{SpDiag\_sqLdiv},\,
\texttt{Xspace\_PiPi2},\,
\texttt{SpDiag\_sqLdim},\,
\texttt{SpHarm\_412}\,
]\,]\,]\,.
\end{equation}
Here, the use of \texttt{SpDiag\_sqLdim} and \texttt{SpDiag\_sqLdiv}
effect multiplication by diagonal matrices $R$ and $R^{-1}$
defined by the expectation values of an operator $\hat R$
having non-zero matrix elements
\begin{equation}\label{Eq:SpDiag}
\StateBraR{\lambda}{a}{\rnu}{\sen}{\alpha}{L}
|{\hat R} |
\StateKetR{\lambda}{a}{\rnu}{\sen}{\alpha}{L}^{\natural}
=(-1)^{L}\sqrt{2L+1}\,.
\end{equation}

The facility described in this section enables, in particular,
the alternative SO(3)-reduced matrix elements of operators
$[\hat\pi\otimes\hat q\otimes\hat q\otimes\cdots
  \otimes\hat q\otimes \hat\pi]_0$
to be obtained.
The commutation relations
$[\hat q_M,\hat\pi_N]= (-1)^M{\text i}\hbar\delta_{-M,N}$
enable this operator to be expressed in terms of the scalar
components of
$\hat\pi \otimes
[\hat q\otimes\hat q\otimes\cdots\otimes\hat q]_L$ for $L=2$ and
$[\hat\pi\otimes\hat\pi]_L \otimes
   [\hat q\otimes\hat q\otimes\cdots\otimes\hat q]_L$
for $L\in\{0,2,4\}$.
In both cases, each $[\hat q\otimes\hat q\otimes\cdots\otimes\hat q]_{LM}$
can be written as a linear combination
of terms $\hat\beta^n\hSphericalu{\sen}{\alpha}LM$.
Then, for the $L=0$ cases, the required matrix elements
are obtained directly after noting that
$[\hat\pi\otimes\hat\pi]_0=-\hbar^2\nabla^2$,
while for the $L\in\{2,4\}$ cases, they are obtained by calculating
those of
$\hat\beta^n\big[\,\hat\pi
        \otimes \hSphericalu \sen{\alpha}2{}\,\big]_0$,
and
$\hat\beta^n\big[\,[\hat\pi\otimes\hat\pi]_L
        \otimes \hSphericalu \sen{\alpha}L{}\,\big]_0$,
using the means described above.

Note that although the operator $[\hat\pi\otimes\hat q\otimes \hat\pi]_0$
could be tackled in this way, more accurate results are obtained
by making use of the symbolic name \texttt{Xspace\_PiqPi},
which directs the ACM code to use the exact expressions for the
matrix elements derived in \elrm{Appendix }\ref{Sec:ExtraMEs4}.


\subsection{Other transition rates}
\label{Sec:ACM_alter}

By default, the procedures \texttt{ACM\_Scale} and \texttt{ACM\_Adapt}
display transition rates and amplitudes of the quadrupole operator $\hat q$
that are calculated
using \eqref{Eq:RateEspace_q} and \eqref{Eq:AmpsXspace_q} respectively.
However, the user might wish to calculate such quantities
for another operator.
For example, computing the matrix elements of $\beta^2$
would furnish information on the beta fluctuations in model states.

In this section, we indicate how to specify that
another operator $\hat W$ be used in place of $\hat q$, and,
in addition, how to change the formulae by which the values displayed
are obtained from the reduced matrix elements of $\hat W$.

\subsubsection{Specifying the operator}
\label{Sec:ACM_alterOps}

Let the operator $\hat W$ be encoded in the Maple variable \argit{TrOp}
as described in Section \ref{Sec:Op_Encode}.
After the procedure call
\begin{equation}\label{Eq:ACM_set_transition}
\texttt{ACM\_set\_transition(}
\argit{TrOp},
\argit{show}
\texttt{):}
\end{equation}
subsequent use of the procedures \texttt{ACM\_Scale} and
\texttt{ACM\_Adapt} will calculate and display transition rates
for the operator $\hat W$.
In addition, the component \argit{Melements} of the value
\eqref{Eq:ACM_Scale_RET} returned by \texttt{ACM\_Scale}
and \texttt{ACM\_Adapt} will then contain the alternative
SO(3)-reduced matrix elements
$\langle{\ctfin\Lfin}\|\hat W\|{\ctini}{\Lini}\rangle^\natural$
exactly as described in Section \ref{Sec:Returns}.

When \eqref{Eq:ACM_set_transition} is invoked, the angular momentum
$L_{\hat W}$ of the operator $\hat W$ is determined and stored.
This is used to limit the range of angular momenta for
which lists of transition rates and amplitudes are displayed
in the procedures \texttt{ACM\_Scale}, \texttt{ACM\_Adapt},
\texttt{Show\_Rats} and \texttt{Show\_Amps}.
The procedures
\texttt{ACM\_Scale}, \texttt{ACM\_Adapt} and \texttt{Show\_Amps}
also make use of $L_{\hat W}$ to calculate the
transition amplitudes they display,
through the formula \eqref{Eq:AmpsXspace}
(\emph{not} \eqref{Eq:AmpsXspace_q}),
unless this formula has been changed
as described in Section \ref{Sec:ACM_alterAmps} below.

\subsubsection{Specifying transition rate formula}
\label{Sec:ACM_alterRats}

The procedures \texttt{ACM\_Scale} and \texttt{ACM\_Adapt} display
two sets of values calculated from the reduced matrix elements
$\langle{\ctfin\Lfin}\|\hat W\|{\ctini}{\Lini}\rangle^\natural$.
In the case of the first set, values are displayed for
various $\Lini$, $\ctini$, $\Lfin$ and $\ctfin$ determined by the
set \argit{ratelist} as described in Section \ref{Sec:Code_Rates}.
In the default implementation, these values are calculated
using \eqref{Eq:RateEspace_TOp}.
Here, we describe how to use, instead, a different expression,
so that values other than transition rates 
can be displayed.

The procedure call
\begin{equation}\label{Eq:ACM_set_rat_form}
\texttt{ACM\_set\_rat\_form(}
\argit{ratfunc},
\argit{ratformat},
\argit{ratdesg},
\argit{show}
\texttt{):}
\end{equation}
specifies how, in subsequent use of the procedures
\texttt{ACM\_Scale}, \texttt{ACM\_Adapt} and \texttt{Show\_Rats},
displayed values are calculated from the matrix elements
$\langle{\ctfin}{\Lfin}\|\hat W\|{\ctini}{\Lini}\rangle^\natural$.
It also specifies the format used to display these values.

In \eqref{Eq:ACM_set_rat_form}, the argument \argit{ratfunc}
should be a Maple procedure which takes three arguments.
The value displayed by \texttt{ACM\_Scale} and \texttt{ACM\_Adapt}
is then
$\argit{ratfunc}(\Lini,\Lfin,\argsub{M}{el})$
divided by $\argsub{scale}{rats}$,
where $\argsub{M}{el}=
\langle{\ctfin}{\Lfin}\|\hat W\|{\ctini}{\Lini}\rangle^\natural$.
The value displayed by \texttt{Show\_Rats} is obtained in the
same way, where as described in Section \ref{Sec:Show_Rats},
the elements of the argument \argit{Melements} in \eqref{Eq:Show_Rats}
are taken to be alternative reduced matrix elements
$\langle{\ctfin}{\Lfin}\|\hat W\|{\ctini}{\Lini}\rangle^\natural$.
The procedure \argit{ratfunc} can be any that takes three numerical
arguments.
The user can define such a procedure,
or make use of one of those that are predefined in the ACM code;
these predefined functions are listed in Table \ref{Tab:Fun_list}.

\begin{table}[ht]
\begin{center}
\begin{tabular}{|l|c|c|}\hline
\rule{0pt}{13pt}%
\:Function Name\: & \:Dependence on $\Lini$, $\Lfin$
                                 \& $\argsub{M}{el}$\: &\:Comment\\[3pt]
\hline
\rule{0pt}{13pt}%
%
\texttt{ quad\_amp\_fun}
 &$\argsub{M}{el}
     \CG{\Lini,}{\Lini,}
                     {L_{\hat W},}{\Lfin-\Lini}
                     {\Lfin,}{\Lfin}$
 &\text{default}\\
\texttt{ mel\_amp\_fun}
 &$\argsub{M}{el}\sqrt{2\Lfin+1}$
 &\\
\texttt{ unit\_amp\_fun}
 &$\argsub{M}{el}$
 &\\
\texttt{ quad\_rat\_fun}
 &$\argsubsup{M}{el}{2}(2\Lfin+1)/(2\Lini+1)$
 &\text{default}\\
\texttt{ mel\_rat\_fun}
 &$\argsubsup{M}{el}{2}(2\Lfin+1)$
 &\\
\texttt{ unit\_rat\_fun}
 &$\argsubsup{M}{el}{2}$
 &\\[3pt]
\hline
\end{tabular}
\caption{Predefined functions of $\Lini$,
             $\Lfin$ and $\argsub{M}{el}$}
\label{Tab:Fun_list}
\end{center}
\end{table}

If the quadrupole operator has been exchanged for another as described
in Section \ref{Sec:ACM_alterOps}, or the function used to calculate
the transition rates is altered, it would be appropriate to change the
format in which each value is displayed.
This is done with the second argument \argit{ratform} in
\eqref{Eq:ACM_set_rat_form} which should be a Maple string
that contains the desired format in the style of
a C programming language format string.%
\footnote{Consult the Maple help entry for `printf',
or any manual on the C programming language.}
Briefly, the string \argit{ratform} should contain two format
specifications `\texttt{\%s}'.
For each required $\Lini$, $\ctini$, $\Lfin$ and $\ctfin$,
the string \argit{ratform} is printed with the first `\texttt{\%s}'
replaced by `$\Lini(\ctini)\to\Lfin(\ctfin)$'
with the appropriate values substituted,
and the second `\texttt{\%s}' replaced by the calculated value
$\argit{ratfunc}(\Lini,\Lfin,\argsub{M}{el})/
\argsub{scale}{rats}$.
As an example, consider the default usage, which is obtained by invoking
\begin{equation}\label{Eq:ACM_set_rat_form_def}
\texttt{ACM\_set\_rat\_form(
quad\_rat\_fun,
"B(E2; \%s) = \%s",
"transition rates",
0 ):}
\end{equation}
Thereupon, the display of a typical transition rate takes the form
\begin{equation}\label{Eq:trans_rat_ex}
\texttt{%
B(E2; 4(1) -> 2(1)\,)\:=\:149.67
}.
\end{equation}

The third argument \argit{ratdesg} in \eqref{Eq:ACM_set_rat_form}
is a Maple string which contains a phrase,
such as \texttt{"transition rates"},
that is used in the output to introduce the values being displayed.
Having an appropriate such phrase in the output would assist anyone
reviewing a calculation, especially if either the quadrupole operator
has been exchanged for a different operator,
or the function used to calculate the transition rates is altered.

In the call \eqref{Eq:ACM_set_rat_form}, if any argument is omitted,
then the previously set value of that parameter is retained.

\subsubsection{Specifying transition amplitude formula}
\label{Sec:ACM_alterAmps}

The second set of values obtained from the reduced matrix elements
$\langle{\ctfin}{\Lfin}\|\hat W\|{\ctini}{\Lini}\rangle^\natural$
that are displayed by the procedures
\texttt{ACM\_Scale} and \texttt{ACM\_Adapt},
are for those $\Lini$, $\ctini$, $\Lfin$ and $\ctfin$ determined by the
set \argit{amplist} as described in Section \ref{Sec:Code_Rates}.
In the default implementation, these values are calculated
using \eqref{Eq:AmpsXspace}.
Here, we describe an analogue to the procedure given in the previous
subsection, which enables the expression used to calculate these
values to be replaced by another.

The procedure call
\begin{equation}\label{Eq:ACM_set_amp_form}
\texttt{ACM\_set\_amp\_form(}
\argit{ampfunc},
\argit{ampformat},
\argit{ampdesg},
\argit{show}
\texttt{):}
\end{equation}
specifies how, in subsequent use of the procedures
\texttt{ACM\_Scale}, \texttt{ACM\_Adapt} and \texttt{Show\_Amps},
displayed values are calculated from the matrix elements
$\langle{\ctfin}{\Lfin}\|\hat W\|{\ctini}{\Lini}\rangle^\natural$.
It also specifies the format used to display these values.

The format of the arguments
\argit{ampfunc}, \argit{ampformat} and \argit{ampdesg}
is precisely that of
\argit{ratfunc}, \argit{ratformat} and \argit{ratdesg}
described in the previous subsection,
and they have the analogous effect on the display
of the values designated by \argit{amplist}.
As an example, consider the default implementation,
which is obtained by invoking
\begin{equation}\label{Eq:ACM_amp_rat_form_def}
\texttt{ACM\_amp\_rat\_form(
quad\_amp\_fun,
"Amp( \%s ) = \%s",
"transition amplitudes",
0 ):}
\end{equation}
A typical value is then displayed in the form
\begin{equation}\label{Eq:trans_amp_ex}
\texttt{%
Amp( 2(1) -> 2(1) )\:=\:5.25
}.
\end{equation}

Note that having the capacity to change the functions through which the
matrix elements designated in the sets \argit{ratelist} and \argit{amplist}
are displayed in the procedures
\texttt{ACM\_Scale} and \texttt{ACM\_Adapt},
means there is little logical distinction between these two sets,
other than that those from \argit{ratelist} are displayed before those
from \argit{amplist}.
In fact, by interchanging the parameters from the default calls
to \texttt{ACM\_set\_rat\_form} and \texttt{ACM\_set\_amp\_form},
transition amplitudes would be displayed before transition rates.
However, when using \texttt{ACM\_Adapt}, there is the distinction
that the scaling factor $\argsub{scale}{rats}$
is chosen so that one of the alternative
reduced matrix elements,
processed by the function \argit{ratfunc}, takes a certain value.
Thus, if the above interchange is made,
this scaling factor would be chosen so that a certain
\emph{transition amplitude} takes a specified value.
To ensure that the correct scaling factor is then applied to the
values designated by \argit{amplist},
the function described in the next subsection would need to be called.

\subsubsection{Specifying dependence between scale factors}

In the default implementation, the procedures
\texttt{ACM\_Scale} and \texttt{ACM\_Adapt}
display the values of the transition rates \eqref{Eq:RateEspace_q}
and transition amplitudes \eqref{Eq:AmpsXspace_q}
after dividing them, respectively, by the scale factors
$\argsub{scale}{rats}$ and $\argsub{scale}{amps}$
in force at that time, as described in Section \ref{Sec:Code_Scale}.
These scale factors are similarly employed by the procedures
\texttt{Show\_Rats} and \texttt{Show\_Amps}.
Because the transition rates vary as the square of the
transition amplitudes, it is appropriate that
$\argsub{scale}{amps}=\sqrt{\argsub{scale}{rats}}$ in this case.
However, given the flexibility offered by the procedures of the
previous section, this relationship may cease to be appropriate.
Here, we provide a means to change it,
although it is expected that this would seldom need to be done.

The procedure call
\begin{equation}\label{Eq:ACM_set_sft_fun}
\texttt{ACM\_set\_sft\_fun(}
\argsub{scalefunc}{amps}, \argit{show}
\texttt{):}
\end{equation}
sets the function by which the scale factor
$\argsub{scale}{amps}$ is obtained from $\argsub{scale}{rats}$,
to the function given in the argument $\argsub{scalefunc}{amps}$.
It would be appropriate to invoke \texttt{ACM\_set\_sft\_fun}
if the arguments \argit{ratfunc} and \argit{ampfunc} supplied to the
most recent calls to
\texttt{ACM\_set\_rat\_form} and
\texttt{ACM\_set\_amp\_form}
are no longer respectively quadratic and linear in the matrix elements 
$\langle{\ctfin\Lfin}||\hat W||{\ctini}{\Lini}\rangle^\natural$.
The default functionality in which
$\argsub{scale}{amps}=\sqrt{\argsub{scale}{rats}}$
is obtained by using \texttt{ACM\_set\_sft\_fun(sqrt\_fun,$0$)}
(the procedure \texttt{sqrt\_fun}, defined in the ACM code,
returns the numerical square root of its argument:
it has the necessary \emph{type} that allows it to be passed
to procedures in the ACM code that expect a procedure argument%
\footnote{Because of its type in recent versions of Maple,
the Maple function \texttt{sqrt} cannot be used directly in this way.}).


\section{Basic procedures for ACM calculations}
\label{Sec:BasicACM}

In this and the following section, we describe a number of
other procedures that are in the ACM code.
Many of these are called by the procedures
\texttt{ACM\_Scale} and \texttt{ACM\_Adapt}
to perform the default implementation of the ACM
described in Sections \ref{Sec:Code} and \ref{Sec:Extend} above.
This additional information about the code is included 
for the benefit of anyone wishing to extend the functionality of the code.

\subsection{Dimensions and labels}
\label{Sec:Code_Dims}
A number of procedures are available to calculate the dimensions
of various spaces used by the model.
Firstly, the procedure calls
\begin{equation}\label{Proc:Basic_Dims}
\begin{split}
&\texttt{dimSO3(}
L
\texttt{):}\\
&\texttt{dimSO5(}
\sen
\texttt{):}
\end{split}
\end{equation}
return the dimensions $2L+1$ and $\frac16(\sen+1)(\sen+2)(2\sen+3)$ of
the angular momentum $L$ irrep of SO(3) and the
seniority $\sen$ irrep of SO(5) respectively.

On restriction from SO(5) to SO(3), the seniority $\sen$ irrep
of the former decomposes into various irreps of the latter,
some of which may have identical SO(3) angular momenta.
The multiplicity $d_{\sen L}$ of the SO(3) irrep of angular momentum $L$
in this restriction is given by \eqref{Eq:DimSO5>SO3}.
This value is returned by the first of the following procedures:
\begin{equation}\label{Proc:Sph_Dims}
\begin{split}
&\texttt{dimSO5r3(}
\sen,L
\texttt{):}\\
&\texttt{dimSO5r3\_allL(}
\sen
\texttt{):}\\
&\texttt{dimSO5r3\_rngVallL(}
\argsub{$\sen$}{min},\argsub{$\sen$}{max}
\texttt{):}\\
&\texttt{dimSO5r3\_rngVvarL(}
\argsub{$\sen$}{min},\argsub{$\sen$}{max},\argsub{L}{min},\argsub{L}{max}
\texttt{):}\\
\end{split}
\end{equation}
The other procedures here take the sum of this value over certain
ranges of $\sen$ and $L$.
%
The second procedure returns
$\sum_{L=0}^\infty d_{\sen L}$,
which is the total number of SO(3) irreps in the SO(5) irrep
of seniority $\sen$.
%
The third procedure returns
$\sum_{\sen=\argsub{$\sen$}{min}}^{\argsub{$\sen$}{max}}
 \sum_{L=0}^\infty d_{\sen L}$,
and the fourth returns
$\sum_{\sen=\argsub{$\sen$}{min}}^{\argsub{$\sen$}{max}}
\sum_{L=\argsub{L}{min}}^{\argsub{L}{max}}
d_{\sen L}$.

The following procedures generate labels for the representations
enumerated above.
\begin{equation}\label{Proc:Sph_Labels}
\begin{split}
&\texttt{lbsSO5r3\_allL(}
\sen 
\texttt{):}\\
&\texttt{lbsSO5r3\_rngVallL(}
\argsub{$\sen$}{min},\argsub{$\sen$}{max}
\texttt{):}\\
&\texttt{lbsSO5r3\_rngVvarL(}
\argsub{$\sen$}{min},\argsub{$\sen$}{max},\argsub{L}{min},\argsub{L}{max}
\texttt{):}\\
\end{split}
\end{equation}
The first of these returns a list of all pairs $[\alpha,L]$
for which $1\le\alpha\le d_{\sen L}$.
Thus, each element of the returned list corresponds to an
SO(3) irrep of angular momentum $L$ in the SO(5) irrep of seniority $\sen$.
In accordance with \eqref{Eq:DimSO5>SO3}, 
$L$ is limited to values $L\leq 2\sen$.
The second procedure produces a list of all triples
$[\sen,\alpha,L]$ for the range
$\argsub{$\sen$}{min}\le\sen\le\argsub{$\sen$}{max}$ of seniorities,
with $1\le\alpha\le d_{\sen L}$.
The third procedure here produces a list of all labels $[\sen,\alpha,L]$ for
$\argsub{$\sen$}{min}\le\sen\le\argsub{$\sen$}{max}$,
$1\le\alpha\le d_{\sen L}$ and 
$L_{\text{min}}\le L\le L_{\text{max}}$.

For the radial space, we have the somewhat trivial
analogues of the above procedures:
\begin{equation}\label{Proc:Rad_Labels}
\begin{split}
&\texttt{dimRadial(}
\argsub{$\rnu$}{min},\argsub{$\rnu$}{max}
\texttt{):}\\
&\texttt{lbsRadial(}
\argsub{$\rnu$}{min},\argsub{$\rnu$}{max}
\texttt{):}\\
\end{split}
\end{equation}
The first of these returns the number of radial states $\rnu$
with $\rnu_{\text{min}}\le \rnu\le \rnu_{\text{max}}$.
This number is simply $\rnu_{\text{max}}-\rnu_{\text{min}}+1$.
The second returns a list of the values $\rnu$ for which
$\argsub{$\rnu$}{min}\le\rnu\le\argsub{$\rnu$}{max}$.

From \eqref{Eq:Hilbert_Trunc}, it is seen that
each truncated Hilbert space $\Hbt$ used in the model is
a direct product of truncated spherical and radial spaces.
The dimensions of the spaces $\Hbt$, and labels for their
basis states, are obtained using the procedure calls:
\begin{equation}\label{Proc:X_Labels}
\begin{split}
&\texttt{dimXspace(}
\argsub{$\rnu$}{min},\argsub{$\rnu$}{max},
\argsub{$\sen$}{min},\argsub{$\sen$}{max},\argsub{L}{min},\argsub{L}{max}
\texttt{):}\\
&\texttt{lbsXspace(}
\argsub{$\rnu$}{min},\argsub{$\rnu$}{max},
\argsub{$\sen$}{min},\argsub{$\sen$}{max},\argsub{L}{min},\argsub{L}{max}
\texttt{):}\\
\end{split}
\end{equation}
The second of these returns a list, each element of which
is a length four list $[\rnu,\sen,\alpha,L]$.
In this list of states,
$L$ varies the slowest, followed by $\sen$, then $\alpha$,
with $\rnu$ varying the fastest.
This order of states is used when constructing the matrices
that represent the Hamiltonian and other operators
on the truncated Hilbert spaces $\Hbt$.

\subsection{Diagonalising Hamiltonians and determining transition rates}
\label{Sec:ACM_Diag}

In this section, we describe the three procedures which
\texttt{ACM\_Scale} and \texttt{ACM\_Adapt} use
to perform the main calculations.
These procedures may, of course, be used independently of one another.

\subsubsection{Obtaining matrix representations}
\label{Sec:ACM_Diag1}

Let the operator $\hat W$ be encoded in the Maple variable \argit{WOp}
as described in Section \ref{Sec:Op_Encode}.
Then, the procedure call
\begin{equation}\label{Eq:RepXspace_OpLC}
\texttt{RepXspace(}
\argit{WOp},
a,\lambda_0,
\argsub{$\rnu$}{min},\argsub{$\rnu$}{max},
\argsub{$\sen$}{min},\argsub{$\sen$}{max},\argsub{L}{min},\argsub{L}{max}
\texttt{):}
\end{equation}
obtains the representation of the operator $\hat W$ on the
truncated Hilbert space $\Hbt$ that is specified as in
Section \ref{Sec:Xspace_Params}.
The return value is a Matrix whose elements are the
alternative SO(3)-reduced matrix elements%
\footnote{%
When evaluating \eqref{Eq:WigEkX2} in the case of a Hamiltonian $\hat W$,
the Clebsch-Gordan coefficient
$\CG{\Lini}{\Mini}{0}{0}{\Lfin}{\Mfin}
  =\delta_{\Lini\Lfin}\delta_{\Mini\Mfin}$,
and thus diagonalising a matrix whose elements are \eqref{Eq:RepXspace_HOp}
gives the energy eigenvalues of $\hat W$ directly.}
\begin{equation}\label{Eq:RepXspace_HOp}
\StateBraR{\lambda_{\sfin}}{a}{\rmu}{\sfin}{\alphafin}{\Lfin}
|\hat W|
\StateKetR{\lambda_{\sini}}{a}{\rnu}{\sini}{\alphaini}{\Lini}^\natural
\equiv
\frac{\StateBraR{\lambda_{\sfin}}{a}{\rmu}{\sfin}{\alphafin}{\Lfin}
|\hat W|
\StateKetR{\lambda_{\sini}}{a}{\rnu}{\sini}{\alphaini}{\Lini}}
{\sqrt{2\Lfin+1}},
\end{equation}
where $\lambda_{\sini}$ and $\lambda_{\sfin}$ are determined
from $\lambda_0$ as described in Section \ref{Sec:Code_Basis}.
The rows and columns of this matrix are labelled by the reduced
states $\StateKetR{\lambda_\sen}{a}{\rnu}{\sen}{\alpha}{L}$
in such a way that $L$ varies the slowest, followed by $\sen$,
then $\alpha$, with $\rnu$ varying the fastest
(this accords with the order returned by the procedure
\texttt{lbsXspace} which was described in Section \ref{Sec:Code_Dims}).
The procedures \texttt{DigXspace} and \texttt{AmpXspeig},
described below, both make use of \texttt{RepXspace}.

\subsubsection{Diagonalising matrix representations}
\label{Sec:ACM_Diag2}

Let the Hamiltonian $\hat H$ be encoded in the Maple variable \argit{HOp}
as described in Section \ref{Sec:Op_Encode}
(\argit{HOp} might have been obtained using the procedure 
\texttt{ACM\_Hamiltonian} that is described
in Section \ref{Sec:Ham_Encode}).
The procedure call
\begin{equation}\label{Eq:DigXspace_OpLC}
\texttt{DigXspace(}
\argit{HOp},
a,\lambda_0,
\argsub{$\rnu$}{min},\argsub{$\rnu$}{max},
\argsub{$\sen$}{min},\argsub{$\sen$}{max},\argsub{L}{min},\argsub{L}{max}
\texttt{):}
\end{equation}
then represents $\hat H$ on the truncated Hilbert space $\Hbt$,
specified as in Section \ref{Sec:Xspace_Params},
and diagonalises it.
Prior to diagonalisation, the truncated matrix for $\hat H$
is made Hermitian by averaging it and its transpose.%
\footnote{This is done because the matrix should be Hermitian but might
not have come out as such due to working with a truncated Hilbert space.
Thus, its Hermiticity is artificially restored to avoid complex eigenvalues.}
The diagonalisation is performed separately on each $L$-space
(making use of Maple's diagonalisation procedure \texttt{Eigenvectors}),
and consequently only produces meaningful results if $\hat H$
is an SO(3)-invariant operator.
This procedure returns a list of four values:
\begin{equation}\label{Eq:DigXspace_OpLC_RET}
[\argit{eigenvals}, \argit{eigenbases}, \argit{Xparams}, \argit{Lvals}].
\end{equation}
The first of these, \argit{eigenvals}, contains the
energy eigenvalues for $\hat H$ acting on $\Hbt$.
Specifically, \argit{eigenvals} is a list of lists, with
$\eigenvals{k}{\countlab}$ the eigenvalue of the $\countlab$th lowest
energy eigenstate $|\countlab\,L\rangle$ of angular momentum $L$, where
$L$ is obtained from the list \argit{Lvals} through $L=\Lvals{k}$.
The component \argit{Lvals} of \eqref{Eq:DigXspace_OpLC_RET} contains,
in ascending order, all values of $L$
in the range $\argsub{L}{min}\le L\le\argsub{L}{max}$
that have non-zero dimension in $\Hbt$.
The eigenvalues in \argit{eigenvals} are most conveniently displayed using
the procedure \texttt{Show\_Eigs} described in Section \ref{Sec:Show_Eigs}.

The second element \argit{eigenbases} of the returned list
\eqref{Eq:DigXspace_OpLC_RET} is a list of Matrices,
one for each angular momentum value given in \argit{Lvals},
that provides the basis transformation from the original basis
\eqref{Eq:Hilbert_Trunc} to the eigenbasis of $\hat H$.
Specifically, if $L=\Lvals{k}$, then the columns of the
Matrix \argit{eigenbases}[k] are the eigenvectors
of the angular momentum $L$ subspace of $\Hbt$.

The third element \argit{Xparams} of the returned list
\eqref{Eq:DigXspace_OpLC_RET} is itself a list which contains the
original arguments $a$, $\lambda_0$, $\argsub{$\rnu$}{min}$,
$\argsub{$\rnu$}{max}$, $\argsub{$\sen$}{min}$, $\argsub{$\sen$}{max}$
to the procedure.
This is useful for passing these parameters to the procedure
described next.

\subsubsection{Transforming basis for matrix representations}
\label{Sec:ACM_Diag3}

Let the operator $\hat W$ be encoded in the Maple variable \argit{WOp}
as described in Section \ref{Sec:Op_Encode}.
Then, the procedure call
\begin{equation}\label{Eq:AmpXspeig_OpLC}
\texttt{AmpXspeig(}
\argit{WOp}, \argit{eigenbases}, \argit{Xparams}, \argit{Lvals}
\texttt{):}
\end{equation}
represents the operator $\hat W$ on the truncated Hilbert space $\Hbt$
specified by the parameters in \argit{Xparams} and \argit{Lvals},
and transforms this to the basis specified by \argit{eigenbases}.
This procedure is designed to use, for its final three arguments,
the final three elements of the list \eqref{Eq:DigXspace_OpLC_RET}
returned by the procedure \texttt{DigXspace},
and thus these arguments have the form described in
Section \ref{Sec:ACM_Diag2}.
The procedure \texttt{AmpXspeig} returns a Matrix,
each element of which is itself a Matrix.
These internal Matrices correspond to pairs $(\Lfin,\Lini)$
of angular momenta from the list \argit{Lvals}.
Specifically, if \argit{Melements} is the value returned by
\eqref{Eq:AmpXspeig_OpLC}, and $\kini$ and $\kfin$ are
such that $\Lini=\Lvals{\kini}$ and $\Lfin=\Lvals{\kfin}$,
then $\Melements{\kfin}{\kini}{\ctfin}{\ctini}$
is the alternative SO(3)-reduced matrix element
$\langle{\ctfin}{\Lfin}||\hat W||{\ctini}{\Lini}\rangle^\natural$
of the operator $\hat W$ between the $\ctini$th and $\ctfin$th states
of angular momenta $\Lini$ and $\Lfin$ respectively in \argit{eigenbases}
(when obtained as part of the return value \eqref{Eq:DigXspace_OpLC_RET}
from \texttt{DigXspace}, called as in \eqref{Eq:DigXspace_OpLC},
these states are eigenstates of the operator $\hat H$).

Note that to obtain the SO(3)-reduced transition rates
\eqref{Eq:RateEspace_TOp}
for the operator $\hat W$,
these reduced matrix elements
$\langle{\ctfin\Lfin}||\hat W||{\ctini}{\Lini}\rangle^\natural$
should be squared and multiplied by $(2\Lfin+1)/(2\Lini+1)$.
This is readily done using the procedure \texttt{Show\_Rats}
described in Section \ref{Sec:Show_Rats}.
Similarly, transition amplitudes \eqref{Eq:AmpsXspace}
are readily obtained using the procedure \texttt{Show\_Amps}
described in Section \ref{Sec:Show_Amps}.
Other values obtained from
$\langle{\ctfin\Lfin}||\hat W||{\ctini}{\Lini}\rangle^\natural$
can be displayed using \texttt{Show\_Rats} or \texttt{Show\_Amps},
as described in Sections \ref{Sec:ACM_alterRats} and
\ref{Sec:ACM_alterAmps} respectively.

\subsection{Further variants on the basis type }
\label{Sec:Code_BasisVar}

The procedure \texttt{ACM\_set\_basis\_type},
described in Section \ref{Sec:Code_Basis},
allows the user to choose between certain ways as to how
$\lambda_\sen$ depends on $\lambda_0$ and $\sen$,
for use in the procedures
\texttt{ACM\_Scale}, \texttt{ACM\_Adapt},
\texttt{RepXspace}, \texttt{DigXspace} and \texttt{AmpXspeig}.
The procedure call
\begin{equation}\label{Eq:ACM_set_lambda_fun}
\texttt{ACM\_set\_lambda\_fun(}
\argsub{func}{btype}, \argit{show}
\texttt{):}
\end{equation}
enables the dependence on $\sen$ to be more general.
Here $\argsub{func}{btype}$ should be a previously defined
Maple procedure that takes a single integer argument,
and returns an integer.
The call \eqref{Eq:ACM_set_lambda_fun} then stipulates that
$\lambda_\sen$ is given by $\lambda_0+\argsub{func}{btype}(\sen)$
(the user should ensure that this sum is always positive).
Table~\ref{Tab:vFun_list} lists the three such procedures that
are predefined in the ACM code, and which are used to
give \eqref{Eq:lambdaDefConst}, \eqref{Eq:lambdaDefSHO} and
\eqref{Eq:lambdaDef} respectively.

The procedure
\begin{equation}\label{Eq:ACM_show_lambda_fun}
\texttt{ACM\_show\_lambda\_fun(}
\argsub{$\sen$}{min}, \argsub{$\sen$}{max}
\texttt{):}
\end{equation}
returns a list of the values of the currently set
function $\argsub{func}{btype}$ for its argument taking the
range from $\argsub{$\sen$}{min}$ to $\argsub{$\sen$}{max}$.
 
\begin{table}[ht]
\begin{center}
\begin{tabular}{|l|c|c|}\hline
\rule{0pt}{13pt}%
\:Function Name\: & \:Dependence on $\sen$\: & \:Comment\\[3pt]
\hline
\rule{0pt}{13pt}%
\texttt{ lambda\_fix\_fun }
 &0
 & \:see \eqref{Eq:lambdaDefConst}\\
\texttt{ lambda\_sho\_fun }
 &$\sen$
 & \:see \eqref{Eq:lambdaDefSHO}\\
\texttt{ lambda\_acm\_fun }
 &$\sen\bmod2$
 & \:see \eqref{Eq:lambdaDef}\\[3pt]
\hline
\end{tabular}
\caption{Predefined functions of $\sen$}
\label{Tab:vFun_list}
\end{center}
\end{table}


\section{Representing operators on the component Hilbert spaces}
\label{Sec:Components}

In this section, we describe procedures by which calculations may
be performed independently on the radial Hilbert space $\RadialHilc$
and the spherical Hilbert space $\SphericalHil$.
Together, they are used by the ACM procedures discussed in
the previous sections.
However, their generality enables them to be used in many other ways,
including the construction of models other than the ACM.

\subsection{Matrix elements of radial operators}
\label{Sec:RadialCode}

\subsubsection{Products of radial operators}
\label{Sec:RadialCode1}
Let
\begin{equation}\label{Eq:ZOp_list}
\hat Z= \hat Z_N \cdots \hat Z_2 \,\hat Z_1
\end{equation}
be a product of the radial operators that appear in
Table \ref{Tab:Op_list_rad},
and let \texttt{RadOpList} be the Maple list assigned by
\begin{equation}\label{Eq:RadOp_list}
\texttt{RadOpList}:=
[op_{N},\ldots,op_{2},op_{1}]:
\end{equation}
where each $op_{k}$ is the symbolic name in
Table~\ref{Tab:Op_list_rad} that corresponds to the operator $\hat Z_{k}$.
Then, the procedure call
\begin{equation}\label{Proc:RepRadialX}
\texttt{RepRadial\_Prod(}
\texttt{RadOpList},
a,\lambda,r,
\argsub{$\rnu$}{min},\argsub{$\rnu$}{max}
\texttt{):}
\end{equation}
returns the Matrix,
having matrix elements $\FOp{a}{\lambda+r,\rmu}{\lambda\rnu}(\hat Z)$,
that represents $\hat Z$ acting between the truncated subspaces
of $\RadialHilc$ spanned by
\begin{equation}\label{Eq:Rad_Trunc}
\{
\Radial{\lambda}{a}{\rnu}
\,|\,
\argsub{$\rnu$}{min}\le\rnu\le\argsub{$\rnu$}{max}\}
\quad\text{and}\quad
\{
\Radial{\lambda+r}{a}{\rmu}
\,|\,
\argsub{$\rnu$}{min}\le\rmu\le\argsub{$\rnu$}{max}\}.
\end{equation}
Note that if \texttt{RadOpList} is an empty list then,
in effect, $\hat Z=\hat 1$ and the returned Matrix
expresses one set of basis elements in terms of the other.

This procedure is implemented by replacing the list \eqref{Eq:ZOp_list} by,
depending on the value of $r$, an equivalent list of operators
$\hat Z= \hat Z'_{M} \cdots \hat Z'_2 \,\hat Z'_1$, and then
forming
\begin{equation}\label{Eq:RepImp}
\FOp{a}{\lambda+r,\rmu}{\lambda\rnu}(\hat Z)
=
\sum_{\rmu_0,\rmu_1,\ldots,\rmu_{M}}
\FOp{a}{\lambda+r,\rmu}{\lambda_{M}\rmu_{M}}(1)\,
\FOp{a}{\lambda_M\rmu_M}{\lambda_{M-1}\rmu_{M-1}}(\hat Z_M')
\cdots
\FOp{a}{\lambda_1\rmu_1}{\lambda_0\rmu_0}(\hat Z_1')
\FOp{a}{\lambda_0\rmu_0}{\lambda\rnu}(1)\,.
\end{equation}
Here, the $\lambda_i$ are chosen so that as many of the matrices
$\FOp{a}{\lambda_i\rmu_i}{\lambda_{i-1}\rmu_{i-1}}(\hat Z_i')$
as possible are obtained directly from the analytic expressions given in
Section \ref{Sec:RadMEs}.
In addition, if $r\le0$ then $\lambda_0=\lambda$, and if
$r\ge0$ then $\lambda_M=\lambda+r$.
This ensures that at least one of the two matrices representing $\hat1$
in \eqref{Eq:RepImp} is the identity matrix,
with a non-identity matrix making use of the appropriate case
of \eqref{Eq:Radial_idsft}.%

For example, the 
call $\texttt{RepRadial\_Prod([Radial\_b,Radial\_Db]},
a,\lambda,2, \argsub{$\sen$}{min},\argsub{$\sen$}{max}
\texttt{)}$
calculates the matrix of elements
$\FOp{a}{\lambda+2,\rmu}{\lambda\rnu}(\hat\beta\,d/d\beta)$
by multiplying together the matrices for
$\FOp{a}{\lambda+2,\rmu}{\lambda+1,\rnu}(\hat\beta)$
and
$\FOp{a}{\lambda+1,\rmu}{\lambda\rnu}(d/d\beta)$
obtained from \eqref{Eq:beta5} and \eqref{Eq:beta9} respectively.
On the other hand,
$\texttt{RepRadial\_Prod([Radial\_b,Radial\_Db]},
a,\lambda,0, \argsub{$\sen$}{min},\argsub{$\sen$}{max}
\texttt{)}$
obtains the matrix of elements
$\FOp{a}{\lambda\rmu}{\lambda\rnu}(\hat\beta\,d/d\beta)$
directly from \eqref{Eq:beta4}.

In some cases, depending on the parity of $r$,
this calculation makes use of the non-analytic
matrix elements of $\hat\beta^{\pm1}$ or $d/d\beta$ obtained as
described in Section \ref{Sec:NonRat}.
Consequently, the truncation can adversely affect the calculation,
and a Hilbert space of larger dimension should be used.


Note that to obtain the matrix representing an operator $\hat X$
acting between the truncated subspaces of $\RadialHil$ spanned
by the states
\begin{equation}\label{Eq:Rad_TruncX}
\{
\RadialKet{\lambda}{a}{\rnu}
\,|\,
\argsub{$\rnu$}{min}\le\rnu\le\argsub{$\rnu$}{max}\}
\quad\text{and}\quad
\{
\RadialKet{\lambda+r}{a}{\rmu}
\,|\,
\argsub{$\rnu$}{min}\le\rmu\le\argsub{$\rnu$}{max}\}.
\end{equation}
we should, in accordance with \eqref{Eq:FOpXX}, obtain the
matrix of elements $\FOp{a}{\lambda+r,\rmu}{\lambda\rnu}(\hat Z)$
for $\hat Z=\hat\beta^2\hat X\hat\beta^{-2}$.

If \argit{RadOp} is the symbolic name for one of the radial operators
$\hat Z$ listed in Table \ref{Tab:Op_list_rad}, then
the procedure call
\begin{equation}\label{Proc:ME_Radial}
\texttt{ME\_Radial(}
\argit{RadOp},
a,\lambda,r,\rmu,\rnu
\texttt{):}
\end{equation}
returns the single matrix element
$\FOp{a}{\lambda+r,\rmu}{\lambda\rnu}(\hat Z)$.
In addition, if the first argument is set to the symbolic name
\texttt{Radial\_id}, then the matrix element
$\FOp{a}{\lambda+r,\rmu}{\lambda\rnu}(1)$ is returned.
Note that, if possible, the value returned by \eqref{Proc:ME_Radial}
is obtained by directly using one of the expressions from
Section \ref{Sec:RadMEs}.
Otherwise, the required matrix element is extracted after using
\texttt{RepRadial\_Prod} to calculate a matrix representing $\hat Z$
on a certain truncated space.
In this latter case, the truncation may affect the accuracy of the
returned value.

\subsubsection{Linear combinations of radial operators}
\label{Sec:RadialCode2}

Linear combinations of radial operators of the type \eqref{Eq:ZOp_list}
can also be readily obtained using the ACM code.
Consider an operator
\begin{equation}\label{Eq:ZOpLC_list}
\hat Z= \sum_{k=1}^N c_k \hat Z_{kM_k} \cdots \hat Z_{k2} \,\hat Z_{k1},
\end{equation}
where $N\ge0$, each $M_k\ge0$, each $c_k$ is a constant,
and each $\hat Z_{ki}$ is an operator from Table \ref{Tab:Op_list_rad}.
Then let \texttt{RadOpLC} be the Maple list assigned by
\begin{equation}\label{Eq:RadOpLC_list}
\texttt{RadOpLC}:=
[\: [co_1, [op_{1M_1},\ldots,op_{12},op_{11}]],
 \: [co_2, [op_{2M_2},\ldots,op_{22},op_{21}]],
 \:\ldots,
 \: [co_N, [op_{NM_N},\ldots,op_{N2},op_{N1}]]
 \:],
\end{equation}
where $op_{ki}$ is the symbolic name in the table
that corresponds to the operator $\hat Z_{ki}$,
and $co_{k}$ corresponds to the constant $c_k$ in \eqref{Eq:ZOpLC_list}.
Then, the procedure call
\begin{equation}\label{Proc:RepRadial_LC}
\texttt{RepRadial\_LC(}
\texttt{RadOpLC},
a,\lambda,r,\argsub{$\rnu$}{min},\argsub{$\rnu$}{max}
\texttt{):}
\end{equation}
returns the Matrix,
having elements $\FOp{a}{\lambda+r,\rmu}{\lambda\rnu}(\hat Z)$,
that represents $\hat Z$ acting between the two sets of basis states
of $\RadialHilc$ given in \eqref{Eq:Rad_Trunc}.


\subsubsection{Improving accuracy}
\label{Sec:RadialCode3}

As explained above, the implementation of \texttt{RepRadial\_Prod}
and \texttt{RepRadial\_LC} sometimes uses matrix multiplication,
and this can result in a loss of accuracy.
To ameliorate this, these two procedures take an optional seventh
parameter which specifies a larger Hilbert space on which
the representations are to be calculated,
before being truncated to the final space required.
Specifically,
\begin{equation}\label{Proc:RepRadial_Lap}
\begin{split}
&\texttt{RepRadial\_Prod(}
\texttt{RadOpList},
a,\lambda,r,
\argsub{$\rnu$}{min},\argsub{$\rnu$}{max},\argsub{$\rnu$}{lap}
\texttt{):}
\\
&\texttt{RepRadial\_LC(}
\texttt{RadOpLC},
a,\lambda,r,
\argsub{$\rnu$}{min},\argsub{$\rnu$}{max},\argsub{$\rnu$}{lap}
\texttt{):}
\end{split}
\end{equation}
each first calculate the matrix representing the operator,
\texttt{RadOpList} or \texttt{RadOpLC} respectively,
between the truncated subspaces of $\RadialHilc$ spanned by
\begin{equation}\label{Eq:Rad_TruncLap}
\{
\Radial{\lambda}{a}{\rnu}
\,|\,
\argsub{$\rnu'\!$}{min}\le\rnu\le\argsub{$\rnu'\!$}{max}\}
\quad\text{and}\quad
\{
\Radial{\lambda+r}{a}{\rmu}
\,|\,
\argsub{$\rnu'\!$}{min}\le\rmu\le\argsub{$\rnu'\!$}{max}\},
\end{equation}
where $\argsub{$\rnu'\!$}{max}=\argsub{$\rnu$}{max}+\argsub{$\rnu$}{lap}$
and $\argsub{$\rnu'\!$}{min}=\max\{0,\argsub{$\rnu$}{min}-\argsub{$\rnu$}{lap}$\}.
The resulting matrix is then truncated to provide a representation
between the subspaces of $\RadialHilc$ spanned by \eqref{Eq:Rad_Trunc}.

\subsection{\texorpdfstring
    {\fivesupthree\ Clebsch-Gordan coefficients and matrix elements
                                         of SO(5) spherical harmonics}
    {SO(5)>SO(3) Clebsch-Gordan coefficients and matrix elements
                                         of SO(5) spherical harmonics}}
\label{Sec:SphericalCode}

The \fivesupthree\ Clebsch-Gordan coefficients
\eqref{Eq:WE_SO5>3b} are obtained using the procedure call
\begin{equation}\label{Proc:CG_SO5}
\texttt{CG\_SO5r3(}
\sini,\alphaini,\Lini,
\sen,\alpha,L,
\sfin,\alphafin,\Lfin
\texttt{):}
\end{equation}
These values are obtained from the datafiles
(for the sake of efficiency, accessing one value from a
particular file loads all the values from that file into memory,
from where they are subsequently obtained).

The SO(5)-reduced matrix elements
$\SphericalBraRR{\sfin} || \hSphericalu{\sen}{}{}{} || \SphericalKetRR{\sini}$,
as specified in \eqref{Eq:SO5red_ME},
are obtained by multiplying by $(4\pi)^{-1}$
the value obtained from the procedure call
\begin{equation}\label{Proc:ME_SO5red}
\texttt{ME\_SO5red(}
\sfin,\sen,\sini
\texttt{):}
\end{equation}

The procedure call
\begin{equation}\label{Proc:ME_SO5}
\texttt{ME\_SO5r3(}
\sfin,\alphafin,\Lfin,
\sen,\alpha,L,
\sini,\alphaini,\Lini
\texttt{):}
\end{equation}
returns the SO(3)-reduced matrix element
\begin{equation}\label{Eq:ME_SO5}
4\pi
\frac{
\SphericalBraR{\sfin}{\alphafin}{\Lfin}
 | \hSphericalu{\sen}{\alpha}{L}{} |
\SphericalKetR{\sini}{\alphaini}{\Lini}
}{
\sqrt{2\Lfin+1}}\,.
\end{equation}
For example, because $\Sphericalu{3}{1}{0}{0} = (3/4\pi)\cos3\gamma$
(see \eqref{Eq:Y310Y620}), the call
$\texttt{ME\_SO5r3(}
\sfin,\alphafin,\Lfin,3,1,0,\sini,\alphaini,\Lini
\texttt{)}$
returns the value
$3\SphericalBraR{\sfin}{\alphafin}{\Lfin}
 | \cos3\gamma |
\SphericalKetR{\sini}{\alphaini}{\Lini}/
\sqrt{2\Lfin+1}$.

Let the operator $\hat Y$ be the product
\begin{equation}\label{Eq:SpOp_list}
\hat Y= \hat Y_N \cdots \hat Y_2 \,\hat Y_1,
\end{equation}
where each $Y_k$ is an SO(5) spherical harmonic,
and at most one has non-zero SO(3) angular momentum.
Then let \texttt{SpOpList} be the Maple list assigned by
\begin{equation}\label{Eq:YOp_list}
\texttt{SpOpList}:=
[op_{N},\ldots,op_{2},op_{1}]:
\end{equation}
where each $op_{k}$ is either the list $[\sen,\alpha,L]$ that labels
(the reduced) $Y_{k}$,
or alternatively, is the corresponding symbolic name
in Table~\ref{Tab:Op_list_sph}.
The procedure call
\begin{equation}\label{Proc:RepSO5}
\texttt{RepSO5r3\_Prod(SpOpList,}
\argsub{$\sen$}{min},\argsub{$\sen$}{max},\argsub{L}{min},\argsub{L}{max}
\texttt{):}
\end{equation}
then returns the Matrix whose elements are the SO(3)-reduced matrix elements
\begin{equation}\label{Eq:RepSO5}
(4\pi)^N\,
\frac{
\SphericalBraR{\sfin}{\alphafin}{\Lfin}
 | \hat Y |
\SphericalKetR{\sini}{\alphaini}{\Lini}
}{
\sqrt{2\Lfin+1}}
\end{equation}
of the operator $\hat Y$ on the truncated subspace of
$\SphericalHil$ spanned by
\begin{equation}\label{Eq:Sph_Trunc}
\{
\SphericalKet{\sen}{\alpha}{L}{M}
\,|\,
1\le\alpha\le d_{\sen L},
\argsub{$\sen$}{min} \le\sen\le \argsub{$\sen$}{max},
\argsub{L}{min} \le{L}\le \argsub{L}{max}\}.
\end{equation}
For example, the procedure call
$\texttt{RepSO5r3\_Prod([\,[3,1,0],\,[3,1,0]\,]},
\argsub{$\sen$}{min},\argsub{$\sen$}{max},\argsub{L}{min},\argsub{L}{max}
\texttt{)}$
returns a Matrix whose entries are the
alternative SO(3)-reduced matrix elements
$9\SphericalBraR{\sfin}{\alphafin}{\Lfin}
 | \cos^23\gamma |
\SphericalKetR{\sini}{\alphaini}{\Lini}/
(2\Lfin+1)$.
Alternatively, because, as indicated in Table~\ref{Tab:Op_list_sph},
the SO(3)-reduced SO(5) spherical harmonic $\Sphericalu{3}{1}{0}{}$
is encoded using \texttt{SpHarm\_310}, we obtain the same Matrix using
$\texttt{RepSO5r3\_Prod([SpHarm\_310,SpHarm\_310]},
\argsub{$\sen$}{min},\argsub{$\sen$}{max},\argsub{L}{min},\argsub{L}{max}
\texttt{)}$.

Note that, because alternative
SO(3)-reduced matrix elements for the
operator $\hat Y$ are obtained from those of the operators
$\hat Y_{k}$ using matrix multiplication, meaningful results are
obtained from \eqref{Proc:RepSO5} only where at most one of factors
$\hat Y_{k}$ of $\hat Y$ has non-zero angular momentum.
The reason for this is the same as that given in the first two sentences
of Section \ref{Sec:Op_Racah}.

\subsection{SO(3) Clebsch-Gordan coefficients}
\label{Sec:CG_SO3}

Finally, we provide a procedure which gives the usual
SO(3) Clebsch-Gordan coefficients.
The procedure call
\begin{equation}\label{Proc:CG_SO3}
\texttt{CG\_SO3(}
\Lini,\Mini,L,M,\Lfin,\Mfin
\texttt{):}
\end{equation}
returns the SO(3) Clebsch-Gordan coefficient
$\CG{\Lini}{\Mini}{L}{M}{\Lfin}{\Mfin}$.
Here, each argument is a half integer.
These Clebsch-Gordan coefficients are calculated using
\cite[eqn.~(3.6.10)]{Edmonds60}.

This procedure is only used in the ACM code to calculate the
transition rates \eqref{Eq:AmpsXspace_q} and \eqref{Eq:AmpsXspace} through
the procedure \texttt{quad\_amp\_fun} given in Table \ref{Tab:Fun_list}.
It can, of course, be used to define other such procedures.


\section{Discussion}
\label{Sec:Dis}

The ACM code makes use of the SU(1,1)$\times$SO(5) dynamical group
structure of the Bohr model to define a basis on which all
operators of interest may be expressed algebraically.
This relies, in particular, on the algebraic expressions of some
operators between different modified oscillator representations
of SU(1,1).
It also relies on the recently available \fivesupthree\
Clebsch-Gordan coefficients and SO(5)-reduced matrix elements
of SO(5) spherical harmonics.
A wide range of SO(3) and time-reversal invariant Hamiltonians
may be defined, and these may be numerically diagonalised on
a user-defined finite-dimensional subspace of the Bohr model.
Their eigenvalues are output, as well as the quadrupole moments,
transition rates and amplitudes between eigenstates of interest to the user.
The efficiency of the calculation may be fine-tuned by altering
the particular modified oscillator representations being used
and the unit $a$ in which the nuclear deformation
parameter $\beta$ is defined.
Various parameter settings enable the user to adjust and
extend the default implementation.

The code is designed to be flexible, enabling calculations to be
implemented in a range of situations, and to be extendable.
For example, calculations are easily done in the rigid-$\beta$ limit,
simply by freezing the $\beta$ degree of freedom, and in the soft-$\beta$
O(5)-symmetric Wilets-Jean limit \cite{WJ56}.
Such calculations were carried out \cite{RWC09} with an early version
of the code as well as calculations which explored the approach to
the Meyer-ter-Vehn \cite{MtV75}
and adiabatic Bohr-Mottelson \cite{BMbooks} limits.
A significant result of these calculations was that, when the
$\beta$ potential was soft enough to give relatively low
$\beta$-vibrational states in the model,
the centrifugal stretching perturbations to the rotational spectra
proved to be stronger than is commonly observed experimentally.
This suggests either that such low-energy $L=0$ excited bands
should not be interpreted as $\beta$-vibrational bands or that
the Bohr model is missing important ingredients.  
Thus, the ACM code provides powerful tools for investigating the
consistency of model interpretations of nuclear data based on
the collective model.

We emphasise that we are well aware of the limitations of the Bohr model
even when applied with Hamiltonians having freely adjustable parameters.
However, we also subscribe to the view that the importance of a model
is to reveal its deficiencies as well as its successes so that more
realistic models can be constructed.

A major limitation of the Bohr model is that it is fundamentally a
liquid drop model.
Moreover, the O(5)-invariance of its standard kinetic energy
(related to the SO(5) Casimir operator)
imposes irrotational-flow relationships between the components of
its moments of inertia \cite{RowanWood}.
To escape from this irrotational-flow limitation, it is necessary
to add vorticity degrees of freedom to the model.
Thus, it is important to understand the extent to which
experimental data favours the addition of such vorticity degrees
of freedom to the Bohr model. 
On the theoretical side, it is also important to understand the
effects of adding such degrees of freedom to the Bohr model.
For such reasons, the ACM code has been designed to admit extensions
as deemed to be desirable.
The primary characteristic of the code is that it is based on an 
SU(1,1)$\times$SO(5) dynamical group.
Thus, we anticipate that it will remain relevant for a model with
a larger dynamical group that contains this group as a subgroup.

In this context, it may be noted that the Interacting Boson Model (IBM)
\cite{AI76,AI78,IA87},
in its ${\rm U(6)} \supset {\rm U}(5) \supset {\rm O}(5)$ and
${\rm U(6)} \supset {\rm O}(6) \supset {\rm O}(5)$
dynamical symmetry limits,
is able to make good use of the SU(1,1)$\times$SO(5) dynamical group 
\cite{Rowe04b}.
Moreover, as shown in \cite{JJD74,GK80,DSI80,RoweThiamova05},
the IBM contracts to the Bohr model in these limits and thus
IBM calculations can be executed in these limits by simple extensions
of the ACM code to include $s$-boson degrees of freedom.
Such calculations have recently been pursued by Thiamova \emph{et al.}
\cite{ThiamovaRowe12}.


\addtocounter{secnumdepth}{-100}  
\section{Acknowledgements}
\addtocounter{secnumdepth}{100}  

This work was supported in part by the Natural Sciences and Engineering
Research Council of Canada.




\appendix

\section{\texorpdfstring
           {Rigid-$\beta$ models}
           {Rigid-beta models}}
\label{Sec:Rigid}

A rigid-$\beta$ model is a sub-model of the Bohr model in which the
$\beta$-vibrational degree of freedom is suppressed and, in effect,
$\beta$ takes a fixed value $\beta_0$.
A rigid-$\beta$ model is also obtained as a limit of a sequence of
Bohr models for which the potential increasingly deepens at
$\beta=\beta_0$.

Rigid-$\beta$ calculations are readily carried out using the
procedures \texttt{ACM\_Scale} and \texttt{ACM\_Adapt},
described in Section \ref{Sec:Code_Main},
after a few minor considerations.
The first is that the Hamiltonian $\hat H$ being analysed
should not contain any reference to $\beta$
(other than its rigid $\beta=\beta_0$ value;
if it does, the results will usually be meaningless).
Thus it will not be possible to use
\texttt{ACM\_Hamiltonian} to obtain the Maple encoding
of $\hat H$ in most cases.
Instead, the ACM code supplies a procedure which is
more convenient for rigid-$\beta$ Hamiltonians of the form
\begin{equation}\label{Eq:RigidHam_OpLC}
x\hat\Lambda^2
+ x_0
+ x_1\cos 3\gamma
+ x_2\cos^2 3\gamma
+ x_3\cos^3 3\gamma
+ x_4\cos^4 3\gamma
+ x_5\cos^5 3\gamma
+ x_6\cos^6 3\gamma
.
\end{equation}
For such a Hamitonian, the Maple encoding is generated by
the procedure call
\begin{equation}\label{Eq:HamRigidBeta}
\texttt{ACM\_HamRigidBeta(}
x,x_0,x_1,x_2,x_3,x_4,x_5,x_6,f
\texttt{):}
\end{equation}
where $f$ is either 0 (the default) or 1,
the former indicating that the encoding should use
only the spherical harmonic $\Sphericalu{3}{1}{0}{0}$,
and the latter indicating that $\Sphericalu{6}{1}{0}{0}$ should be
used as much as possible.%
\footnote{Expressing the Hamiltonian in terms of
$\Sphericalu{6}{1}{0}{0}$ is preferable because, when forming
the representation matrices on the truncated Hilbert space,
the fewer matrix multiplications required gives greater accuracy.
However, in the supplied data files, the
\fivesupthree\ Clebsch-Gordan coefficients for $\Sphericalu{6}{1}{0}{0}$
are not available to as high seniority as those
for $\Sphericalu{3}{1}{0}{0}$, and thus for some calculations it
might be necessary to restrict usage to the latter.}%

As when using \texttt{ACM\_Hamiltonian},
the value returned by this procedure should be assigned to a
Maple variable, such as in
\begin{equation}\label{Eq:RigidHOp}
\texttt{HOp := ACM\_HamRigidBeta(1,0,11.75,0.8)}:
\end{equation}
This variable is then used as the first argument to the
procedures \texttt{ACM\_Scale} or \texttt{ACM\_Adapt}.
Note that arguments to \texttt{ACM\_HamRigidBeta} that
are not specified are taken to be 0.

The second consideration is that in selecting the basis
\eqref{Eq:Hilbert_Trunc} of the truncated Hilbert space $\Hbt$,
the range of radial states should be suppressed.
This is done by setting $\argsub{$\rnu$}{min}=\argsub{$\rnu$}{max}$
(${}=0$, for example).
With the Hamiltonian not referencing $\beta$,
the values of $a$ and $\lambda_0$
become irrelevant.
Consequently, for the procedures \texttt{ACM\_Scale} and \texttt{ACM\_Adapt},
any values can be used for the second and third arguments.

The third and final consideration is that, in order to properly
calculate transition rates, the quadrupole operator \eqref{Eq:q_Encode1}
should be replaced by one in which the $\beta$ variable is suppressed.
To readily accommodate this, the ACM code contains the assignment
\begin{equation}\label{Eq:qRigid_Encode1}
\texttt{quadRigid\_op:=}
[\, [\texttt{Convert\_112},\,
[\,\texttt{SpHarm\_112}\,]\,]\,]:
\end{equation}
To ensure that transition rates are calculated using this operator,
it is necessary, as described in Section \ref{Sec:ACM_alterOps},
to invoke
\begin{equation}\label{Eq:qRigid_Encode2}
\texttt{ACM\_set\_transition(quadRigid\_op):}
\end{equation}
before using \eqref{Eq:RigidHOp}.
Note that the transition rates calculated using \eqref{Eq:qRigid_Encode1}
should be multiplied by $\beta_0$ to make them compatible with
those calculated using \eqref{Eq:q_Encode1}.
If required, this could be accomplished by defining an operator
similar to \eqref{Eq:qRigid_Encode1} with the value of $\beta_0$
included, and then invoking \texttt{ACM\_set\_transition}
for this operator.

\section{Hilbert space bases parameters}
\label{Sec:BasisStuff}

To determine a basis \eqref{Eq:BasisVar} of the  Hilbert space $\Hb$
completely, it is necessary
to specify values of the parameters $a$ and $\lambda_\sen$ for $\sen\ge0$.
In the ACM, $\lambda_\sen$ is chosen to depend on $\lambda_0$ in a certain
convenient way,
whereupon it is necessary to specify the values of $a$ and $\lambda_0$.
These values can have a severe effect on the efficiency of a
calculation performed using the basis \eqref{Eq:BasisVar}.
Here, for Hamiltonians of the form
\begin{equation}\label{Eq:HamGen}
\hat H_{\text{gen}}
=
{}-\frac1{2B}\nabla^2 + \pot_{\text{gen}}(\beta,\gamma),
\end{equation}
we discuss means of obtaining near-optimal values of $a$ and $\lambda_0$
based on the energy minimisation variational principle.%
\footnote{This variational principle is described in most elementary
books on Quantum Mechanics.}

\subsection{Two parameter minimisation}
\label{Sec:TwoParam}

Define $\WaveFun{\lambda_\sen}{a}{\rnu}{\sen}{\alpha}{L}{M}(\beta,\gamma,\Omega)
        = \beta^{-2}\Radial{\lambda_\sen}{a}{\rnu}(\beta)
          \Sphericalu{\sen}{\alpha}{L}{M}(\gamma,\Omega)$,
so that $\WaveFun{\lambda_\sen}{a}{\rnu}{\sen}{\alpha}{L}{M}$ is
the wave function corresponding to the state
$\StateKet{\lambda_\sen}{a}{\rnu}{\sen}{\alpha}{L}{M}$.
Then, on using \eqref{Eq:LaplacianDef} and \eqref{Eq:CasimirDef},
we obtain
\begin{equation}\label{Eq:HamAct}
\hat H_{\text{gen}}
\WaveFun{\lambda_\sen}{a}{\rnu}{\sen}{\alpha}{L}{M}
=
\frac1{\beta^2}
\Sphericalu{\sen}{\alpha}{L}{M}
\left[
-\frac{1}{2B}\frac{d^2}{d\beta^2}
+\frac{ (\sen+\afrac3/2)^2-\afrac1/4}{2B\beta^2}
+\pot_{\text{gen}}
\right]
\Radial{\lambda_\sen}{a}{\rnu}.
\end{equation}
The expectation value of $\hat H_{\text{gen}}$ on the
basis state for which $\rnu=\sen=0$ is then given by
(using the volume element \eqref{Eq:d5x})
\begin{equation}\label{Eq:genHexpt0}
\begin{split}
\StateBra{\lambda_0}{a}00100
\hat H_{\text{gen}}
\StateKet{\lambda_0}{a}00100
&=
\int_{\text{SO(3)}}\!\!\!d\Omega
\int_{0}^{\infty}\!\!\beta^4\,d\beta
\int_{0}^{{\pi}/3}\!\!\sin3\gamma\,d\gamma\,
\WaveFun{\lambda_0}{a}{0}{0}{1}{0}{0}{}^*\,
\hat H_{\text{gen}}\,
\WaveFun{\lambda_0}{a}{0}{0}{1}{0}{0}
\\
&\hskip -25mm=
\frac{3}{2}
\int_{0}^{\infty}\!\!d\beta
\int_{0}^{{\pi}/3}\!\!\sin3\gamma\,d\gamma\,
\Radial{\lambda_0}{a}{0}
\left[
  -\frac{1}{2B}\frac{d^2}{d\beta^2} + \frac{1}{B\beta^2}
  + \pot_{\text{gen}}
\right]
\Radial{\lambda_0}{a}{0},
\end{split}
\end{equation}
having used the first case of \eqref{Eq:Y310Y620}.
After defining the function $\tilde \pot_{\text{gen}}(\beta)$ of $\beta$ by
\begin{equation}\label{Eq:VradDef}
\tilde \pot_{\text{gen}}(\beta)
=\frac32\int_{0}^{{\pi}/3}\!
 \pot_{\text{gen}}(\beta,\gamma)\,\sin3\gamma\,d\gamma,
\end{equation}
and using the definition \eqref{Eq:DefFOp}, we then obtain
\begin{equation}\label{Eq:genHexpt1}
\begin{split}
\StateBra{\lambda_0}{a}00100
\hat H_{\text{gen}}
\StateKet{\lambda_0}{a}00100
&=
\FOpb{a}{\lambda_0,0}{\lambda_0,0}{-\frac{1}{2B}\frac{d^2}{d\beta^2}
                          + \frac{1}{B\beta^2} +\tilde \pot_{\text{gen}}}\\
&=
\frac{a^2}{2B}\left[ 1+\frac{9}{4(\lambda_0-1)} \right]
+
\FOp{a}{\lambda_0,0}{\lambda_0,0}(\tilde \pot_{\text{gen}}),
\end{split}
\end{equation}
after also making use of \eqref{Eq:beta2} and \eqref{Eq:beta3}.
Note that if $\pot_{\text{gen}}(\beta,\gamma)$ is independent
of $\gamma$ then \eqref{Eq:VradDef} yields
$\tilde \pot_{\text{gen}}(\beta)=\pot_{\text{gen}}(\beta,\gamma)$.
Also note that if $\tilde \pot_{\text{gen}}(\beta)$ is a polynomial
in $\beta^{\pm2}$,
then the term $\FOp{a}{\lambda_0,0}{\lambda_0,0}(\tilde \pot_{\text{gen}})$
in \eqref{Eq:genHexpt1}
is readily evaluated using the expressions of Section \ref{Sec:RadMEs}.

In accordance with the variational principle,
optimal values of $a$ and $\lambda_0$ are now obtained by determining where
\eqref{Eq:genHexpt1} is a minimum with respect to these two parameters.

\subsection{Parameter choices based on physical considerations}
\label{Sec:OneParam}

In this section, we use physical interpretations of the parameters
of $a$ and $\lambda_0$, which define the basis \eqref{Eq:BasisVar},
to simplify the process of approximating optimal values
of these parameters.

Use of \eqref{Eq:S0_ME} and \eqref{Eq:S0_realise}
(or \cite[(22.6.18)]{AbramowitzStegun68}
with $\alpha\to\lambda_\sen-1$, $n\to\rnu$ and $x\to a\beta$)
shows that the basis wave functions
$\Radial{\lambda_\sen}{a}{\rnu}(\beta)$ of $\RadialHilc$,
defined in \eqref{Eq:DefRadial}, satisfy
\begin{equation}\label{Eq:EigRadial}
\frac{1}{2a^2}
\left[
-\frac{d^2}{d\beta^2}
+\frac{(\lambda_\sen-1)^2-\afrac1/4} {\beta^2}
+   a^4\beta^2  
\right]
\Radial{\lambda_\sen}{a}{\rnu}
=
(\lambda_\sen+2\rnu)\,
\Radial{\lambda_\sen}{a}{\rnu}.
\end{equation}
Consequently, if we set
\begin{equation}\label{Eq:EigLambda}
\lambda_\sen=1+\sqrt{(\sen+\afrac3/2)^2+(a\beta_0)^4},
\end{equation}
then the Hamiltonian
\begin{equation}\label{Eq:HamRad}
\hat H^{(\sen)}_{\text{rad}} =
-\frac{d^2}{d\beta^2}
+\frac{ (\sen+\afrac3/2)^2-\afrac1/4}{\beta^2}
+\pot_{\text{rad}},
\end{equation}
with
$\pot_{\text{rad}}(\beta)=\pot_{\text{DM}}(\beta)$ given by
\begin{equation}\label{Eq:DMPotential}
\pot_{\text{DM}}(\beta)=
a^4
\left(
\beta^2+\frac{\beta_0^4}{\beta^2}
\right),
\end{equation}
has eigenfunctions
$\Radial{\lambda_\sen}{a}{\rnu}(\beta)$ for $\rnu=0,1,2,3,\ldots$.
Note that this holds for all $a>0$.
Also note that the potential $\pot_{\text{DM}}(\beta)$ has a minimum at
$\beta = \beta_0$.

The form of \eqref{Eq:EigRadial} shows that,
in effect, the parameter $a$ defines the scale of the system
determined by $\hat H^{(\sen)}_{\text{rad}}$.
This is also reflected in the fact that,
apart from an overall normalisation,
each eigenfunction $\Radial{\lambda_\sen}{a}{\rnu}(\beta)$,
defined by \eqref{Eq:DefRadial},
is a function of $a$ and $\beta$ through the combination $a\beta$ alone.
The parameter $a$ is therefore a measure of the \emph{width}
of $\Radial{\lambda_\sen}{a}{\rnu}(\beta)$ and,
correspondingly, is also related to the width of the
potential $\pot_{\text{DM}}(\beta)$ defined by \eqref{Eq:DMPotential}.
For this form of potential, the parameter $\lambda_\sen$
and the scaled value $a\beta_0$ of the potential's minimum
determine one another through \eqref{Eq:EigLambda}.
The dependence of $\Radial{\lambda_\sen}{a}{\rnu}(\beta)$ on $\beta_0$
is illustrated in Fig.~1 of \cite{RWC09}.

For a Hamiltonian $\hat H^{(\sen)}_{\text{rad}}$
of the form \eqref{Eq:HamRad} with an arbitrary potential
$\pot_{\text{rad}}(\beta)$ that has a minimum at $\beta=\beta_0$,
we may exploit these properties to obtain estimates of
the optimal values of $a$ and $\lambda_\sen$.
A robust means of doing this is to use the value of
$\beta_0$ to define $\lambda_\sen$ as the function of $a$
given by \eqref{Eq:EigLambda},
and then to use the variational method to optimize the value of $a$.
Thus, with the dependence \eqref{Eq:EigLambda} imposed,
we seek the value of the single parameter $a$ at which
\begin{equation}\label{Eq:ExptRad}
\begin{split}
\int_{0}^{\infty}
\Radial{\lambda_\sen}{a}{0}\,
\hat H^{(\sen)}_{\text{rad}}\,
\Radial{\lambda_\sen}{a}{0}
\,d\beta
&
=\FOpb{a}{\lambda_\sen,0}{\lambda_\sen,0}{-\frac{d^2}{d\beta^2}
                            +\frac{ (\sen+\afrac3/2)^2-\afrac1/4}{\beta^2}
                            +\pot_{\text{rad}}}
\\
&=
{a^2}\left[ 1+\frac{(\sen+\afrac3/2)^2}{\lambda_\sen-1} \right]
+
\FOp{a}{\lambda_\sen,0}{\lambda_\sen,0}(\pot_{\text{rad}})
\end{split}
\end{equation}
is a minimum.

For the Hamiltonian $\hat H_{\text{gen}}$ of \eqref{Eq:HamGen}
acting on $\Hb$, in the case that $\pot_{\text{gen}}(\beta,\gamma)$
is independent of $\gamma$,
comparison of \eqref{Eq:HamAct} with \eqref{Eq:HamRad} shows that
in order to apply this method, we should use
$\pot_{\text{rad}}(\beta) =2B\,\tilde \pot_{\text{gen}}(\beta)$
in \eqref{Eq:ExptRad}.
In fact, in the $\sen=0$ case,
the same substitution also applies when
$\pot_{\text{gen}}(\beta,\gamma)$ is dependent on $\gamma$,
because according to \eqref{Eq:genHexpt1},
the optimal values depend only on $\tilde \pot_{\text{gen}}(\beta)$.
In this case, we then obtain values of $a$ and $\lambda_0$ such that 
$\WaveFun{\lambda_0}{a}{0}{0}{1}{0}{0}
=\beta^{-2}\Radial{\lambda_0}{a}{0} \Sphericalu{0}{1}{0}{0}$
is a good approximation
to the ground state wave function of $\hat H_{\text{gen}}$
by treating $\lambda_0$ as dependent on $a$ via
\begin{equation}\label{Eq:EigLambda0}
\lambda_0(a)=1+\sqrt{\afrac9/4+(a\beta_0)^4},
\end{equation}
and then minimizing \eqref{Eq:genHexpt1} with respect to $a$.

The above optimisation strategy is used in \cite{RWC09}
for Hamiltonians of the form
\begin{equation}\label{Eq:HamRWC}
\hat H_{\text{RWC}} (B,c_1,c_2,\chi,\kappa)
=
{}-\frac1{2B}\nabla^2
+\frac{B}2 (c_1\beta^2+c_2\beta^4)
-\chi\beta\cos3\gamma
+\kappa\cos^2 3\gamma
\end{equation}
for $c_2\ge0$, with $c_1>0$ if $c_2=0$.
Here, the potential
\begin{equation}\label{Eq:PotRWC}
\pot_{\text{RWC}} (\beta,\gamma)
=
\frac{B}2 (c_1\beta^2+c_2\beta^4)
-\chi\beta\cos3\gamma
+\kappa\cos^2 3\gamma,
\end{equation}
leads to
\begin{equation}\label{Eq:tPotRWC}
\tilde \pot_{\text{RWC}} (\beta)
=
\frac{B}2 (c_1\beta^2+c_2\beta^4)
+\frac{\kappa}3,
\end{equation}
via \eqref{Eq:VradDef}.
Note that the potential $\tilde \pot_{\text{RWC}}(\beta)$ has a minimum
at $\beta=\beta_0$, where
\begin{equation}\label{Eq:MinRWC}
\beta_0=
\begin{cases}
  \displaystyle
  \sqrt{-\frac{c_1}{2c_2}}&\text{if $c_1<0$,}\\
  0&\text{if $c_1\ge0$.}
\end{cases}
\end{equation}

In this case where $\tilde \pot_{\text{gen}}=\tilde \pot_{\text{RWC}}$,
given by \eqref{Eq:tPotRWC},
use of \eqref{Eq:genHexpt1} and \eqref{Eq:beta1}
yields the expectation value
\begin{equation}\label{Eq:ExptRWC}
\StateBra{\lambda_0}{a}00100
\hat H_{\text{RWC}} (B,c_1,c_2,\chi,\kappa)
\StateKet{\lambda_0}{a}00100
=
\frac{a^2}{2B}\left[ 1+\frac{9}{4(\lambda_0-1)} \right]
+\frac{B}{2a^2} c_1\lambda_0
+\frac{B}{2a^4} c_2\lambda_0(\lambda_0+1)
+\frac{\kappa}{3}.
\end{equation}
Expressions \eqref{Eq:EigLambda0} and \eqref{Eq:MinRWC} are then used
to define $\lambda_0$ as a function of $a$ (and $c_1$ and $c_2$),
and their optimal values are obtained from where the
expectation value \eqref{Eq:ExptRWC} is minimal with respect to $a$.%
\footnote{The equations (81) and (76) of \cite{RWC09},
corresponding to \eqref{Eq:MinRWC} and \eqref{Eq:ExptRWC} here,
each contain typographical errors.
The account in Section 4.7 of \cite{RowanWood} deals
(correctly) with the special case in which $c_1=1-2\alpha$ and
$c_2=\alpha$, although little explicit about the optimal value of
$a$ is given.
The expressions given here should replace those in \cite{RWC09}.}
The results given in \cite{RWC09} are obtained,
after using this method, with $\lambda_\sen$
determined from $\lambda_0$ using \eqref{Eq:lambdaDef}.

\subsection{Further fine tuning}

Having determined optimal values of $a$ and $\lambda_0$,
the above analysis also suggests that for general seniority $\sen$,
a more optimal value of $\lambda_\sen$ is given by \eqref{Eq:EigLambda}.
However, efficient use of the ACM requires that
the values of $\lambda_\sen$ should satisfy \eqref{Eq:lambdapm1}.
To achieve this we can thus approximate the optimal value of
$\lambda_\sen$ by using
\begin{equation}\label{Eq:lambdaDefDavi}
\lambda_\sen=\lambda_0
+\left[
\sqrt{(\sen+\afrac3/2)^2+\beta_*^4}
-\sqrt{\afrac9/4+\beta_*^4}\,
\right]_\sen
\end{equation}
for $\sen>0$, where $\beta_*=a\beta_0$ and by $[x]_\sen$,
we mean the nearest integer to $x$ with the same parity of $\sen$
(in the event that 
$x-\sen$ is an odd integer, we set $[x]_\sen=x+1$ for definiteness).
That $\lambda_\sen$ defined in this way satisfies \eqref{Eq:lambdapm1}
is guaranteed because $0<d\lambda^{\text{DM}}_\sen/d\sen<1$.

Note that bases obtained using \eqref{Eq:lambdaDefDavi}
interpolate between those for which $\lambda_\sen=\lambda_0+\sen$
and those for which $\lambda_\sen=\lambda_0+(\sen\bmod2)$.

\subsection{Example computation}
\label{Sec:OpEx}

Here, we illustrate the efficiencies that can be achieved by
optimising the basis parameters.
We do this with the Hamiltonian
\begin{equation}\label{Eq:HamEX}
\hat H_{\text{EX}}
=
{}-\frac1{100}\nabla^2
-100\beta^2+25\beta^4
-\frac1{10}\beta\cos3\gamma.
\end{equation}
This is of the form $\hat H_{\text{RWC}}$ in \eqref{Eq:HamRWC} with
$B=50$, $c_1=-4.0$, $c_2=1.0$, $\chi=0.1$ and $\kappa=0.0$.
According to \eqref{Eq:MinRWC}, the spherically averaged potential
\eqref{Eq:tPotRWC} has a minimum at $\beta_0=\sqrt{2}$.
Then, with $\lambda_0$ the function of $a$ given by \eqref{Eq:EigLambda0},
the expectation value \eqref{Eq:ExptRWC} is minimal when $a=8.47$
(this is computed using the procedure \texttt{RWC\_alam}
described in Section \ref{Sec:Code_Params}).
Correspondingly, $\lambda_0=144.42$.

After encoding the Hamitonian $\hat H_{\text{EX}}$ using \texttt{RWC\_Ham},
we compute its eigenvalues using the procedure \texttt{ACM\_Scale}
with arguments
$\argsub{$\sen$}{min}=0$,
$\argsub{$\sen$}{max}=15$,
$\argsub{L}{min}=0$,
$\argsub{L}{max}=2$,
along with $\argsub{$\rnu$}{min}=0$ and
various values of $\argsub{$\rnu$}{max}$.
When the basis type \eqref{Eq:lambdaDef} (the default) is used with
the optimal values of $a$ and $\lambda_0$ given above,
the computations produce the lowest four eigenvalues for
$L=0$ and $L=2$ listed in Table~\ref{Tab:EigsPar}.
Note that the values are given relative to that of the $0(1)$ state and
have been scaled so that the relative value of the converged $2(1)$
state takes the value $6.0$.
This has been achieved by first using \texttt{ACM\_Adapt} for a succession
of values $\argsub{$\rnu$}{max}$ until convergence is apparent.
As explained in Section \ref{Sec:ACM_Adapt},
use of \texttt{ACM\_Adapt} sets scaling factors that are
applied to the output of subsequent uses of \texttt{ACM\_Scale}.
In this example, we see that using $\argsub{$\rnu$}{max}=2$
achieves values better than within 1\% of the converged values.

\begin{table}[ht]
\begin{center}
\begin{tabular}{|c|c|c|c|c|c|c|c|c|}\hline
\rule{0pt}{13pt}%
\:$\argsub{$\rnu$}{max}$\:&\multicolumn{4}{|c|}{$L=0$}
                         &\multicolumn{4}{|c|}{$L=2$}\\[3pt]
\hline
\rule{0pt}{13pt}%
0 &0.00 &76.27 &149.76 &271.21 &8.03 &45.62 &90.54 &121.56\\
1 &0.00 &72.60 &152.22 &268.61 &3.64 &45.40 &90.43 &116.73\\
2 &0.00 &74.44 &150.91 &270.40 &6.01 &45.43 &90.36 &119.28\\
${}\ge5$ &0.00 &74.43 &150.93 &270.42 &6.00 &45.44 &90.37 &119.27\\
\hline
\end{tabular}
\caption{Eigenvalues obtained using basis type \eqref{Eq:lambdaDef}
         and optimised parameters}
\label{Tab:EigsPar}
\end{center}
\end{table}

We contrast this with the values obtained using the standard
harmonic oscillator basis \eqref{Eq:lambdaDefSHO}
with $\lambda_0=2.5$, and the corresponding scaling $a=\sqrt{B}$.
The basis type \eqref{Eq:lambdaDefSHO} is enforced by
invoking \texttt{ACM\_set\_basis\_type(1)},
as described in Section \ref{Sec:Code_Basis}.
Use of \texttt{ACM\_Scale} for $\argsub{$\rnu$}{max}=20,25,30,32,33$
then yields the eigenvalues listed in Table \ref{Tab:EigsSHO}
(the scaling factors have been retained from the calculations above).
We see that here we require $\argsub{$\rnu$}{max}=33$
in order that the displayed values are within 1\% of the converged values.

\begin{table}[ht]
\begin{center}
\begin{tabular}{|c|c|c|c|c|c|c|c|c|}\hline
\rule{0pt}{13pt}%
\:$\argsub{$\rnu$}{max}$\:&\multicolumn{4}{|c|}{$L=0$}
                         &\multicolumn{4}{|c|}{$L=2$}\\[3pt]
\hline
\rule{0pt}{13pt}%
20 &0.00 &83.48 &362.11 &901.48 &3.98 &26.53 &147.21 &240.59\\
25 &0.00 &96.57 &118.84 &280.33 &0.67 &\phantom07.66 &\phantom064.82
                                                     &\phantom075.55\\
30 &0.00 &76.76 &154.48 &273.94 &8.12 &48.37 &\phantom093.77 &122.85\\
32 &0.00 &74.28 &150.71 &270.20 &5.86 &45.29 &\phantom090.15 &119.05\\
33 &0.00 &74.43 &150.94 &270.42 &5.97 &45.41 &\phantom090.37 &119.28\\
\hline
\end{tabular}
\caption{Eigenvalues obtained using the standard harmonic oscillator basis}
\label{Tab:EigsSHO}
\end{center}
\end{table}

For this example, we see that the basis type \eqref{Eq:lambdaDef}
with optimal parameters converges using far fewer basis states
(by a factor of at least 11) than in using the standard harmonic
oscillator basis, and is thus significantly quicker.
Of course, the contrast in efficiency between the two bases
varies significantly between examples.
For some calculations, such as for Hamiltonians
of the form \eqref{Eq:HamRWC} with $c_1\ge0$ so that $\beta_0=0$,
the standard harmonic oscillator basis performs as well as
the basis type \eqref{Eq:lambdaDef}.
A systematic study of the efficiencies obtained in computing
the ground state eigenvalue for a certain range of model
Hamiltonians is given in \cite{ACM2}.

\section{Identity radial matrix elements}
\label{Sec:Laguerre}

The generalised Laguerre polynomials
$\{{\text L}_\rmu^{(\alpha)}(x),\rmu=0,1,2,3,\ldots\}$
satisfy the recurrences
\cite{AbramowitzStegun68}:
\begin{align}
\label{Eq:LaguerreRec3}
x\,{\text L}_\rmu^{(\alpha)}(x)
&=
(\alpha+\rmu) {\text L}_{\rmu}^{(\alpha-1)}(x)
   - (\rmu+1) {\text L}_{\rmu+1}^{(\alpha-1)}(x),
\\
\label{Eq:LaguerreRec4}
{\text L}_\rmu^{(\alpha)}(x)
&=
{\text L}_{\rmu}^{(\alpha-1)}(x)
   + {\text L}_{\rmu-1}^{(\alpha)}(x),
\end{align}
 with ${\text L}_{-1}^{(\alpha)}(x)=0$.
For a non-negative integer $r$,
the first of these leads (by induction) to
\begin{equation}\label{Eq:LaguerreRec3N}
x^r{\text L}_\rmu^{(\alpha)}(x)=
\sum_{j=0}^r
(-1)^j \binom{r}{j}
\frac
{\Gamma(\alpha+\rmu+1)\,\Gamma(\rmu+j+1)}
{\Gamma(\alpha+\rmu+1-r+j)\,\Gamma(\rmu+1)}
{\text L}_{\rmu+j}^{(\alpha-r)}(x).
\end{equation}
The second \eqref{Eq:LaguerreRec4} implies that
\begin{equation}\label{Eq:LaguerreRec5}
{\text L}_\rmu^{(\alpha)}(x)=
\sum_{\sumer=0}^{\rmu}
{\text L}_{\sumer}^{(\alpha-1)}(x).
\end{equation}
Recursively applying this $r$ times leads (by induction) to
\begin{equation}\label{Eq:LaguerreRec5N}
{\text L}_\rmu^{(\alpha)}(x)=
\sum_{\sumer=0}^{\rmu}
\binom{\rmu-\sumer+r-1}{\rmu-\sumer}
{\text L}_{\sumer}^{(\alpha-r)}(x).
\end{equation}
(The case $r=0$ requires us to use
 an extended version of the binomial coefficient for which
$\binom{n-1}{n}=\delta_{n,0}$.)

Together, \eqref{Eq:LaguerreRec3} and \eqref{Eq:LaguerreRec5N} give:
\begin{equation}\label{Eq:LaguerreRec6}
x^r{\text L}_\rmu^{(\alpha)}(x)=
\sum_{j=0}^r
(-1)^j \binom{r}{j}
\frac
{\Gamma(\alpha+\rmu+1)\,\Gamma(\rmu+j+1)}
{\Gamma(\alpha+\rmu+1-r+j)\,\Gamma(\rmu+1)}
\sum_{\sumer=0}^{\rmu+j}
\binom{\rmu-\sumer+j+r-1}{\rmu-\sumer+j}
{\text L}_{\sumer}^{(\alpha-2r)}(x).
\end{equation}
Thus, for $\alpha=\lambda+2r-1$, we obtain
\begin{equation}\label{Eq:LaguerreExp}
x^r{\text L}_\rmu^{(\lambda+2r-1)}(x)=
\sum_{\sumer=0}^\infty
c^{ (2r)}_{\rmu,\sumer}\,
{\text L}_{\sumer}^{(\lambda-1)}(x),
\end{equation}
where we define
\begin{equation}\label{Eq:LaguerreRecC}
c^{(2r)}_{\rmu,\sumer}
=
\sum_{j=\max\{0,\sumer-\rmu\}}^r
(-1)^j \binom{r}{j}
\frac
{\Gamma(\lambda+2r+\rmu)\,\Gamma(\rmu+j+1)}
{\Gamma(\lambda+r+\rmu+j)\,\Gamma(\rmu+1)}
\binom{\rmu-\sumer+j+r-1}{\rmu-\sumer+j}.
\end{equation}
Note, in particular, that $c^{(2r)}_{\rmu,\sumer}=0$ if $\sumer>\rmu+r$.

By virtue of the definition \eqref{Eq:DefRadial},
we obtain
\begin{equation}\label{Eq:Radial_pl}
\begin{split}
&\int_0^\infty
\Radial{\lambda+2r}{a}{\rmu}(\beta)\,
\Radial{\lambda}{a}{\rnu}(\beta)\,d\beta\\
&\qquad
 = (-1)^{\rnu-\rmu}2a
  \sqrt{\frac{\rmu!\,\rnu!}{\Gamma(\lambda+\rmu+2r)\,\Gamma(\lambda+\rnu)}}
  \int_0^\infty
  (a\beta)^{2\lambda+2r-1} \: e^{-a^2\beta^2}
   \: {\text L}_\rmu^{(\lambda+2r-1)}(a^2\beta^2)
   \: {\text L}_\rnu^{(\lambda-1)}(a^2\beta^2)\,d\beta\\
&\qquad
 = (-1)^{\rnu-\rmu}
  \sqrt{\frac{\rmu!\,\rnu!}{\Gamma(\lambda+\rmu+2r)\,\Gamma(\lambda+\rnu)}}
  \int_0^\infty
  x^{\lambda+r-1} \: e^{-x}
   \: {\text L}_\rmu^{(\lambda+2r-1)}(x)
   \: {\text L}_\rnu^{(\lambda-1)}(x)\,dx,
\end{split}
\end{equation}
where the last line follows on making the substitution $x=a^2\beta^2$,
so that $dx=2a^2\beta\,d\beta$.
Then, on using \eqref{Eq:LaguerreExp}
and applying the Laguerre orthogonality relationship
\cite{AbramowitzStegun68}
\begin{equation}\label{Eq:Laguerre_Ortho}
  \int_0^\infty
  x^{\lambda-1} \: e^{-x}
   \: {\text L}_{\sumer}^{(\lambda-1)}(x)
   \: {\text L}_\rnu^{(\lambda-1)}(x)\,dx
   = \delta_{\sumer,\rnu} \frac{\Gamma(\lambda+\rnu)}{\rnu!},
\end{equation}
we obtain:
\begin{equation}\label{Eq:Radial_pl2}
\begin{split}
&\int_0^\infty
\Radial{\lambda+2r}{a}{\rmu}(\beta)\,
\Radial{\lambda}{a}{\rnu}(\beta)\,d\beta\\
&\qquad
 = (-1)^{\rnu-\rmu}
  \sqrt{\frac{\rmu!\,\rnu!}{\Gamma(\lambda+\rmu+2r)\,\Gamma(\lambda+\rnu)}}
  \int_0^\infty
  x^{\lambda-1} \: e^{-x}
   \sum_{\sumer=0}^\infty c^{(2r)}_{\rmu,\sumer}
   \: {\text L}_{\sumer}^{(\lambda-1)}(x)
   \: {\text L}_\rnu^{(\lambda-1)}(x)\,dx\\
&\qquad
 = (-1)^{\rnu-\rmu}
  \sqrt{\frac{\rmu!\,\Gamma(\lambda+\rnu)}{\rnu!\,\Gamma(\lambda+\rmu+2r)}}\:
   c^{ (2r)}_{\rmu,\rnu}.
\end{split}
\end{equation}

In fact, similar means could be used to obtain \emph{precise} analytic
values of the matrix elements
$\FOp{a}{\lambda+r,\rmu}{\lambda,\rnu}(\hat Z)$ of all the radial
operators considered in Section \ref{Sec:RadMEs},
for $r$ of the same parity as $\hat Z$.
This would avoid truncation errors that currently arise through
using matrix multiplication.
Moreover, this would provide, in particular, alternative proofs of
the expressions given in Section \ref{Sec:RadMEs}.


\section{\texorpdfstring
     {Reduced matrix elements of tensors coupled from $\hat q$ and $\hat\pi$}
     {Reduced matrix elements of tensors coupled from q and pi}}
\label{Sec:ExtraMEs}

From \eqref{Eq:XQME} and \eqref{Eq:XPiME},
it is evident that analytical expressions exist, in principal,
for the matrix elements of any polynomial in
the basic $\hat q_m$ and $\hat \pi_n$ operators.
We illustrate how this is done for the SO(3)-coupled tensor
operators $[\hat q\otimes \hat q]_2$,
          $[\hat q\otimes \hat q]_4$,
             $[\hat \pi\otimes \hat \pi]_2$,
             $[\hat \pi\otimes \hat \pi]_4$ and
          $[\hat\pi\otimes\hat q\otimes\hat\pi]_0$.
We also show that the method readily extends to rational
operators such as $\beta\cos3\gamma$.

\subsection{\texorpdfstring
     {Reduced matrix elements of various coupled tensors formed from $\hcalQ$}
     {Various coupled tensors formed from Q}}

For what follows, we require SO(3)-reduced matrix elements of
    $\hcalQ$,
    $[\hcalQ\otimes \hcalQ]_2$,
    $[\hcalQ\otimes \hcalQ]_4$ and
    $[\hcalQ\otimes \hcalQ\otimes \hcalQ]_0$.
For the first of these, \eqref{Eq:WE_SO5>3a} and \eqref{Eq:QME} gives
\begin{equation}\label{Eq:Qto1}
\begin{split}
&\frac{
\SphericalBraR{\sfin}{\alphafin}{\Lfin} |
         \hcalQ
         | \SphericalKetR{\sini}{\alphaini}{\Lini}
}{\sqrt{2\Lfin+1}}
=
(\sini\alphaini \Lini,\, 112 || \sfin\alphafin \Lfin)
\left(
  \delta_{\sfin,\sini+1}\,\sqrt{\frac{\sini+1}{2\sini+5}}
  +\delta_{\sfin,\sini-1}\,\sqrt{\frac{\sini+2}{2\sini+1}}
\right)
.
\end{split}
\end{equation}
To obtain the others, we use \eqref{Eq:Qexp} to infer that
(see \cite[Table I]{CRW09}
\footnote{The fifth entry in \cite[Table I]{CRW09} contains a
typographical error in that the coefficient of $\xi^{(6)}_6$ should be
$\frac{3}{16}\sqrt{\frac{35}{2}}(3\sin\gamma-\sin3\gamma)$.})
\begin{subequations}\label{Eq:Ys}
\begin{align}
\label{Eq:Y2}
[\calQ\otimes\calQ]_{LM} &= (-1)^{\frac L2}
                              4\pi\sqrt{\frac{2}{105}}\, \Sphericalu21LM
\hskip -20mm 
&(L=2,4),
\\
\label{Eq:Y3}
[\calQ\otimes\calQ\otimes\calQ]_{LM}
 &= -(-1)^{\frac L2}\frac{4\pi}{3}\sqrt{\frac{2}{35}}\, \Sphericalu31LM
\hskip -20mm 
&(L=0,6).
\end{align}
\end{subequations}
For $L\in\{2,4\}$, using first \eqref{Eq:Y2} and \eqref{Eq:WE_SO5>3a},
and then \eqref{Eq:SO5red_ME}, we obtain
\begin{equation}\label{Eq:Qto2}
\begin{split}
&\frac{
\SphericalBraR{\sfin}{\alphafin}{\Lfin} |
         [\hcalQ\otimes\hcalQ]_{L}
         | \SphericalKetR{\sini}{\alphaini}{\Lini}
}{\sqrt{2\Lfin+1}}
=
(-1)^{\frac L2}4\pi\sqrt{\frac{2}{105}}
(\sini\alphaini \Lini,\, 21L || \sfin\alphafin \Lfin)
\SphericalBraRR{\sfin}|| \hSphericalu{2}{}{}{} ||\SphericalKetRR{\sini}\\
&\hskip8mm
=(-1)^{\frac L2}
(\sini\alphaini \Lini,\, 21L || \sfin\alphafin \Lfin)
\left(
\delta_{\sfin,\sini+2}\sqrt{\frac{(\sini+1)(\sini+2)}{(2\sini+5)(2\sini+7)}}
\right.\\
&\hskip60mm
\left.
+\delta_{\sfin,\sini}\sqrt{\frac{6\sini(\sini+3)}{5(2\sini+1)(2\sini+5)}}
+\delta_{\sfin,\sini-2}\sqrt{\frac{(\sini+1)(\sini+2)}{4\sini^2-1}}
\right)
.
\end{split}
\end{equation}
Similarly, for $L\in\{0,6\}$, using first \eqref{Eq:Y3} and
\eqref{Eq:WE_SO5>3a}, and then \eqref{Eq:SO5red_ME}, we obtain
\begin{equation}\label{Eq:Qto3}
\begin{split}
&\frac{
\SphericalBraR{\sfin}{\alphafin}{\Lfin} |
         [\hcalQ\otimes\hcalQ\otimes\hcalQ]_{L}
         | \SphericalKetR{\sini}{\alphaini}{\Lini}
}{\sqrt{2\Lfin+1}}
=
-(-1)^{\frac L2}\frac{4\pi}3\sqrt{\frac{2}{35}}
(\sini\alphaini \Lini,\, 31L || \sfin\alphafin \Lfin)
\SphericalBraRR{\sfin}|| \hSphericalu{3}{}{}{} ||\SphericalKetRR{\sini}\\
&\hskip8mm
=-(-1)^{\frac L2}
(\sini\alphaini \Lini,\, 31L || \sfin\alphafin \Lfin)
\left(
\delta_{\sfin,\sini+3}
   \sqrt{\frac{(\sini+1)(\sini+2)(\sini+3)}{(2\sini+5)(2\sini+7)(2\sini+9)}}
\right.\\
&\hskip60mm
+3\delta_{\sfin,\sini+1}
   \sqrt{\frac{\sini(\sini+1)(\sini+4)}{7(2\sini+1)(2\sini+5)(2\sini+7)}}
\\
&\hskip60mm
\left.
+3\delta_{\sfin,\sini-1}\sqrt{\frac{(\sini-1)(\sini+2)(\sini+3)}
                                 {7(4\sini^2-1)(2\sini+5)}}
+\delta_{\sfin,\sini-3}\sqrt{\frac{\sini(\sini+1)(\sini+2)}
                                 {(2\sini-3)(4\sini^2-1)}}
\right)
.
\end{split}
\end{equation}

\subsection{\texorpdfstring
              {Reduced matrix elements of $[\hat q\otimes \hat q]_2$,
               $[\hat q\otimes \hat q]_4$ and $\beta\cos3\gamma$}
              {[q x q]\_2, [q x q]\_4 and beta.cos(3.gamma)}}

Because $q=\beta\calQ$, we obtain
\begin{equation}
\StateBraR{\lambda'}{a}{\rmu}{\sfin}{\alphafin}{\Lfin}
| [\hat q\otimes\hat q]_L |
\StateKetR{\lambda}{a}{\rnu}{\sini}{\alphaini}{\Lini}
=\FOp{a}{\lambda'\rmu}{\lambda\rnu}(\beta^2)\,
\SphericalBraR{\sfin}{\alphafin}{\Lfin}| [\hcalQ\otimes\hcalQ]_L |
\SphericalKetR{\sini}{\alphaini}{\Lini},
\end{equation}
for which explicit analytic expressions are 
obtained, when $\lambda'-\lambda$ is even, from
\eqref{Eq:Qto2} and the expressions of Section \ref{Sec:RadMEs}.

In addition, \eqref{Eq:Y3} and \eqref{Eq:Y310Y620} imply that
\begin{equation}\label{Eq:cos3g1}
\cos3\gamma= -\sqrt{\tfrac{35}{2}} [\hcalQ\otimes\hcalQ\otimes\hcalQ]_0,
\end{equation}
whereupon
\begin{equation}\label{Eq:cos3g2}
\StateBraR{\lambda'}{a}{\rmu}{\sfin}{\alphafin}{\Lfin}
| \beta\cos3\gamma |
\StateKetR{\lambda}{a}{\rnu}{\sini}{\alphaini}{\Lini}
= -\sqrt{\tfrac{35}{2}}\, \FOp{a}{\lambda'\rmu}{\lambda\rnu}(\beta)\,
\SphericalBraR{\sfin}{\alphafin}{\Lfin}| 
[\hcalQ\otimes\hcalQ\otimes\hcalQ]_0
|\SphericalKetR{\sini}{\alphaini}{\Lini},
\end{equation}
for which explicit analytic expressions 
are obtained when $\lambda'-\lambda$ is odd,
using \eqref{Eq:Qto3} and the expressions of Section \ref{Sec:RadMEs}.

\subsection{\texorpdfstring
     {Reduced matrix elements of $[\hat \pi\otimes \hat \pi]_2$
                             and $[\hat \pi\otimes \hat \pi]_4$}
     {[pi x pi]\_2 and [pi x pi]\_4}}

To obtain results involving the momentum operators $\hat\pi_m$,
first note that, in view of \eqref{Eq:QME}, we can define
operators $\hcalQ^{\pm}_M$ such that
$\hcalQ_M=\hcalQ^{+}_M+\hcalQ^{-}_M$ and
\begin{equation}\label{Eq:Qpm_red}
\SphericalBraRR{\sen'}|| \hcalQ^+ ||\SphericalKetRR{\sen}
 =\delta_{\sen',\sen+1}\sqrt{\frac{\sen+1}{2\sen+5}}\, ,
\qquad
\SphericalBraRR{\sen'}|| \hcalQ^- ||\SphericalKetRR{\sen}
 =\delta_{\sen',\sen-1}\sqrt{\frac{\sen+2}{2\sen+1}}\, .
\end{equation}
Let $\WaveFun{\lambda}{a}{\rnu}{\sen}{\alpha}{L}{M}$ denote
the wave function corresponding to
$\StateKet{\lambda}{a}{\rnu}{\sen}{\alpha}{L}{M}$,
so that
$\WaveFun{\lambda}{a}{\rnu}{\sen}{\alpha}{L}{M}=
\beta^{-2}\Radial{\lambda}{a}{\rnu}
\Sphericalu{\sen}{\alpha}{L}{M}$.
Then, from \eqref{Eq:Qpm_red} and \eqref{Eq:XPiME}, we obtain
\begin{equation}
\label{Eq:I.14}
\hat\pi_m \WaveFun{\lambda}{a}{\rnu}{\sen}{\alphaini}{\Lini}{\Mini}
= - \frac{{\text i} \hbar}{\beta^2} 
\left[\left( \frac{d}{d\beta}-\frac{\sen+2}{\beta}\right) \hcalQ^+_m +
\left( \frac{d}{d\beta}+\frac{\sen+1}{\beta}\right) \hcalQ^-_m \right] 
\Radial{\lambda}{a}{\rnu}\,\Sphericalu{\sen}{\alphaini}{\Lini}{\Mini}.
\end{equation}
It follows that
\begin{equation}\label{Eq:PiPiExpansion}
\begin{split}
\beta^2 \hat\pi_m \hat\pi_n
 \WaveFun{\lambda}{a}{\rnu}{\sen}{\alphaini}{\Lini}{\Mini}
&=
-\hbar^2 \left( \frac{d^2}{d\beta^2}+\frac{(\sen+2)(\sen+4)}{\beta^2}
-\frac{2\sen+5}{\beta}\frac{d}{d\beta}\right)
 \Radial{\lambda}{a}{\rnu}\,
 \hcalQ^+_m \hcalQ^+_n \Sphericalu{\sen}{\alphaini}{\Lini}{\Mini}\\
&\quad
-\hbar^2 \left( \frac{d^2}{d\beta^2}-\frac{(\sen+1)(\sen+2)}{\beta^2}\right) 
 \Radial{\lambda}{a}{\rnu}\,
 \hcalQ^+_m \hcalQ^-_n \Sphericalu{\sen}{\alphaini}{\Lini}{\Mini}\\
&\quad
-\hbar^2 \left( \frac{d^2}{d\beta^2}-\frac{(\sen+1)(\sen+2)}{\beta^2}\right) 
 \Radial{\lambda}{a}{\rnu}\,
 \hcalQ^-_m \hcalQ^+_n \Sphericalu{\sen}{\alphaini}{\Lini}{\Mini}\\
&\quad
-\hbar^2 \left( \frac{d^2}{d\beta^2}+\frac{(\sen-1)(\sen+1)}{\beta^2}
+\frac{2\sen+1}{\beta}\frac{d}{d\beta}\right) 
 \Radial{\lambda}{a}{\rnu}\,
 \hcalQ^-_m \hcalQ^-_n \Sphericalu{\sen}{\alphaini}{\Lini}{\Mini}.
\end{split}
\end{equation}
This leads to the following expressions for the non-zero SO(3)-reduced
matrix elements of $[\hat\pi\otimes\hat\pi]_L$:
\begin{subequations}\label{Eq:PixPip}
\begin{align}
&
\StateBraR{\lambda'}{a}{\rmu}{\sen+2,}{\alphafin}{\Lfin}
| [\hat\pi\otimes\hat\pi]_{L} |
\StateKetR{\lambda}{a}{\rnu}{\sen}{\alphaini}{\Lini}
\nonumber\\
&\hskip20mm
= -\hbar^2 \FOpb{a}{\lambda'\rmu}{\lambda\rnu}
                   {\frac{d^2}{d\beta^2} +\frac{(\sen+2)(\sen+4)}{\beta^2}
                                 -\frac{2\sen+5}{\beta}\frac{d}{d\beta}}\,
\SphericalBraR{\sen+2,}{\alphafin}{\Lfin}
| [\hcalQ\otimes\hcalQ]_{L} |
\SphericalKetR{\sen}{\alphaini}{\Lini},
\label{Eq:PixPip2}\\
&
\StateBraR{\lambda'}{a}{\rmu}{\sen-2,}{\alphafin}{\Lfin}
| [\hat\pi\otimes\hat\pi]_{L} |
\StateKetR{\lambda}{a}{\rnu}{\sen}{\alphaini}{\Lini}
\nonumber\\
&\hskip20mm
= -\hbar^2 \FOpb{a}{\lambda'\rmu}{\lambda\rnu}
                   {\frac{d^2}{d\beta^2} +\frac{(\sen-1)(\sen+1)}{\beta^2}
                                 +\frac{2\sen+1}{\beta}\frac{d}{d\beta}}\,
\SphericalBraR{\sen-2,}{\alphafin}{\Lfin}
| [\hcalQ\otimes\hcalQ]_{L} |
\SphericalKetR{\sen}{\alphaini}{\Lini},
\label{Eq:PixPim2}\\
\noalign{\noindent and}
&
\StateBraR{\lambda'}{a}{\rmu}{\sen}{\alphafin}{\Lfin}
| [\hat\pi\otimes\hat\pi]_{L} |
\StateKetR{\lambda}{a}{\rnu}{\sen}{\alphaini}{\Lini}
\nonumber\\
&\hskip20mm
= -\hbar^2 \FOpb{a}{\lambda'\rmu}{\lambda\rnu}
                {\frac{d^2}{d\beta^2}-\frac{(\sen+1)(\sen+2)}{\beta^2}}\,
\SphericalBraR{\sen}{\alphafin}{\Lfin}
| [\hcalQ\otimes\hcalQ]_{L} |
\SphericalKetR{\sen}{\alphaini}{\Lini}.
\label{Eq:PixPi0}
\end{align}
\end{subequations}
Explicit analytic expressions for these matrix elements
in the cases where $\lambda'-\lambda$ is even are then obtained using
\eqref{Eq:Qto2} and the expressions of Section \ref{Sec:RadMEs}.

\subsection{\texorpdfstring
     {Reduced matrix elements of $[\hat \pi\otimes\hat q\otimes\hat\pi]_0$}
     {[pi x q x pi]\_0}}
\label{Sec:ExtraMEs4}

To derive matrix elements of the scalar-coupled product
$[\hat \pi\otimes\hat q\otimes\hat\pi]_0$,
first note that the momenta $\hat\pi_N$ conjugate to the
quadrupole moments $\hat q_M$
are defined to satisfy the commutation relations
$[\hat q_M,\hat\pi_N]=
(-1)^M{\text i}\hbar\delta_{-M,N}$.
This implies that
$[\hat \pi\otimes\hat q\otimes\hat\pi]_0
=[\hat q\otimes\hat\pi\otimes\hat\pi]_0$,
whereupon
equations \eqref{Eq:PixPip2} and \eqref{Eq:PixPim2}
(or \eqref{Eq:PiPiExpansion})
immediately lead to
\begin{subequations}\label{Eq:qpipi3}
\begin{align}
\nonumber
&
\StateBraR{\lambda'}{a}{\rmu}{\sen+3,}{\alphafin}{L}
| [\hat\pi\otimes\hat q\otimes\hat\pi]_0 |
\StateKetR{\lambda}{a}{\rnu}{\sen}{\alphaini}{L}
\\
\label{Eq:qpipi_p3}
&\hskip20mm
= -\hbar^2 \FOpb{a}{\lambda'\rmu}{\lambda\rnu}
                  {\beta\frac{d^2}{d\beta^2}+\frac{(\sen+2)(\sen+4)}{\beta}
                                              -(2\sen+5)\frac{d}{d\beta}}\,
\SphericalBraR{\sen+3,}{\alphafin}{L}
| [\hcalQ\otimes\hcalQ\otimes\hcalQ]_0 |
\SphericalKetR{\sen}{\alphaini}{L}
\end{align}
and
\begin{align}
\nonumber
&
\StateBraR{\lambda'}{a}{\rmu}{\sen-3,}{\alphafin}{L}
| [\hat\pi\otimes\hat q\otimes\hat\pi]_0 |
\StateKetR{\lambda}{a}{\rnu}{\sen}{\alphaini}{L}
\\
\label{Eq:qpipi_m3}
&\hskip20mm
= -\hbar^2 \FOpb{a}{\lambda'\rmu}{\lambda\rnu}
                  {\beta\frac{d^2}{d\beta^2}+\frac{(\sen-1)(\sen+1)}{\beta}
                                              +(2\sen+1)\frac{d}{d\beta}}\,
\SphericalBraR{\sen-3,}{\alphafin}{L}
| [\hcalQ\otimes\hcalQ\otimes\hcalQ]_0 |
\SphericalKetR{\sen}{\alphaini}{L}.
\end{align}
\end{subequations}
Explicit analytic expressions for these \emph{stretched}
matrix elements of $[\hat\pi\otimes \hat q\otimes\hat\pi]_0$
in the cases where $\lambda'-\lambda$ is odd are then obtained using
\eqref{Eq:Qto3} and the expressions of Section \ref{Sec:RadMEs}.
Expressions for the other non-zero matrix elements of 
$[\hat\pi\otimes\hat q\otimes \hat\pi]_0$
are obtained,
after again noting that
$[\hat \pi\otimes\hat q\otimes\hat\pi]_0
=[\hat q\otimes\hat\pi\otimes\hat\pi]_0$,
by combining
\begin{equation}
\SphericalBraR{\sen\pm1,}{\alphafin}{L}
| [\hcalQ\otimes\hcalQ\otimes\hcalQ]_0 |
\SphericalKetR{\sen}{\alphaini}{L}
=
\SphericalBraR{\sen\pm1,}{\alphafin}{L}
| [(\hcalQ^{+}+\hcalQ^{-})\otimes\hcalQ\otimes\hcalQ]_0 |
\SphericalKetR{\sen}{\alphaini}{L},
\end{equation}
with
\eqref{Eq:PixPip}.
This yields
\begin{subequations}\label{Eq:qpipi1}
\begin{align}
\nonumber
&
\StateBraR{\lambda'}{a}{\rmu}{\sen+1,}{\alphafin}{L}
| [\hat\pi\otimes\hat q\otimes\hat\pi]_0 |
\StateKetR{\lambda}{a}{\rnu}{\sen}{\alphaini}{L}\\
\nonumber
&\hskip15mm
=  -\hbar^2 \FOpb{a}{\lambda'\rmu}{\lambda\rnu}
                 {\beta\frac{d^2}{d\beta^2}-\frac{(\sen+1)(\sen+2)}{\beta}}\,
      \SphericalBraR{\sen+1,}{\alphafin}{L}
      | [\hcalQ^+\otimes\hcalQ\otimes\hcalQ]_0 |
      \SphericalKetR{\sen}{\alphaini}{L}\\
\nonumber
&\hskip25mm
  -\hbar^2 \FOpb{a}{\lambda'\rmu}{\lambda\rnu}
                {\beta\frac{d^2}{d\beta^2}+\frac{(\sen+2)(\sen+4)}{\beta}
                      - (2\sen+5)\frac{d}{d\beta}}\,
      \SphericalBraR{\sen+1,}{\alphafin}{L}
      | [\hcalQ^-\otimes\hcalQ\otimes\hcalQ]_0 |
      \SphericalKetR{\sen}{\alphaini}{L}\\
\nonumber
&\hskip15mm
=  -\hbar^2 \FOpb{a}{\lambda'\rmu}{\lambda\rnu}
                 {\beta\frac{d^2}{d\beta^2}-\frac{(\sen+1)(\sen+2)}{\beta}}\,
      \SphericalBraR{\sen+1,}{\alphafin}{L}
      | [\hcalQ\otimes\hcalQ\otimes\hcalQ]_0 |
      \SphericalKetR{\sen}{\alphaini}{L}\\
\label{Eq:qpipi_p1}
&\hskip25mm
  +\hbar^2 (2\sen+5) \FOpb{a}{\lambda'\rmu}{\lambda\rnu}
                       {\frac{d}{d\beta} -\frac{\sen+2}{\beta}}\,
      \SphericalBraR{\sen+1,}{\alphafin}{L}
      | [\hcalQ^-\otimes\hcalQ\otimes\hcalQ]_0 |
      \SphericalKetR{\sen}{\alphaini}{L}
\end{align}
and
\begin{align}
\nonumber
&
\StateBraR{\lambda'}{a}{\rmu}{\sen-1,}{\alphafin}{L}
| [\hat\pi\otimes\hat q\otimes\hat\pi]_0 |
\StateKetR{\lambda}{a}{\rnu}{\sen}{\alphaini}{L}\\
\nonumber
&\hskip15mm
=  -\hbar^2 \FOpb{a}{\lambda'\rmu}{\lambda\rnu}
                 {\beta\frac{d^2}{d\beta^2}-\frac{(\sen+1)(\sen+2)}{\beta}}\,
      \SphericalBraR{\sen-1,}{\alphafin}{L}
      | [\hcalQ^-\otimes\hcalQ\otimes\hcalQ]_0 |
      \SphericalKetR{\sen}{\alphaini}{L}\\
\nonumber
&\hskip25mm
  -\hbar^2 \FOpb{a}{\lambda'\rmu}{\lambda\rnu}
                {\beta\frac{d^2}{d\beta^2}+\frac{(\sen-1)(\sen+1)}{\beta}
                      + (2\sen+1)\frac{d}{d\beta}}\,
      \SphericalBraR{\sen-1,}{\alphafin}{L}
      | [\hcalQ^+\otimes\hcalQ\otimes\hcalQ]_0 |
      \SphericalKetR{\sen}{\alphaini}{L}\\
\nonumber
&\hskip15mm
=  -\hbar^2 \FOpb{a}{\lambda'\rmu}{\lambda\rnu}
                 {\beta\frac{d^2}{d\beta^2}-\frac{(\sen+1)(\sen+2)}{\beta}}\,
      \SphericalBraR{\sen-1,}{\alphafin}{L}
      | [\hcalQ\otimes\hcalQ\otimes\hcalQ]_0 |
      \SphericalKetR{\sen}{\alphaini}{L}\\
\label{Eq:qpipi_m1}
&\hskip25mm
  -\hbar^2 (2\sen+1) \FOpb{a}{\lambda'\rmu}{\lambda\rnu}
                       {\frac{d}{d\beta} +\frac{\sen+1}{\beta}}\,
      \SphericalBraR{\sen-1,}{\alphafin}{L}
      | [\hcalQ^+\otimes\hcalQ\otimes\hcalQ]_0 |
      \SphericalKetR{\sen}{\alphaini}{L}.
\end{align}
\end{subequations}

To evaluate these, we require, in addition to expressions for the
reduced matrix elements 
$\SphericalBraR{\sen\pm1,}{\alphafin}{L}
      | [\hcalQ\otimes\hcalQ\otimes\hcalQ]_0 |
      \SphericalKetR{\sen}{\alphaini}{L}$
provided by \eqref{Eq:Qto3}, expressions for the reduced matrix elements 
$\SphericalBraR{\sen\pm1,}{\alphafin}{L}
      | [\hcalQ^{\mp}\otimes\hcalQ\otimes\hcalQ]_0 |
      \SphericalKetR{\sen}{\alphaini}{L}$.
To obtain the latter, first note that \eqref{Eq:WE_SO5>3a} implies that
\begin{equation}
\label{Eq:QpmxQxQmp}
\frac{ \SphericalBraR{\sen\pm1,}{\alphafin}{L} |
         [\hcalQ^\mp\otimes\hcalQ\otimes\hcalQ]_{0}
         | \SphericalKetR{\sen}{\alphaini}{L}
       }{\sqrt{2L+1}\,(\sen\alphaini L,\, 310 || \sen\pm1,\alphafin L)}
=
\SphericalBraRR{\sen\pm1} ||\,[[\hcalQ^\mp\otimes\hcalQ\otimes\hcalQ]]^3
    || \SphericalKetRR{\sen},
\end{equation}
where 
$\SphericalBraRR{\sen\pm1} ||\,[[\hcalQ^\mp\otimes\hcalQ\otimes\hcalQ]]^3
    || \SphericalKetRR{\sen}$ is an SO(5)-reduced matrix element.
To evaluate this, we consider the special cases of the left side
with $L=\text{min}\{2\sen,2(\sen\pm1)\}$ and $\alphaini=\alphafin=1$.
For this, we use the standard expression for factoring an
SU(2) or SO(3)-reduced matrix element of angular momentum zero
(see \eqref{Eq:NewRacah}).
In a form applicable here, this reads
\begin{equation}
\label{Eq:AxB}
\SphericalBraR{\sfin}{\alphafin}{L} |
         [A_{L_1} \otimes B_{L_1} ]_0 
         | \SphericalKetR{\sini}{\alphaini}{L}
= \sum_{\sen',\alpha',L'} (-1)^{L+L_1-L'}
\frac{
\SphericalBraR{\sfin}{\alphafin}{L} |
         A_{L_1} | \SphericalKetR{\sen'}{\alpha'}{L'}
\SphericalBraR{\sen'}{\alpha'}{L'} |
         B_{L_1} | \SphericalKetR{\sini}{\alphaini}{L}
}
{\sqrt{(2L_1+1)(2L+1)}}.
\end{equation}
For
$A_{2}=\hcalQ^\mp$,
$B_{2}=[\hcalQ\otimes\hcalQ ]_2$, 
$\sini=\sen$, $\sfin=\sen\pm1$ and
$L=\text{min}\{2\sini,2\sfin\}$, this yields
\begin{subequations}\label{Eq:QpmxQxQ}
\begin{align}
&
\SphericalBraR{\sen+1,}{1,}{2\sen} |
         [\hcalQ^-\otimes\hcalQ\otimes\hcalQ]_{0}
         | \SphericalKetR{\sen,}{1,}{2\sen}
\nonumber\\
&\hskip20mm
\label{Eq:QmxQxQp}
= \sum_{\alpha',L'}
 (-1)^{L'} \frac{
\SphericalBraR{\sen+1,}{1,}{2\sen} | \hcalQ | \SphericalKetR{\sen+2,}{\alpha',}{L'}
\SphericalBraR{\sen+2,}{\alpha',}{L'} | [\hcalQ\otimes\hcalQ ]_2
    | \SphericalKetR{\sen,}{1,}{2\sen}
}{\sqrt{5(4\sen+1)}}\\
\noalign{\noindent and}
&
\SphericalBraR{\sen-1,}{1,}{2\sen-2} |
         [\hcalQ^+\otimes\hcalQ\otimes\hcalQ]_{0}
         | \SphericalKetR{\sen,}{1,}{2\sen-2}
\nonumber\\
&\hskip20mm
\label{Eq:QpxQxQm}
=\frac{
\SphericalBraR{\sen-1,}{1,}{2\sen-2} | \hcalQ
    | \SphericalKetR{\sen-2,}{1,}{2\sen-4}
\SphericalBraR{\sen-2,}{1,}{2\sen-4} | [\hcalQ\otimes\hcalQ ]_2
    | \SphericalKetR{\sen,}{1,}{2\sen-2}
}{\sqrt{5(4\sen-3)}}.
\end{align}
\end{subequations}
In the second case here, use has been made of the fact that,
in accordance with \eqref{Eq:DimSO5>SO3},
all states of seniority $\sen$ have SO(3) angular momentum at most $2\sen$,
and this angular momentum has multiplicity one.
Thus, because the values of $L'$ being summed over
may be restricted to the range $L-2\le L'\le L+2$,
with $L=2\sen-2$ in this case,
just one state contributes to the sum from \eqref{Eq:AxB}.

Given that the value of \eqref{Eq:QpmxQxQmp} is independent of
$L$, $\alphaini$ and $\alphafin$, we conclude that
\begin{subequations}\label{Eq:QpmxQxQ1}
\begin{align}
&
\SphericalBraR{\sen+1,}{\alphafin}{L} |
         [\hcalQ^-\otimes\hcalQ\otimes\hcalQ]_{0}
         | \SphericalKetR{\sen}{\alphaini}{L}
\nonumber\\
&\hskip20mm
=\frac{1}{4\sen+1}\sqrt{\frac{2L+1}{5}}
\frac{ (\sen\alphaini L,\, 310 || \sen+1,\alphafin L)}
     { (\sen,1,2\sen,\, 310 || \sen+1,1,2\sen)}
\nonumber\\
&\hskip30mm
\label{Eq:QmxQxQp1}
\times \sum_{\alpha',L'}
 (-1)^{L'}
\SphericalBraR{\sen+1,}{1,}{2\sen} | \hcalQ | \SphericalKetR{\sen+2,}{\alpha'}{L'}
\SphericalBraR{\sen+2,}{\alpha'}{L'} | [\hcalQ\otimes\hcalQ ]_2
    | \SphericalKetR{\sen,}{1,}{2\sen}\\
\noalign{\noindent and}
&
\SphericalBraR{\sen-1,}{\alphafin}{L} |
         [\hcalQ^+\otimes\hcalQ\otimes\hcalQ]_{0}
         | \SphericalKetR{\sen}{\alphaini}{L}
\nonumber\\
&\hskip20mm
=\frac{1}{4\sen-3}\sqrt{\frac{2L+1}{5}}
\frac{ (\sen\alphaini L,\, 310 || \sen-1,\alphafin L)}
     { (\sen,1,2\sen-2,\, 310 || \sen-1,1,2\sen-2)}
\nonumber\\
&\hskip30mm
\label{Eq:QpxQxQm1}
\times
\SphericalBraR{\sen-1,}{1,}{2\sen-2} | \hcalQ
    | \SphericalKetR{\sen-2,}{1,}{2\sen-4}
\SphericalBraR{\sen-2,}{1,}{2\sen-4} | [\hcalQ\otimes\hcalQ ]_2
    | \SphericalKetR{\sen,}{1,}{2\sen-2}.
\end{align}
\end{subequations}
 Analytic expressions for the SO(3)-reduced matrix elements
$\StateBraR{\lambda'}{a}{\rmu}{\sen\pm1,}{\alphafin}{L}
| [\hat\pi\otimes\hat q\otimes\hat\pi]_0 |
\StateKetR{\lambda}{a}{\rnu}{\sen}{\alphaini}{L}$
in the cases where $\lambda'-\lambda$ is odd are then obtained
from \eqref{Eq:qpipi1},
making use of \eqref{Eq:Qto3}, \eqref{Eq:QpmxQxQ1} via \eqref{Eq:Qto1}
and \eqref{Eq:Qto2}, and the expressions of Section \ref{Sec:RadMEs}.




\section{Summary of computer implementation}
\label{Sec:Imp}

\noindent
{\em Program Title:} ACM                                      \\
{\em Programming language:} Maple 18 (or versions 17, 16, 15) \\
{\em Operating system:} Any which supports Maple; tested under
Linux, Max OSX, Windows 7              \\
{\em RAM:} $\ge500$Mb                                              \\
{\em Supplementary material (supplied with code):}

\noindent
  1. The code makes use of \fivesupthree\ Clebsch-Gordan coefficients
     which must be installed by the user.
     These are supplied in three zip files:
     \texttt{so5cg-data13.zip},
     \texttt{so5cg-data24.zip} and
     \texttt{so5cg-data56.zip}.

\noindent
  2. A Maple worksheet \texttt{acm-examples.mw} that gives various
     example calculations and tests carried out using procedures
     from the code.

\noindent
  3. A 162 page PDF file \texttt{acm-examples.pdf} containing everything
     displayed in the worksheet
     (input, output and comments, and making use of colour).
                                                              \\
{\em Nature of problem:}
  The calculation of energy eigenvalues, transition rates and
  amplitudes of user specified Hamiltonians in the Bohr model
  of the atomic nucleus.
   \\
{\em Solution method:}
  Exploit the model's SU(1,1)$\times$SO(5) dynamical group to
  calculate analytic (as far as possible) expressions for
  matrix elements, making use of extensive files (supplied) of
  \fivesupthree\ Clebsch-Gordan coefficients.
  Diagonalisation of the resulting matrices
  (once the entries are converted to floating point) is carried out
  using the Maple library procedure \texttt{Eigenvectors}.
  (Maple \cite{Maple} makes use of the NAG \cite{NAG}
   and CLAPACK \cite{CLAPACK} linear algebra libraries.)
   \\
{\em Additional comments:}

\noindent
  1.\ The dimension of the Hilbert space that can be handled is
     limited only by the available computer memory and
     the available \fivesupthree\ Clebsch-Gordan coefficients
     $(\sen_1\alpha_1 L_1 \sen_2\alpha_2 L_2 || \sen_3\alpha_3 L_3)$.

\noindent
  2.\ The supplied data files provide coefficients
     $(\sen_1\alpha_1 L_1 \sen_2\alpha_2 L_2 || \sen_3\alpha_3 L_3)$
     for $1\le\sen_2\le6$,
     and contain all non-zero coefficients for
     $\sen_1<\sen_3\le50$ when $\sen_2\in\{1,3\}$,
     for $\sen_1\le\sen_3\le30$ when $\sen_2\in\{2,4\}$, and
     for $\sen_1\le\sen_3\le25$ when $\sen_2\in\{5,6\}$.
     (Once calculated, further coefficients can be readily made
     available to the code without changing the code.%
       \footnote{The authors will post such coefficients at
                \url{www.physics.utoronto.ca/~twelsh}.})
     Thus, depending on the model Hamiltonian being analysed,
     the states in the Hilbert space used are limited in their seniority.
     For analysis of the more typical types of model Hamiltonian,
     only the coefficients with $\sen_2\in\{1,3\}$ are required,
     and therefore, with the supplied files, the seniority limit is 50.
     More exotic Hamiltonians having terms with seniority
     $\sen_2\in\{2,4,5,6\}$ would have the seniority limited to
     30 or 25 accordingly.

\noindent
  3.\ The code provides lower level procedures that give ready
     access to the Clebsch-Gordan coefficients and
     the SU(1,1) and SO(5) matrix elements.
     These procedures are described in the manuscript and
     enable extensions to the code and model to be made easily.

\noindent
  4.\ The accuracy to which Maple performs numerical calculations
     is determined by the Maple parameter \texttt{Digits},
     which specifies the number of significant decimal digits used.
     The default value of $10$ is more than adequate for most
     ACM calculations.
     Note, however, that if \texttt{Digits} is increased beyond a
     certain value
     (obtained from the Maple command \texttt{evalhf(Digits)},
      and usually $15$ on modern computers)
     then the code can no longer take advantage of hardware
     mathematical operations, and is significantly slower.
   \\
{\em Running time:}
  For a fixed value of the parameter \texttt{Digits},
  the running time depends on the dimension of the Hilbert space
  on which the diagonalisation is performed,
  and this in turn is governed by the number of eigenvalues
  required and the accuracy required.
  Note that diagonalisation is performed separately in each $L$-space.
  For typical ACM calculations
  (such as those carried out in \cite{RWC09}),
  the matrices being diagonalised are usually of dimension at most
  a few hundred, and often much smaller.
  On a modest personal computer, the computation for the
  smallest cases takes at most a few seconds.
  The worksheet contains a range of examples for which the
  calculation time varies between a few seconds and 750s.
  In the latter case, diagonalisation is performed
  on $L$-spaces for $0\le L\le 8$,
  the dimensions of these spaces being between 154 and 616.
   \\

\end{document}